\documentclass[a4paper,11pt]{article}
\pdfoutput=1 

\usepackage{jcappub} 

\usepackage[T1]{fontenc} 

\usepackage{hyperref}
\usepackage{amsmath}
\usepackage{amssymb}
\usepackage{enumerate}   

\title{When precise inference is wrong: Diagnosing model misspecification with applications to cosmic dawn and reionisation}

\author[a,b,1]{D. Breitman,\note{Corresponding author.}}

\affiliation[a]{University of Tokyo,\\Department of Physics, Graduate School of Science, The University of Tokyo, 7-3-1 Hongo, Bunkyo, Tokyo 133-0033, Japan}
\affiliation[b]{Research Center for the Early Universe,\\Graduate School of Science, The University of Tokyo, 7-3-1 Hongo, Bunkyo, Tokyo 113-0033, Japan}

\emailAdd{breitmandaniela@g.ecc.u-tokyo.ac.jp}

\abstract{
As observations of the cosmic dawn (CD) and epoch of reionisation (EoR) rapidly improve in precision, the approximate forward models used to interpret them might struggle to keep pace. Bayesian inference with a misspecified model, one that cannot represent the processes generating the observations, can yield tightly constrained, yet misleading and biased astrophysics, with different parts of the observations favouring incompatible parameter values. Model misspecification can evade a $\chi^2$ goodness-of-fit test, and Bayesian model selection requires a better model, which is not always available. We therefore apply the suspiciousness, a statistic developed to quantify tensions between datasets, to CD/EoR inference as a model criticism tool that tests whether these parts can be explained by the same parameters. We first apply it to HERA 21-cm power spectrum (PS) upper limits, which are too weak to reveal misspecification, as expected, and to HST and JWST UV luminosity functions (LFs). Under our galaxy model, the UV LFs at $z\leq10$ and $z>10$ are in tension at $5.0\sigma$ in the Gaussian limit, a failure visible in the fits, although doubling the uncertainties of the $z>10$ observations reduces it to $2.1\sigma$. Finally, we analyse a mock SKA-Low AA$^\ast$ 1080\,h 21-cm PS with the model that generated it and with a simpler, misspecified model. We find that the misspecified model silently fails: it fits the PS well, passing visual inspection and, for most noise realisations, a $\chi^2$ test, yet its inferred astrophysics is wrong. The Bayesian evidence disfavours it, but only relative to the true model, which is unavailable for real observations. The suspiciousness finds no tension for the well-specified model, but a $\sim3.4\sigma$ tension for the misspecified one, strongly ruling it out \textit{without requiring a more adequate candidate model}.
}

\begin{document}
\maketitle
\flushbottom

\section{Introduction}
\label{sec:intro}

The cosmic dawn (CD), when the first stars and galaxies formed, and the epoch of reionisation (EoR), when their radiation transformed the intergalactic medium (IGM) from neutral to ionised, are now probed by a growing set of increasingly precise observations. The abundance of galaxies is measured beyond $z=10$ with HST and JWST \citep[e.g.][]{Bouwens15,Harikane23}; quasar damping wings \citep[e.g.][]{Greig19,Greig22} and the Ly$\alpha$ forest \citep[e.g.][]{Bosman22,Qin25}, from ground-based spectroscopy of high-redshift quasars, constrain the late stages of reionisation; cosmic microwave background (CMB) polarisation from Planck constrains its integrated history \citep[e.g.][]{Planck18,Qin20}; and upper limits on the 21-cm power spectrum (PS) from radio interferometers continue to tighten. The Square Kilometre Array (SKA)~\citep{Koopmans15} is expected to detect this signal and map it tomographically. As observational precision improves, approximations in the models that were adequate for earlier data may no longer be.

Bayesian inference is central to interpreting these observations. Analyses of 21-cm PS upper limits from the Hydrogen Epoch of Reionisation Array (HERA)~\citep{DeBoer17}, LOFAR \citep{Haarlem13}, and MWA \citep{Tingay13} experiments, for example, have already constrained the properties of the first galaxies and the IGM \citep{HERA22a,HERA23,Mertens25,Nunhokee25}. Similar analyses are also used to forecast the constraining power of upcoming SKA observations \citep[e.g.][]{Greig15,Breitman26}. In both applications, a forward-modelling pipeline maps astrophysical parameters to simulated observables, which are compared with real or mock observations through a likelihood. Combining this likelihood with parameter priors yields a posterior distribution quantifying the parameter values supported by the observations and their associated uncertainties. However, forecasts typically use mock observations generated and analysed with the same physical model. 
Successful parameter recovery in this idealised setting does not establish whether the inferred astrophysics remains reliable when the processes generating the observations cannot be represented by the assumed model---a situation known as \textit{model misspecification}.
Crucially, such a mismatch can lead to incorrect scientific interpretation, even when the inferred parameters are tightly constrained. Recognising whether the inferred astrophysics reflects limitations of the model, rather than the physical processes that generated the observations, motivates our study of model misspecification and the assessment of \textit{model adequacy}.

Bayesian parameter estimation does not, by itself, establish whether the assumed model family adequately describes the observations. Here, the \textit{assumed model family} comprises the probability distributions for the observations obtained by varying the allowed parameters within a fixed modelling pipeline. This family depends not only on the adopted physical description, but also on the numerical methods, instrument modelling, and likelihood used to relate predictions to observations. A model is \textit{well specified} when the true data-generating distribution lies within this family and \textit{misspecified} when it lies outside it \citep{Kleijn12,Walker13}. Bayesian parameter inference answers the question “What parameter values are supported by the observations \textit{within the assumed model family}?”, not “Does this model family \textit{adequately describe} the observations?”. The established ways of comparing models with observations answer the second question only in part. Bayesian model selection compares the evidence of competing model families, and frequentist analyses compare their best-fitting $\chi^2$: both rank the candidates, but require an alternative model and do not establish that the preferred one is adequate. A $\chi^2$ goodness-of-fit test does assess a single model, but only through its misfit at the best-fitting parameters. A flexible misspecified model can absorb much of this misfit with biased parameter values, and the remaining misfit, spread over many data points, can be difficult to distinguish from noise. 

In many areas of physics, inference is naturally performed in an \textit{open-model} setting: forward models are increasingly sophisticated approximations of reality rather than competing hypotheses expected to contain the literal truth \citep{BernardoSmith94,VehtariOjanen12}. Under model misspecification, the posterior still describes which parameter values are supported within the assumed model family, but those values need not yield a faithful interpretation of the physical processes that generated the observations.

In CD/EoR inference, the \textit{assumed family} is determined by cosmology and the adopted descriptions of galaxies and their effects on the IGM, together with the rest of the inference pipeline. Adjusting the available parameters may compensate for limitations in those descriptions. For example, a galaxy model without stochastic star formation may need different star formation efficiencies to reproduce the UV luminosity functions above and below $z=10$ (Section~\ref{sec:UVLFs}), and a model without molecularly-cooling galaxies may compensate for their missing X-ray emission with overly X-ray-luminous atomically-cooling galaxies (Section~\ref{sec:ska}). Simplifications elsewhere in the pipeline, such as in the halo mass function \citep{Greig24a}, the numerical algorithm \citep{Berklas25}, or the likelihood \citep{Prelogovic23}, can similarly bias the inferred parameters. Consequently, an apparently satisfactory fit to the measured observable does not imply a faithful interpretation of galaxy properties or IGM histories. We could mistake the model’s compensating behaviour for evidence about underlying astrophysics and cosmology, potentially misdirecting subsequent theoretical development. Model misspecification is increasingly recognised as a problem for 21-cm inference \citep[e.g.][]{Prelogovic23,Cooper25,Pietschke26}. Together, greater observational sensitivity and increasingly efficient inference pipelines create both the need and the opportunity to make model criticism a standard component of CD/EoR Bayesian workflows.

Assessing whether an assumed model family is adequate is the goal of Bayesian \textit{model criticism} \citep[e.g.][]{Box80,GelmanShalizi13,Gelman20}. Here, adequacy is judged relative to the available observations and the scientific conclusions being drawn, rather than by literal correctness. For real observations, this assessment is challenging because the true data-generating process is not available as a benchmark. Moreover, noisy observations may differ from the predictions of an adequate model simply because of statistical fluctuations. Diagnostics must therefore assess whether these discrepancies are consistent with the variation expected under the assumed model or provide evidence of inadequacy. Unlike model selection, model criticism does not require an explicit alternative model: it can identify reasons to question the assumed modelling setup without specifying what should replace it \citep[see also][]{Starkman08}, potentially using additional observables to help identify the source of the discrepancy \citep{Gelman20}. Conversely, finding no evidence of inadequacy does not establish that the model is correct. A misspecified model may have sufficient flexibility to accommodate the available observations without producing detectable discrepancies.

In this work, we assess model adequacy through Bayesian \textit{internal} consistency diagnostics, asking: \textit{can different parts of the observations be explained by the same physical model, with mutually consistent posteriors?} These parts may correspond to different redshift ranges, spatial scales, or observables. We analyse the corresponding likelihood contributions separately and jointly, and quantify the tension between the resulting posteriors relative to that expected under the assumed model. Model misspecification can cause different contributions to favour incompatible parameter values, even when the joint posterior appears tightly constrained.

This approach builds on tension metrics developed in cosmology, where combining datasets under the same model raises the same question. The suspiciousness, which we adopt here, was introduced to quantify the tension of Planck CMB data with Dark Energy Survey weak lensing and galaxy clustering, and with the distance-ladder measurement of the Hubble constant \citep{HandleyLemos19}. It has since been compared with other tension metrics, including ones based on the loss of goodness of fit \citep{Lemos21}, applied to the internal consistency of a single survey \citep{Lemos20} and to grids of cosmological datasets and models \citep{Ong25}, and used to test whether DESI baryon acoustic oscillation, CMB, and supernova data can be combined, a question central to the reported preference for dynamical dark energy \citep[e.g.][]{YeLin25,Ong25b,Ong26}. There, a tension signals unmodelled systematics or physics beyond $\Lambda$CDM, both forms of misspecification; here, we use it to test the approximate astrophysical models of CD/EoR inference. Our approach also complements recent assessments of model misspecification using scale-dependent Bayesian evidence \citep{Akhmetzhanova25}. Some machine-learning approaches train models to generalise across, or to ignore, differences between simulators \citep[e.g.][]{Jo25,Breitman25}. We instead aim to detect misspecification, treating inconsistencies between parts of the data as evidence that the assumed model may be inadequate \citep[see][for a review of both strategies in physics]{CruzMartinez26}.

We implement this assessment using the \texttt{21cmEMU} emulator suite \citep{Breitman24,Breitman26}, which enables hour-scale parameter inferences and makes the required repeated analyses feasible. Alongside applications to existing observations, we evaluate the diagnostic in a controlled experiment using a \texttt{21cmEMUv3} mock SKA 21-cm PS observation. We analyse the same mock using both \texttt{21cmEMUv3} and the simpler \texttt{21cmEMUv1} model, providing well-specified and misspecified cases, respectively. This comparison demonstrates how internal inconsistency can reveal misspecification even when the inferred model provides an apparently satisfactory fit to the observed PS. Knowledge of the generating model allows us to evaluate the performance of the diagnostic in this controlled setting.

The paper is organised as follows. In Section~\ref{sec:inf}, we introduce \texttt{21cmEMU}, the inference pipeline, and the internal consistency diagnostic. In Section~\ref{sec:HERA}, we apply the diagnostic to current HERA 21-cm PS upper limits. In Section~\ref{sec:UVLFs}, we examine the consistency between the highest-redshift JWST UV luminosity functions and lower-redshift measurements. Section~\ref{sec:ska} presents the controlled mock-SKA analysis, and we conclude in Section~\ref{sec:conclusion}.

\section{Forward model}
\label{sec:forwardmodel}

The forward models in this work are based on \texttt{21cmFAST} \citep[e.g.][]{Mesinger07, Mesinger11, Murray20, Davies25}, an efficient semi-numerical simulator of galaxies and their radiation fields during the CD/EoR. Simulations begin by sampling initial density and velocity fields according to the input cosmological parameters. These fields are evolved gravitationally to lower redshifts using second-order Lagrangian perturbation theory (2LPT)~\citep{Scoccimarro98}. Galaxy properties are then assigned through semi-empirical relations connecting them to their host dark matter haloes, with these relations controlled by the input astrophysical parameters \citep{Park19}. The resulting galaxy populations generate radiation fields that heat and ionise the IGM. The simulated three-dimensional fields are used to construct lightcones and can be used to calculate lower-dimensional summary statistics corresponding to observables such as the 21-cm PS.

Even an hour-scale simulation becomes computationally prohibitive when an inference requires $\sim 10^5$--$10^6$ forward-model evaluations \citep[e.g.][]{HERA22a,Breitman24}. Machine-learning methods, including emulation and simulation-based inference~\citep[e.g.][]{Cranmer19}, offer ways to improve inference as well as reduce its computational cost. In the emulation approach adopted here, simulations are first used to construct a training database, on which a neural network is trained to approximate the mapping from input parameters to simulated summaries. Once trained, the emulator can replace the simulator in subsequent analyses, amortising the initial computational investment across multiple inferences \citep{Breitman24,Breitman26}. Here, we use the \texttt{21cmEMU} suite to perform these repeated analyses under different physical models, making our model-criticism assessment computationally feasible.

\subsection{\texttt{21cmEMU}} 

We use two versions of \texttt{21cmEMU}. The third and most recent, \texttt{21cmEMUv3} \citep{Breitman26}, emulates the \texttt{21cmFASTv3.0.4} \citep{Mesinger07,Mesinger11,Murray20} simulator, with a galaxy model comprising two distinct populations: (i) atomically-cooling galaxies (ACGs) hosting Pop II stars, and (ii) molecularly-cooling galaxies (MCGs), expected to host primarily Pop III stars and their remnants. The model allows these populations to have different stellar and radiative properties, with both contributing to the radiation fields that heat and ionise the IGM. The physical prescriptions are described in detail in Qin et al.\ (2020)~\cite{Qin20a} and Mu\~noz et al.\ (2022)~\cite{Munoz22}, and summarised in Breitman et al.\ (2026)~\cite{Breitman26}.

Table~\ref{tab:paramsv3} summarises the eleven \texttt{21cmEMUv3} input parameters: ten astrophysical parameters and the cosmological parameter $\sigma_8$. The stellar-to-halo mass relations (SHMRs) are controlled by the normalisations $f_{\ast,10}$ and $f_{\ast,7}$ for ACGs and MCGs, respectively, and their common power-law index $\alpha_{\ast}$. The subscripts indicate the reference halo masses of $10^{10}$ and $10^7\,M_\odot$ for ACGs and MCGs, respectively. The characteristic star-formation timescale, $t_{\ast}$, expressed in units of the Hubble time, relates stellar mass to star-formation rate (SFR). Similarly, $f_{\rm esc,10}$ and $f_{\rm esc,7}$ set the ionising escape fractions at these reference masses, with their halo-mass dependence controlled by the common index $\alpha_{\rm esc}$. The soft-band X-ray luminosities per unit SFR, $L_{\rm X<2keV}^{\textsc{acg}}/{\rm SFR}$ and $L_{\rm X<2keV}^{\textsc{mcg}}/{\rm SFR}$, determine the X-ray output of each population, while $E_0$ sets the minimum energy of X-ray photons that escape their host galaxy. Finally, $\sigma_8$ controls the normalisation of the matter power spectrum. The table also lists the upper and lower bounds for each parameter in the emulator training database.

\begin{table*}[ht]
\begin{tabular}{lll}
Parameter            & Training range & Description \\ \hline
$\log_{10} f_{*,10}$       & $\in [-2, 1]$     & SHMR normalisation for ACGs          \\
$\alpha_{\ast}$   & $\in[0,1.2]$   & power-law index of the SHMR       \\
$t_{\ast}$         & $\in[0.01,1]$    & characteristic star-formation timescale         \\
$\log_{10} f_{\rm esc, 10}$     & $\in[-3,0]$   & normalisation of the $f_{\rm esc}-\rm{M}_h$ relation for ACGs         \\
$\alpha_{\rm esc}$     & $\in[-1,1]$       & power-law index of the $f_{\rm esc}-\rm{M}_h$ relation          \\
$\log_{10} f_{\ast,7}$ &  $\in[-4,-1]$      & SHMR normalisation for MCGs   \\
$\log_{10} f_{\rm esc, 7}$ & $\in[-3,-1]$& normalisation of the $f_{\rm esc}-\rm{M}_h$ relation for MCGs \\
$\log_{10}\!\left(L_{\rm X<2keV}^{\textsc{acg}}/{\rm SFR}\right)$ & $\in[38,43]$ & soft-band X-ray luminosity per unit SFR for ACGs\\
$\log_{10}\!\left(L_{\rm X<2keV}^{\textsc{mcg}}/{\rm SFR}\right)$ & $\in[39,44]$ & soft-band X-ray luminosity per unit SFR for MCGs \\
$E_{\rm 0}$ & $\in[100,1500]$ eV & minimum energy of X-ray photons that can escape their host galaxy\\
$\sigma_8$ & $\in[0.75,0.85]$& matter power spectrum normalisation\\
\end{tabular}
\caption{\label{tab:paramsv3}
Input parameters of \texttt{21cmEMUv3}: ten astrophysical parameters and the cosmological parameter $\sigma_8$. The second column lists the parameter-wise bounds of the training database \citep{Breitman26} to show training coverage. Note that the luminosity-to-SFR ratios are expressed in units of $\mathrm{erg}\,\mathrm{s}^{-1}(M_\odot\,\mathrm{yr}^{-1})^{-1}$ before taking their base-10 logarithms.}
\end{table*}

The first version, \texttt{21cmEMUv1}, in contrast, emulates a \texttt{21cmFAST} model containing only ACGs, with nine astrophysical inputs and fixed cosmological parameters \citep{Breitman24}. Its inputs comprise the seven parameters describing ACGs in Table~\ref{tab:paramsv3}, together with two additional parameters: (i) $M_{\rm turn}$, the characteristic halo mass below which haloes become exponentially less likely to host a star-forming galaxy, and (ii) $\alpha_X$, the power-law index of the X-ray spectral energy distribution. Note that the training bounds in Table~\ref{tab:paramsv3} apply only to \texttt{21cmEMUv3}; the corresponding ranges for \texttt{21cmEMUv1} are described in Breitman et al.\ (2024)~\cite{Breitman24}. All \texttt{21cmEMUv1} parameters are listed, together with the priors we adopt for both emulators, in Table~\ref{tab:priors} (Appendix~\ref{sec:priors}).

For their respective input parameters, the two emulators predict the summary statistics listed below. Both achieve median emulation errors at the $\lesssim1\%$ level under the summary-specific validation metrics reported in Breitman et al.\ (2024, 2026)~\cite{Breitman24,Breitman26}. Unless otherwise stated, each summary is available in both versions.

\begin{enumerate}[(i)]

    \item $\boldsymbol{\overline{x}_{\rm HI}(z)}$ --- the volume-averaged neutral hydrogen fraction as a function of redshift, describing the reionisation history.

    \item $\boldsymbol{\overline{T}_{\rm S}(z)}$ --- the mean spin temperature of the neutral IGM as a function of redshift. In both emulator versions, the volume average is restricted to simulation cells with $x_{\rm HI}\geq0.95$.

    \item $\boldsymbol{\overline{\delta T}_{\rm b}(z)}$ --- the global 21-cm brightness temperature, defined as the volume average of the local differential brightness temperature, $\overline{\delta T}_{\rm b}(z) =\langle\delta T_{\rm b}(\boldsymbol{x},z)\rangle_V$, where the local field is \citep[e.g.][]{Madau97,Furlanetto06,Pritchard12}:
    \begin{align}
        \delta T_{\rm b}(\boldsymbol{x},z)
        &= \frac{T_{\rm S}-T_{\rm R}}{1+z}
           \left(1-e^{-\tau_{21}}\right)
           \label{eq:brightness_temperature}\\
        &\approx 27\,x_{\rm HI}(1+\delta_b)
           \left(\frac{\Omega_b h^2}{0.023}\right)
           \left(
               \frac{0.15}{\Omega_m h^2}\frac{1+z}{10}
           \right)^{1/2}
           \nonumber\\
        &\quad\times
           \left(1-\frac{T_{\rm R}}{T_{\rm S}}\right)
           \left[
               \frac{H(z)/(1+z)}
                    {\partial v_\parallel/\partial r_\parallel}
           \right]\,
           \mathrm{mK}.
           \nonumber
    \end{align}
    Here, $\tau_{21}$ is the 21-cm optical depth, $\delta_b\equiv\rho_b/\overline{\rho}_b-1$ is the baryon overdensity, and $T_{\rm S}$ and $T_{\rm R}$ are the spin and background radiation temperatures, respectively. We assume that the radio background is provided by the CMB, so $T_{\rm R}=T_{\rm CMB}$. The velocity $v_\parallel$ includes both the Hubble flow and peculiar velocity, while $r_\parallel$ is the comoving line-of-sight coordinate. Note that the approximate expression on the second line applies in the optically thin limit, $\tau_{21}\ll1$, and is shown just to illustrate the physical dependencies of the signal.

    \item $\boldsymbol{\Delta^2_{21}(k,z)}$ --- the spherically-averaged 21-cm power spectrum as a function of wavemode $k$ and redshift. We define the three-dimensional power spectrum through
    \begin{align}
        &\left\langle
            \widetilde{T}_{\rm b}(\boldsymbol{k},z)
            \widetilde{T}_{\rm b}^{\ast}(\boldsymbol{k}',z)
        \right\rangle
        \nonumber\\
        &\qquad =
        (2\pi)^3\delta_{\rm D}^{(3)}
        (\boldsymbol{k}-\boldsymbol{k}')
        P_{21}(\boldsymbol{k},z),
        \label{eq:power_spectrum}
    \end{align}
    where $\widetilde{T}_{\rm b}$ is the Fourier transform of the brightness-temperature fluctuations about their spatial mean, and $\delta_{\rm D}^{(3)}$ is the three-dimensional Dirac delta function. The spherically-averaged summary is
    \begin{equation}
        \Delta^2_{21}(k,z)
        = \frac{k^3}{2\pi^2}P_{21}(k,z),
    \end{equation}
    where $P_{21}(k,z)$ is the spherical average of $P_{21}(\boldsymbol{k},z)$ and $k=|\boldsymbol{k}|$. We express $k$ in $\mathrm{Mpc}^{-1}$ and $\Delta^2_{21}$ in $\mathrm{mK}^2$. Both emulator versions predict this summary directly.

    \item $\boldsymbol{\Delta^2_{21}(k_\perp,k_\parallel,z)}$ --- the cylindrically-averaged 21-cm power spectrum, which retains separate dependence on transverse wavemode $k_\perp$ and line-of-sight wavemode $k_\parallel$ and is emulated in \texttt{21cmEMUv3} only:
    \begin{equation}
        \Delta^2_{21}(k_\perp,k_\parallel,z)
        = \frac{k^3}{2\pi^2}
          P_{21}(k_\perp,k_\parallel,z),
    \end{equation}
    where $k=(k_\perp^2+k_\parallel^2)^{1/2}$ and $P_{21}(k_\perp,k_\parallel,z)$ is the cylindrical average of the three-dimensional power spectrum.
    Note that this summary is not used in this work.

    \item $\boldsymbol{\phi(M_{\rm UV},z)}$ --- the non-ionising ($h\nu<13.6\,\mathrm{eV}$) UV luminosity function, defined as the number density of galaxies per unit absolute UV magnitude at rest-frame wavelength $\simeq1500\,\mathrm{\AA}$. The UV luminosity is assumed to be proportional to the SFR:
    \begin{equation}
        L_{\rm UV}
        = \frac{\dot{M}_{\ast}(M_h,z)}{\mathcal{K}_{\rm UV}},
    \end{equation}
    where $\mathcal{K}_{\rm UV} =1.15\times10^{-28}\, M_\odot\,\mathrm{yr}^{-1} (\mathrm{erg}\,\mathrm{s}^{-1}\,\mathrm{Hz}^{-1})^{-1}$ assumes a Salpeter initial mass function \citep[e.g.][]{Madau14,Sun16}. The luminosity is related to the absolute AB magnitude by:
    \citep{Oke83}
    \begin{equation}
        \log_{10}\!\left[
            \frac{L_{\rm UV}}
                 {\mathrm{erg}\,\mathrm{s}^{-1}\,\mathrm{Hz}^{-1}}
        \right]
        =0.4\left(51.63-M_{\rm UV}\right).
    \end{equation}

    \item $\boldsymbol{\tau_e}$ --- the Thomson scattering optical depth of CMB photons, calculated from the adopted ionisation history:
    \begin{equation}
        \tau_e
        = \sigma_{\rm T}
          \int_0^{z_{\rm LSS}}
          c\,n_e(z)
          \left|\frac{\mathrm{d}t}{\mathrm{d}z}\right|
          \mathrm{d}z .
    \end{equation}
    Here, $\sigma_{\rm T}$ is the Thomson scattering cross section, $z_{\rm LSS}\simeq1100$ is the redshift of last scattering, and $n_e(z)$ is the proper free-electron number density. Hydrogen and helium are assumed to be singly ionised at a fraction $1-\overline{x}_{\rm HI}$, with helium doubly ionised at $z<3$.

\end{enumerate}

\section{Inference pipeline}
\label{sec:inf}

Inference begins by sampling a parameter vector $\tilde{\boldsymbol{\theta}}$ from the prior distribution $p(\boldsymbol{\theta}\mid\mathcal{M})$ specified for the chosen model, $\mathcal{M}$. The sampled parameter vector is passed to \texttt{21cmEMU}, which predicts observables such as those listed in the previous section. These predictions are compared with the corresponding observations $d$ through the likelihood $\mathcal{L}(d\mid\tilde{\boldsymbol{\theta}},\mathcal{M})$. Bayes' theorem then combines the prior and likelihood to give the posterior distribution,
\begin{equation}
    p(\boldsymbol{\theta}\mid d,\mathcal{M}) =
    \frac{\mathcal{L}(d\mid\boldsymbol{\theta},\mathcal{M})
        p(\boldsymbol{\theta}\mid\mathcal{M})}{\mathcal{Z}},
    \label{eq:bayes}
\end{equation}
which describes the parameter values supported by the observations within the assumed model family. The normalising constant, $\mathcal{Z}\equiv p(d\mid\mathcal{M})$, is the Bayesian evidence of the model:
\begin{equation}
    \mathcal{Z} = \int \mathcal{L}(d\mid\boldsymbol{\theta},\mathcal{M}) p(\boldsymbol{\theta}\mid\mathcal{M})\, \mathrm{d}\boldsymbol{\theta}.
    \label{eq:evidence}
\end{equation}
The evidence is therefore the likelihood averaged over the parameter prior, rather than evaluated only at the preferred parameter values. For two models, the ratio of their evidences, $\mathcal{Z}_1/\mathcal{Z}_2=p(d\mid\mathcal{M}_1)/p(d\mid\mathcal{M}_2)$, is the Bayes factor used in Bayesian model selection (Section~\ref{sec:intro}): it ranks the two models, but does not establish that either is adequate. Every quantity below refers to a single model, so we omit $\mathcal{M}$ from here on.

Every inference with a given emulator uses the same parameter prior, listed in Table~\ref{tab:priors} in Appendix~\ref{sec:priors}. Choosing priors for the MCG parameters is particularly challenging, since MCGs are difficult to observe directly, leaving their properties largely unconstrained. For \texttt{21cmEMUv3}, we therefore adopt the informed prior of Breitman et al.\ (2026)~\cite[Section~5.1]{Breitman26}, based on the Renaissance simulations \citep{Xu16}, which model the formation of the first galaxies and their Pop~III stars in detail: a truncated normal prior on $\log_{10}f_{\ast,7}$, together with truncated normal priors on $\log_{10}f_{\ast,10}$ and $\alpha_\ast$. All other parameters have flat priors, as do all \texttt{21cmEMUv1} parameters. Where the observations constrain a parameter only weakly, its posterior follows this prior; we point out such cases below.

Here we estimate the posterior and evidence using \texttt{nautilus} \citep{Lange23}, an importance nested-sampling algorithm that uses neural networks to construct efficient sampling proposals. The sampler repeatedly proposes parameter vectors and evaluates their likelihoods, producing weighted posterior samples and an estimate of the evidence.

\subsection{Internal consistency diagnostic}
\label{sec:model_criticism}

\begin{figure}[tbp]
    \centering
    \includegraphics[width=\linewidth]{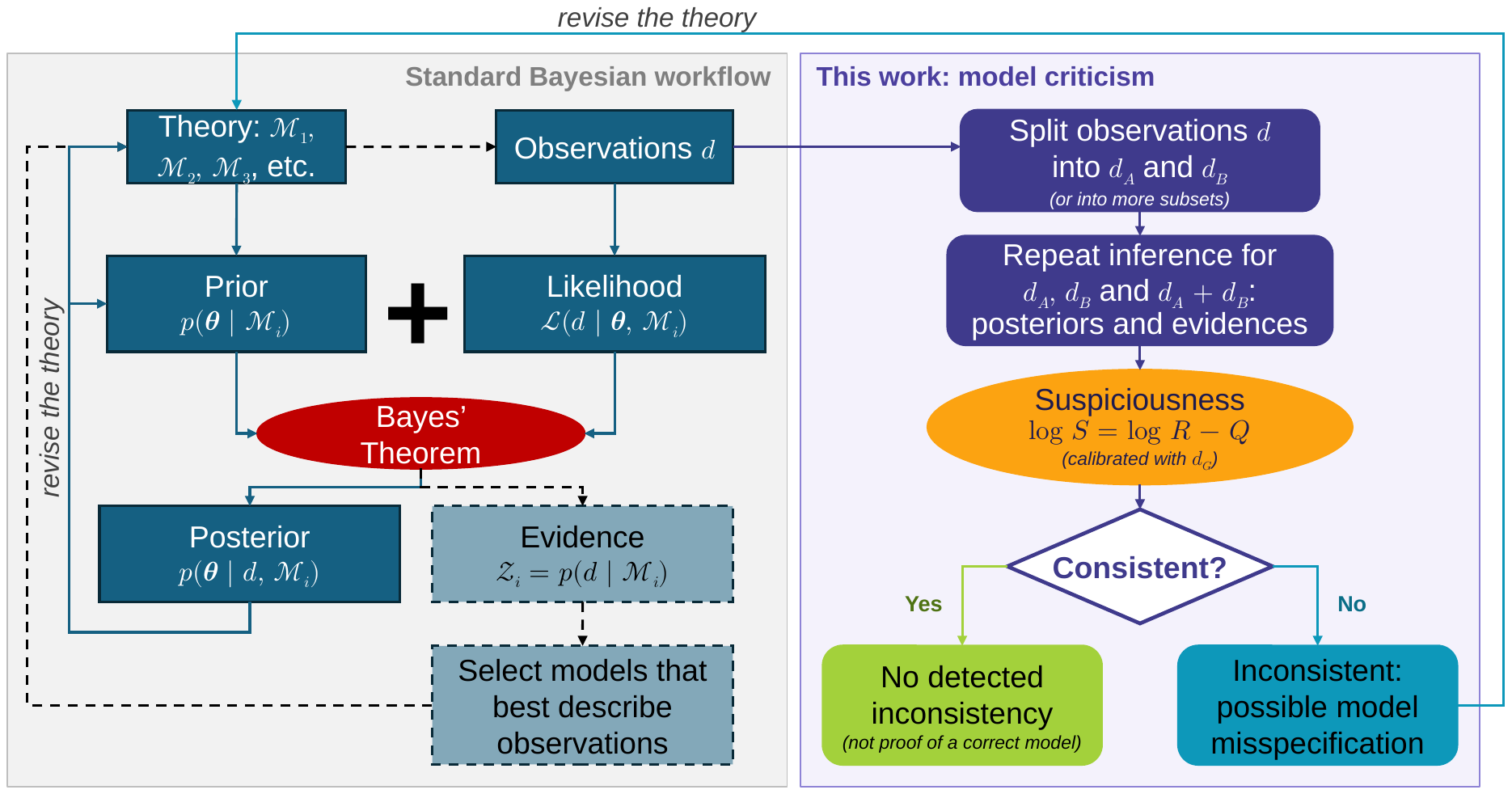}
    \caption{Bayesian inference workflow, extended with the model-criticism stage introduced in this work. \textit{Left:} the standard workflow. For each model $\mathcal{M}_i$, Bayes' theorem combines the prior and likelihood into the posterior, which constrains the model parameters, and the evidence, which is used to compare and select models. \textit{Right:} model criticism. The observations are split into subsets, shown here for the simplest case of two, $d_A$ and $d_B$, and the inference is repeated on each subset and on their combination. The suspiciousness, $\log S=\log R-Q$, and the number of parameter directions constrained by both subsets, $d_G$ (Section~\ref{sec:sus}), quantify whether the model can describe both subsets with the same parameter values. Here, $R$ is the ratio of the evidence of the joint fit to the product of the evidences of the separate fits, and $Q$ removes the dependence of $\log R$ on the prior volume, using the information gained in each fit. An inconsistency indicates possible model misspecification and motivates revising the theory, whereas consistency only means that no inconsistency was detected, not that the model is correct.}
    \label{fig:inference_pipeline}
\end{figure}

Figure~\ref{fig:inference_pipeline} summarises the standard Bayesian workflow described above (left) and the model-criticism stage with which we extend it in this work (right). Rather than constraining parameters within a model, or comparing it with alternative models, this stage tests whether the model can describe different subsets of the observations with the same parameter values. We motivate this test with an illustrative example and then define the statistic that quantifies it.

\subsubsection{An illustrative example}
\label{sec:mm_ex}

\begin{figure}[tbp]
    \centering
    \includegraphics[width=0.49\linewidth]{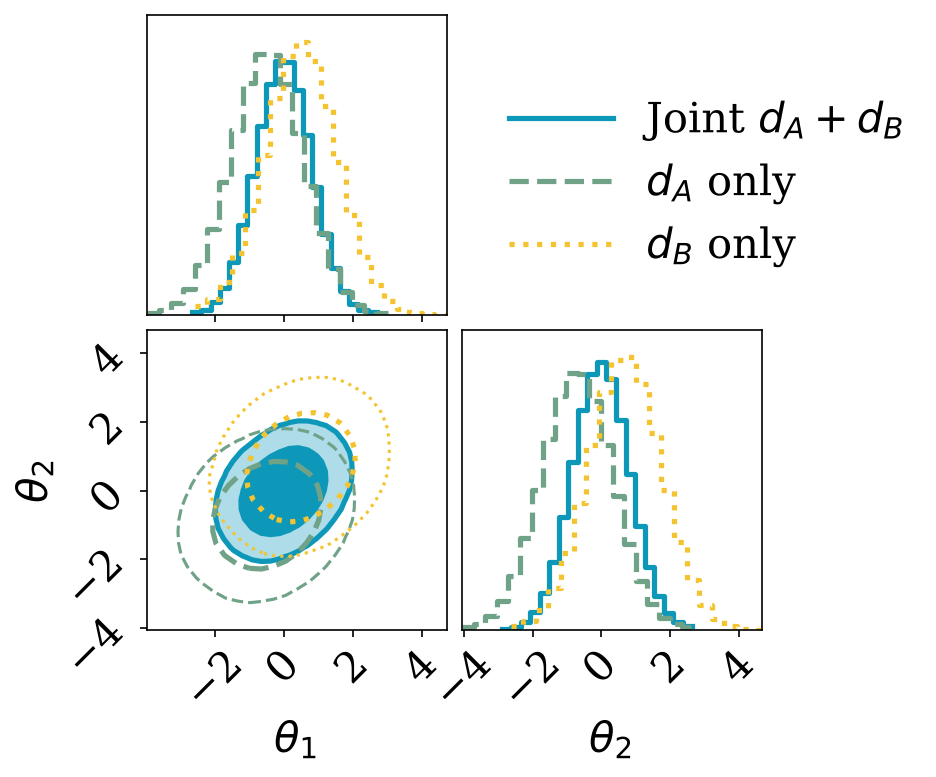}\hfill
    \includegraphics[width=0.49\linewidth]{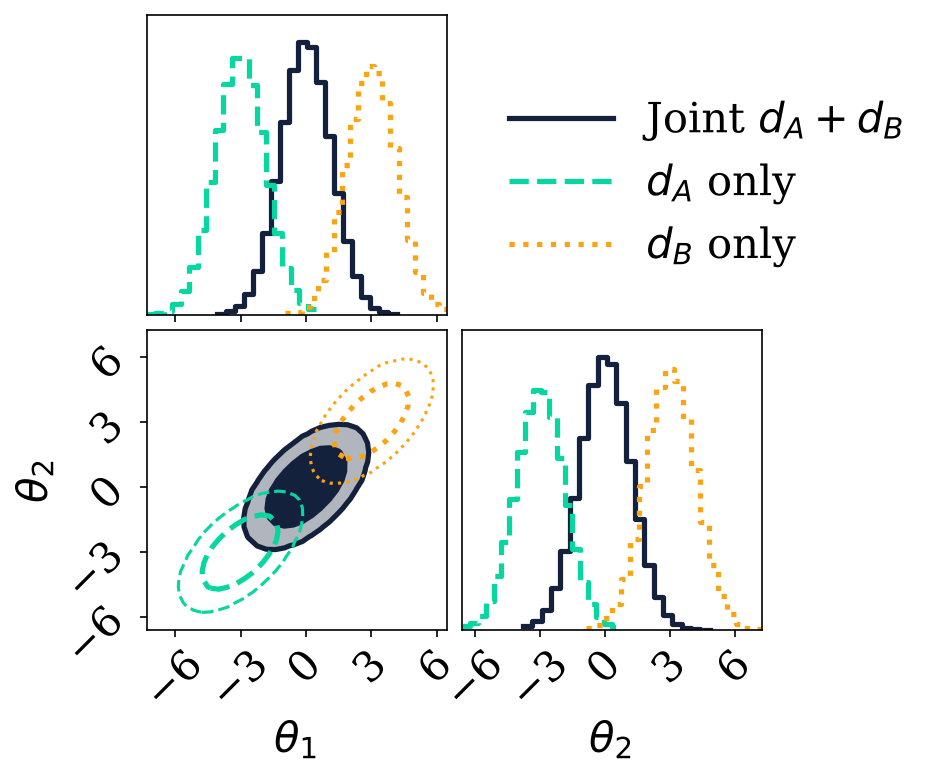}
    \caption{Illustration of the internal consistency comparison for a model with two parameters, $\theta_1$ and $\theta_2$. The posteriors from analysing $d_A$ alone (dashed) and $d_B$ alone (dotted) are shown together with the joint posterior (solid, filled); contours enclose 68\% and 95\% of the posterior mass. \textit{Left:} consistent datasets, whose separate posteriors overlap, so that the joint posterior lies in the region favoured by both. \textit{Right:} inconsistent datasets, whose separate posteriors favour largely disjoint regions; the joint posterior is a compromise between them that neither dataset supports.}
    \label{fig:consistency_illustration}
\end{figure}

Suppose we have a dataset $d_A$, such as a set of 21-cm PS measurements, and a second, statistically independent dataset $d_B$, which could contain measurements at different redshifts or of a complementary observable, such as UV luminosity functions. We interpret both using the same physical model and parameter prior. We first analyse the datasets separately to establish which parameter values each one favours, and then analyse the two together. Figure~\ref{fig:consistency_illustration} illustrates this comparison for a model with two parameters, $\theta_1$ and $\theta_2$, showing the individual posteriors
alongside the posterior from the joint analysis.

In the left panel of Figure~\ref{fig:consistency_illustration}, the green and yellow 68\% and 95\% contours from the $d_A$-only and $d_B$-only inferences significantly overlap in the $(\theta_1,\theta_2)$ plane, and their marginals are fairly similar. The joint posterior, shown in blue, is then concentrated within the region favoured by both datasets: combining them refines a common interpretation. Conversely, in the right panel of Figure~\ref{fig:consistency_illustration}, the individual contours favour largely disjoint regions of parameter space. In this example, the joint posterior, shown in dark blue, is concentrated in the small overlap between the regions preferred by the individual datasets, a region that neither dataset strongly favours.

Viewed alone, the dark-blue posterior from the right panel could appear to provide a well-determined answer when in reality it is a compromise, rather than a parameter region strongly supported by both datasets. The example in Figure~\ref{fig:consistency_illustration} therefore illustrates that a consistent joint inference requires the model to account for all datasets simultaneously with the same parameter values.

The disagreement observed in the right panel may arise when the model is misspecified, that is, when it cannot represent the processes generating the observations: parameter adjustments that compensate for its limitations in one dataset may not provide a satisfactory description of the other. The apparent precision of the joint posterior does not resolve this conflict.

These examples motivate our internal consistency test, which asks whether different parts of the observations can be explained by the same physical model, with mutually consistent posteriors. The requirement is not that the separate posteriors be identical. Even under an adequate model, statistical fluctuations can shift the parameter values favoured by each dataset. Visual inspection of the contours alone is therefore insufficient to establish inconsistency. We need a quantitative comparison of how well the model explains each dataset separately and how well it explains both simultaneously \citep{HandleyLemos19,Ong25}. We describe the test for two subsets, the simplest case. The observations can also be divided into more subsets, for example by redshift or by observable, and tested in the same way.

\subsubsection{Evidence ratio and suspiciousness}
\label{sec:sus}

The distinction between the two panels of Figure~\ref{fig:consistency_illustration} is not whether the joint posterior is narrow, but whether the individual datasets support the answer it gives. To quantify their agreement, we use the Bayesian evidence, which averages the likelihood over the parameter prior. The evidence ratio compares the joint analysis with the two separate analyses
\citep{HandleyLemos19,Ong25}:
\begin{equation}
    \log R = \log\mathcal{Z}_{AB} -\log\mathcal{Z}_A -\log\mathcal{Z}_B.
    \label{eq:logR}
\end{equation}
Each evidence is calculated using the same parameter prior,
\begin{equation}
    \mathcal{Z}_X = \int \mathcal{L}_X(\boldsymbol{\theta}) p(\boldsymbol{\theta})\, \mathrm{d}\boldsymbol{\theta}, \qquad X\in\{A,B,AB\},
    \label{eq:subset_evidence}
\end{equation}
where $AB$ denotes the combination of $d_A$ and $d_B$. For datasets that are statistically independent conditional on $\boldsymbol{\theta}$, the joint likelihood is $\mathcal{L}_{AB}=\mathcal{L}_A\mathcal{L}_B$. We denote the corresponding posteriors by $\mathcal{P}_X(\boldsymbol{\theta}) \equiv p(\boldsymbol{\theta}\mid d_X)$.

Since $\mathcal{Z}_X=p(d_X)$ and $\mathcal{Z}_{AB}=p(d_A,d_B)=p(d_B\mid d_A)\,p(d_A)$, the ratio has a useful predictive interpretation: $R=\mathcal{Z}_{AB}/(\mathcal{Z}_A\mathcal{Z}_B)=p(d_B\mid d_A)/p(d_B)$. A positive $\log R$ means that learning from one dataset makes the other more probable under the assumed model; a negative value means that it becomes less probable. This asks whether the parameter constraints supplied by one dataset help the model explain the other, rather than simply whether their contours overlap.

\begin{figure}[tbp]
    \centering
    \includegraphics[width=0.49\linewidth]{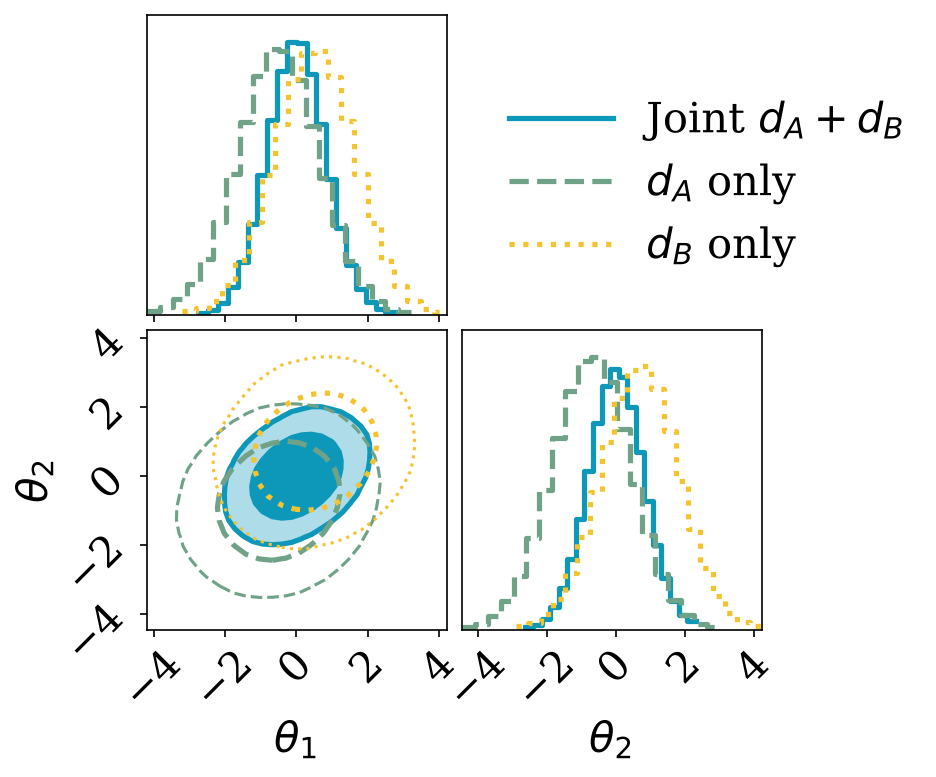}\hfill
    \includegraphics[width=0.49\linewidth]{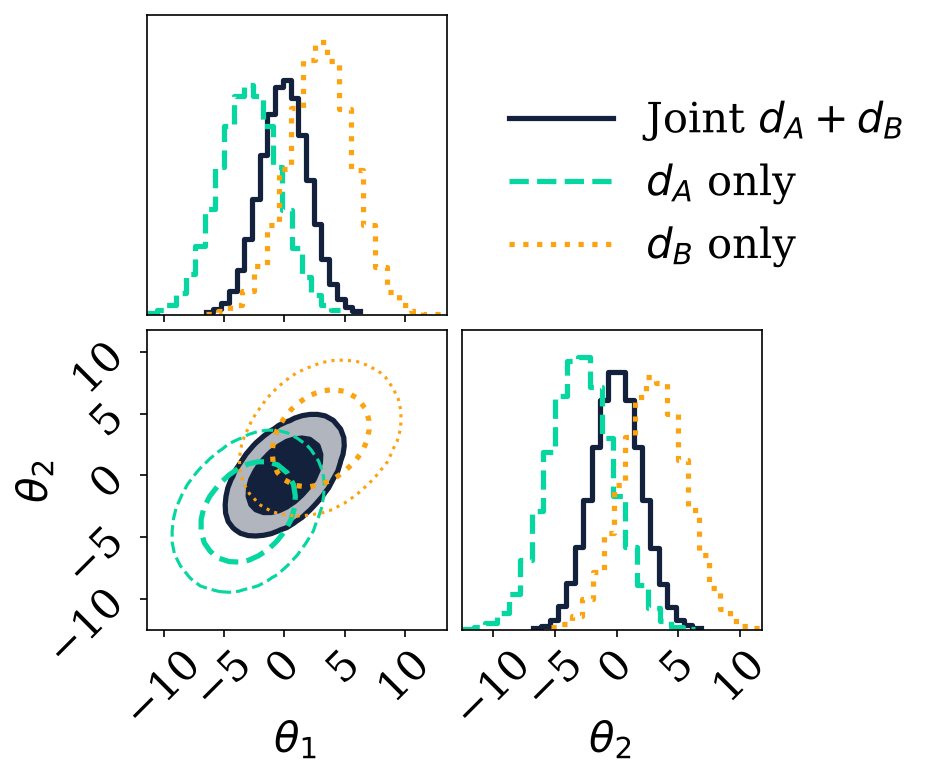}
    \caption{As Figure~\ref{fig:consistency_illustration}, but with a wider parameter prior (note the wider axis range in the right panel). The separate posteriors broaden and overlap more, so the datasets in the right panel appear less discordant, although the means of the separate posteriors are unchanged.}
    \label{fig:wide_prior}
\end{figure}

However, this assessment also depends on the prior. Figure~\ref{fig:wide_prior} repeats the example of Figure~\ref{fig:consistency_illustration} with a wider prior. Where the data do not dominate the posteriors, the separate constraints broaden and overlap more, and the datasets in the right panel appear less discordant, although the means of the separate posteriors are unchanged. The evidence ratio is sensitive to the prior even when the posteriors do not change at all: broadening a uniform prior well beyond the regions supported by the posteriors leaves every contour in place, yet $R$ increases with the prior volume. The datasets can therefore appear more concordant simply because the initially allowed parameter space is larger \citep{HandleyLemos19}. We want to distinguish this prior-volume effect from agreement between the constraints themselves.

To account for it, we quantify the information gained in each analysis through the Kullback--Leibler divergence,
\begin{align}
    \mathcal{D}_X
    &=
    \int
    \mathcal{P}_X(\boldsymbol{\theta})
    \log\!\left[
        \frac{\mathcal{P}_X(\boldsymbol{\theta})}
             {p(\boldsymbol{\theta})}
    \right]
    \mathrm{d}\boldsymbol{\theta}
    \nonumber\\
    &=
    \left\langle
        \log\mathcal{L}_X
    \right\rangle_{\mathcal{P}_X}
    -\log\mathcal{Z}_X,
    \label{eq:KL}
\end{align}
where the angle brackets denote a posterior average. For uniform prior and posterior distributions, this reduces to the logarithm of the ratio of their volumes: it measures the compression from the initially allowed parameter space to the region supported by the observations.
The information correction combines these compressions as
\begin{equation}
    Q = \mathcal{D}_A+\mathcal{D}_B-\mathcal{D}_{AB}.
    \label{eq:information_ratio}
\end{equation}
Subtracting it from the evidence ratio gives the
\textit{suspiciousness} \citep{HandleyLemos19,Ong25},
\begin{equation}
    \log S=\log R-Q.
    \label{eq:suspiciousness}
\end{equation}
Unlike $\log R$, suspiciousness is unchanged when a uniform prior is widened well beyond the region where the likelihood is non-negligible, so that the posteriors themselves do not change. It therefore removes the prior-volume sensitivity illustrated by the thought experiment without claiming independence from every possible choice of prior. When the prior also shapes the posteriors, as in Figure~\ref{fig:wide_prior}, $\log S$ changes as well.

Substituting Equation~\ref{eq:KL} makes the connection to the separate and joint fits explicit:
\begin{align}
    \log S
    &=
    \left\langle
        \log\mathcal{L}_{AB}
    \right\rangle_{\mathcal{P}_{AB}}
    \nonumber\\
    &\quad-
    \left\langle
        \log\mathcal{L}_A
    \right\rangle_{\mathcal{P}_A}
    -
    \left\langle
        \log\mathcal{L}_B
    \right\rangle_{\mathcal{P}_B}.
    \label{eq:suspiciousness_loglike}
\end{align}
Each separate inference can concentrate on the parameter values preferred by its own dataset. The joint inference must instead account for both datasets at once. Suspiciousness compares the posterior-averaged log-likelihoods of these fits, with more negative values indicating a greater loss of fit in the joint analysis. In the right-hand panel of Figure~\ref{fig:consistency_illustration}, this measures the cost of the compromise represented by the dark-blue posterior, rather than its apparent precision.

CD/EoR analyses often combine several complementary observations, so testing the consistency of two datasets need not require discarding the others. For example, we may compare two sets of 21-cm PS measurements while retaining constraints from UV luminosity functions in both analyses. Denoting the common observations by $d_C$, the question becomes whether $d_A$ and $d_B$ are consistent within the parameter space supported by this additional information. We therefore use
\begin{equation}
    \mathcal{P}_C(\boldsymbol{\theta})
    =
    p(\boldsymbol{\theta}\mid d_C)
    =
    \frac{
        \mathcal{L}_C(\boldsymbol{\theta})
        p(\boldsymbol{\theta})
    }{\mathcal{Z}_C}
    \label{eq:conditioning_distribution}
\end{equation}
as the reference prior for the comparison. Conditioning on $d_C$ is therefore only a change of prior: the statistics above apply unchanged, with $\mathcal{P}_C$ in place of $p(\boldsymbol{\theta})$, the posteriors $\mathcal{P}_{XC}=p(\boldsymbol{\theta}\mid d_X,d_C)$ in place of $\mathcal{P}_X$, for $X\in\{A,B,AB\}$, and the likelihoods $\mathcal{L}_X$ of the tested datasets alone. We denote the resulting quantities by $\log R_{AB\mid C}$, $Q_{AB\mid C}$, and $\log S_{AB\mid C}$. In Equation~\ref{eq:suspiciousness_loglike}, for example, the complementary observations then shape the posteriors being compared, while the likelihood terms belong only to the two datasets whose consistency we are testing. In practice, we sample with the original prior, and Appendix~\ref{sec:conditional} expresses the conditional statistics in terms of the evidences and posterior samples of these runs.

A numerical measure of disagreement does not yet tell us how surprising that disagreement is. Even when both datasets are generated by the assumed model, different noise realisations shift the parameter values each favours. The likelihood describes this noise, and the posteriors are correspondingly broad, but a single noise realisation still yields one draw of $\log S$, which scatters around zero from one realisation to the next (Figure~\ref{fig:ska_calibration} shows this scatter for 150 noisy realisations of a well-specified analysis of a mock SKA PS). To interpret a measured $\log S$, we therefore need to determine how often an equally strong disagreement would arise under that model.

An analytic calibration for unconditioned datasets is available in the Gaussian, likelihood-dominated limit \citep{HandleyLemos19,Ong25}; we apply it to the conditional statistic by treating $\mathcal{P}_C$ as the prior. It accounts for the number of parameter directions constrained by the datasets through the Bayesian model dimensionalities \citep{HandleyLemos19b},
\begin{equation}
    d_{X}
    =
    2\,\operatorname{Var}_{\mathcal{P}_{X}}
    \!\left(\log\mathcal{L}_X\right),
    \qquad X\in\{A,B,AB\},
    \label{eq:dimensionality}
\end{equation}
and their combination
\begin{equation}
    d_G
    =
    d_{A}+d_{B}-d_{AB}.
    \label{eq:dG}
\end{equation}
In this limit, $d_G$ represents the effective number of parameter directions constrained by both datasets. With conditioning observations, the variances are taken over $\mathcal{P}_{XC}$ instead (Appendix~\ref{sec:conditional}).

For Gaussian parameter constraints and a sufficiently broad reference prior, the null distribution satisfies
\begin{equation}
    d_G-2\log S
    \sim\chi^2_{d_G}.
    \label{eq:suspiciousness_chi2}
\end{equation}
Under conditioning, the reference prior is $\mathcal{P}_C$, which need not be broad compared with the tested constraints, so this limit should be checked with simulations where possible (Appendix~\ref{sec:calibration}; see Bevins et al.\ (2025)~\cite{Bevins24} for a simulation-based calibration of $\log R$). For $d_G>0$, the probability of obtaining at least as much disagreement as observed is therefore approximated by
\begin{equation}
    p_{\rm tens}
    =
    1-F_{\chi^2_{d_G}}\!\left(
        d_G-2\log S
    \right),
    \label{eq:tension_probability}
\end{equation}
where $F_{\chi^2_{d_G}}$ is the chi-squared cumulative distribution function. Small $p_{\rm tens}$ indicates that the observed disagreement would be unusual under this null distribution. We express it as an equivalent Gaussian significance,
\begin{equation}
    n_\sigma
    =
    \sqrt{2}\,
    \operatorname{erfc}^{-1}(p_{\rm tens}),
    \label{eq:tension_significance}
\end{equation}
using the two-sided Gaussian-tail convention \citep{Ong25}. This calibration plays the role of the Jeffreys scale used to interpret Bayes factors \citep[e.g.][]{Trotta08}, but its thresholds follow from the null distribution and depend on $d_G$: a $3\sigma$ tension requires $\log S\lesssim-4.0$ for $d_G=1$ and $\lesssim-4.9$ for $d_G=2$. A positive $d_G$ is necessary for this conversion but does not, by itself, validate the Gaussian approximation. A non-positive value prevents this conversion and requires an alternative calibration to assign a significance. Non-positive values can arise when the posteriors are non-Gaussian or truncated by the prior, or when an informative $\mathcal{P}_C$ pulls them away from the maximum of $\mathcal{L}_X$; the variance of $\log\mathcal{L}_X$ then no longer counts only constrained parameter directions. In practice, we quote $n_\sigma$ only when $d_G\geq0.5$, and mark smaller positive values as indeterminate (--) in the tables.\footnote{\label{fn:dG}Under the null distribution of Equation~\ref{eq:suspiciousness_chi2}, $\langle\log S\rangle=0$. For $d_G\geq1$, this expected value corresponds to $n_\sigma\leq1$, but as $d_G\to0$ the $\chi^2_{d_G}$ distribution concentrates at zero and the same value falls in its tail: $n_\sigma=1.1$, $1.6$, and $2.3$ for $d_G=0.5$, $0.1$, and $0.01$. Below $d_G\simeq0.5$, Equation~\ref{eq:tension_significance} would therefore assign a significance to negligible disagreements.}

We apply these diagnostics in three settings. Section~\ref{sec:results} analyses current observations: HERA 21-cm PS upper limits (Section~\ref{sec:HERA}), whose one-sided constraints give the test little leverage, and HST and JWST UV LFs (Section~\ref{sec:UVLFs}), for which we expect our pre-JWST galaxy model to fail, as it cannot simultaneously reproduce the HST measurements and the overabundance of UV-bright galaxies reported by JWST \citep[e.g.][]{Harikane23,Finkelstein23,Donnan24}. Section~\ref{sec:ska} analyses a mock SKA 21-cm PS detection, generated with the \texttt{21cmEMUv3} physical model and analysed with both \texttt{21cmEMUv3} and \texttt{21cmEMUv1}, which omits MCGs, to test whether the diagnostic can flag a misspecification that is not apparent from the posterior predictive PS.
\section{Application to current observations}
\label{sec:results}

\subsection{HERA 21-cm PS upper limits}
\label{sec:HERA}

HERA is a zenith-pointing interferometer in South Africa designed to detect the 21-cm PS \citep{DeBoer17}. It has observed in two phases. Phase~I, which culminated in the 2017--2018 observing season, used fewer than 60 antennas with dipole feeds reused from the PAPER experiment, covering 100--200\,MHz \citep{Parsons10,DeBoer17,HERA25}. Phase~II replaced them with broadband Vivaldi feeds and an upgraded signal chain covering 50--250\,MHz \citep{Fagnoni21a,Berkhout24,HERA25}. We use the Phase~I limits from 94 nights of the 2017--2018 season \citep[hereafter H23;][]{HERA23} and the first Phase~II limits from 14 nights (129 unflagged hours) of the 2022--2023 season \citep[hereafter H25;][]{HERA25}, which extend the measurements to $z\simeq6$--25. HERA limits have previously been interpreted with a range of astrophysical models \citep[e.g.][]{HERA22a,HERA23,Lazare23,Dhandha25,HERA25,Breitman26}. From H25, we use the six bands at $z\simeq7.0$, 7.6, 9.9, 10.8, 16.8, and 19.8, which lie within the redshift range of the \texttt{21cmEMUv1} PS predictions \citep[$5.9\leq z\leq21.2$;][]{Breitman24}.\footnote{As in previous HERA analyses \citep{HERA23,HERA25}, we use the noise-only uncertainties and set negative band powers to zero while keeping their uncertainties.} With each emulator, we perform two tests:
\begin{enumerate}[Test 1.]
    \item \textit{Cross-release consistency:} H23 versus H25;
    \item \textit{Redshift consistency:} the two H25 bands at $z<9$ versus the four at $z>9$.
\end{enumerate}

Both tests are conditional on complementary constraints on galaxies and reionisation, $d_C$, i.e. their prior is the posterior from these constraints, $\mathcal{P}_C$ (Section~\ref{sec:sus}):
\begin{enumerate}[(i)]
    \item UV LFs \citep{Bouwens15,Bouwens17,Oesch18}, with a Gaussian likelihood;
    \item the Thomson scattering optical depth, $\tau_e = 0.0569^{+0.0081}_{-0.0066}$, with an asymmetric likelihood \citep{Planck18,Qin20};
    \item the dark-fraction bound on the neutral fraction at $z=5.9$ \citep{Mesinger10,McGreer15}, with a likelihood of unity for $\overline{x}_{\rm HI}\leq0.06$ that decreases as a one-sided Gaussian of width $0.05$ above it.
\end{enumerate}
Their product, $\mathcal{L}_C= \mathcal{L}_{\rm UV}\mathcal{L}_{\tau_e}\mathcal{L}_{\rm HI}$, enters every run, so each test asks whether the two HERA subsets impose compatible requirements within the parameter space allowed by $d_C$. Table~\ref{tab:hera_inferences} lists the inferences of both tests: the H25 and conditioning-only inferences are shared, so each model requires six inferences.

For the HERA limits, we use the standard upper-limit likelihood, which marginalises over a positive, unbounded systematic contribution \citep[e.g.][]{Li19,Ghara20,HERA22a}:
\begin{equation}
    \mathcal{L}_{\rm HERA}(\boldsymbol{\theta}) =
    \prod_i \frac{1}{2} \left[1+\operatorname{erf}\!\left( \frac{d_i-\mu_i(\boldsymbol{\theta})}{\sqrt{2}\sigma_i}\right)\right],
    \label{eq:hera_ul_likelihood}
\end{equation}
where $d_i$, $\mu_i(\boldsymbol{\theta})$, and $\sigma_i$ are the measured band power, the model prediction, and the measurement uncertainty. We neglect the emulator error, which is much smaller than $\sigma_i$ \citep{Breitman24,Breitman26}.

\begin{table}[tbp]
\centering
\begin{tabular}{ll}
\hline
Likelihood & Role \\
\hline
\multicolumn{2}{l}{\textit{Test 1: H23 versus H25, conditional on $d_C$}}\\
$\mathcal{L}_{\rm H23}\mathcal{L}_{\rm H25}\mathcal{L}_C$ & Joint \\
$\mathcal{L}_{\rm H23}\mathcal{L}_C$ & Subset $A$ \\
$\mathcal{L}_{\rm H25}\mathcal{L}_C$ & Subset $B$ \\
$\mathcal{L}_C$ & Conditioning data only \\
\hline
\multicolumn{2}{l}{\textit{Test 2: H25 below versus above $z=9$, conditional on $d_C$}}\\
$\mathcal{L}_{\rm H25}\mathcal{L}_C$ & Joint \\
$\mathcal{L}_{\rm H25,\,low}\mathcal{L}_C$ & Subset $A$ \\
$\mathcal{L}_{\rm H25,\,high}\mathcal{L}_C$ & Subset $B$ \\
$\mathcal{L}_C$ & Conditioning data only \\
\hline
\end{tabular}
\caption{
Inferences performed for the HERA consistency tests. The H25 and conditioning-only inferences are shared by both tests, so each model requires six inferences. Each is run with \texttt{21cmEMUv1} and with \texttt{21cmEMUv3}, the latter with the same X-ray luminosity per unit star formation rate for ACGs and MCGs, using the same parameter prior within each model. $\mathcal{L}_C$ contains the UV LF, optical-depth, and dark-fraction constraints.}
\label{tab:hera_inferences}
\end{table}

Neither test reveals tension with either model: $|\log S_{AB\mid C}|\leq0.14$ in all four cases (Table~\ref{tab:hera_results}; Figures~\ref{fig:hera_summaries} and~\ref{fig:hera_xray} in Appendix~\ref{sec:hera_app}). We do not quote an equivalent Gaussian significance because $d_G<0.5$ in every case (see footnote~\ref{fn:dG}). Regardless, $\log S_{AB\mid C}\simeq0$ indicates agreement between the two subsets, not tension. This absence of tension does not mean that the limits are uninformative. As in previous analyses of HERA Phase~I limits \citep{HERA22a,HERA23}, H23 disfavours weak X-ray heating of the IGM: with v1, it raises the 97.5\% lower bound on the soft-band X-ray luminosity per unit star formation rate from $\log_{10}[L_{\rm X<2\,keV}/{\rm SFR}\,({\rm erg\,s^{-1}\,M_\odot^{-1}\,yr})]=38.1$ for $d_C$ alone, which is set by the prior, to 39.7, and H25 gives a weaker bound in the same direction (Figure~\ref{fig:hera_xray}). With v3, MCGs provide an additional source of X-rays \citep[e.g.][]{Qin20a,Munoz22}. As in Lazare et al.\ (2024)~\cite{Lazare23} and Breitman et al.\ (2026)~\cite{Breitman26}, we assign ACGs and MCGs the same X-ray luminosity per unit star formation rate in the HERA analyses. H23 then raises the v3 bound from 38.1 to 39.1, a weaker bound than with v1, as MCGs supply part of the required heating. This bound is similar to that of Breitman et al.\ (2026)~\cite{Breitman26}, whose priors on star formation we adopt in this work, whereas Lazare et al.\ (2024)~\cite{Lazare23} report a weaker bound, due to a prior allowing lower MCG stellar-to-halo mass ratios, though both results remain prior-dependent.

The two HERA subsets, however, constrain the same parameter direction from the same side. An upper limit excludes models with too much 21-cm power rather than favouring particular parameter values, so two sets of limits can only be in significant tension if their excluded regions together cover nearly all of the parameter space allowed by $d_C$. Here, H23 and H25 exclude overlapping, weakly heated models, giving $\log R_{AB\mid C}\gtrsim0$. Excess power in one release, from residual systematics for example, cannot create a tension either, since the upper-limit likelihood marginalises over positive systematic contributions. Because the upper-limit likelihood is nearly flat within the allowed region, these one-sided constraints also contribute little to the Bayesian model dimensionality, so $d_G\simeq0$. Table~\ref{tab:hera_results} lists $\log R_{AB\mid C}$, $Q_{AB\mid C}$, $\log S_{AB\mid C}$, and $d_G$ for each test and model, computed with Equations~\ref{eq:conditional_logR}--\ref{eq:shared_dimensionality} from the nested-sampling runs of Table~\ref{tab:hera_inferences}. With $|\log R_{AB\mid C}|\leq0.27$ and $|Q_{AB\mid C}|\leq0.13$, the suspiciousness remains close to zero in every case. The evidences do not distinguish the two models either: $\log\mathcal{Z}$ differs between v1 and v3 by at most 0.3 in every inference of Table~\ref{tab:hera_inferences}. Current upper limits therefore carry important astrophysical information, but give the internal consistency test little leverage. A detection, which favours specific signal amplitudes rather than only excluding large ones, would change this.

\begin{table}[tbp]
\centering
\begin{tabular}{clcrrrrc}
\hline
Test & Partition, $d_A$ vs $d_B$ & Model & $\log R$ & $Q$ & $\log S$ & $d_G$ & $n_\sigma$ \\
\hline
1 & H23 vs H25 & v1 & $0.27$ & $0.13$ & $0.14$ & $0.14$ & -- \\
  &            & v3 & $0.07$ & $-0.02$ & $0.08$ & $0.06$ & -- \\
2 & H25 at $z<9$ vs $z>9$ & v1 & $0.04$ & $-0.02$ & $0.05$ & $0.00$ & -- \\
  &                       & v3 & $-0.01$ & $-0.02$ & $0.01$ & $-0.01$ & -- \\
\hline
\end{tabular}
\caption{
Consistency of the HERA upper limits, conditional on the UV LF, $\tau_e$, and dark-fraction constraints ($d_C$). v1 and v3 denote \texttt{21cmEMUv1} and \texttt{21cmEMUv3}, the latter with the same X-ray luminosity per unit star formation rate for ACGs and MCGs. $\log R$, $Q$, $\log S$, and $d_G$ are the conditional quantities of Equations~\ref{eq:conditional_logR}--\ref{eq:shared_dimensionality}. With $d_G<0.5$ in all cases, no equivalent Gaussian significance $n_\sigma$ is quoted.}
\label{tab:hera_results}
\end{table}

\subsection{HST and JWST UV luminosity functions}
\label{sec:UVLFs}
JWST \citep{Rigby23} has extended UV LF measurements well beyond the reach of HST, with spectroscopic confirmations now beyond $z=14$ \citep{Carniani24,Naidu26}. These measurements reveal more UV-luminous galaxies at $z>10$ \citep[e.g.][]{Harikane23,Finkelstein23,Adams24,Donnan24,Whitler25} than predicted by extrapolating galaxy models calibrated to HST observations \citep[e.g.][]{Mason15,Tacchella18,Yung19a,Park19}. Proposed explanations fall into two classes. Either galaxies are brighter at fixed halo mass, through more efficient star formation \citep[e.g.][]{Dekel23,Somerville25,Kar26}, a top-heavy initial mass function \citep[e.g.][]{Yung24,Hutter25,Jeong25}, weaker dust attenuation \citep[e.g.][]{Ferrara23,Narayanan26}, or emission from active galactic nuclei (AGN) \citep[e.g.][]{Pacucci22,Trinca24,Hegde24}; or stochastic star formation scatters galaxies in abundant low-mass haloes to high luminosities \citep[e.g.][]{Mason23,Mirocha23,Shen23,Sun23,Gelli24,Nikolic24,NikolicMason26}. The galaxy model underlying \texttt{21cmEMUv3}, designed for pre-JWST observations and without stochastic star formation \citep{Park19,Qin20a}, is therefore expected to fail: without scatter in the galaxy--halo connection, raising its star formation efficiency to match the abundance of UV-bright galaxies at $z>10$ also raises the abundances at lower redshifts.\footnote{The latest version of \texttt{21cmFAST} \citep[v4;][]{Davies25} instead samples discrete galaxy populations with scatter in their stellar masses and star formation rates. Such scatter raises the abundance of UV-bright galaxies by upscattering the more numerous faint ones, although the fiducial level of scatter of Nikoli\'c et al.\ (2024)~\cite{Nikolic24} does not fully explain the JWST measurements. A Bayesian inference of the JWST UV LFs and galaxy clustering with an extended version of this model will be presented in Nikoli\'c et al.\ (in prep.), and their interpretation with \texttt{21cmFASTv4.3.1} in the upcoming EOS26 (Breitman et al., in prep.).} The UV LFs thus provide an observational test in which the model is expected to fail: \textit{can suspiciousness detect the failure of this model to reconcile the lower- and higher-redshift measurements?}

We combine HST UV LFs \citep{Bouwens15,Bouwens17} with two JWST datasets. Donnan et al.\ (2024)~\cite{Donnan24} combine PRIMER \citep{Dunlop21}, JADES \citep{Eisenstein23}, and NGDEEP \citep{Bagley24} imaging over $z\simeq9$--12.5. The GLIMPSE measurements of Chemerynska et al.\ (2026)~\cite{Chemerynska26} use deep NIRCam \citep{Rieke23} imaging behind the lensing cluster Abell~S1063 \citep{Atek25} to reach fainter galaxies, and indicate a slow evolution of the UV LF, with particularly high abundances in their highest-redshift interval, $11.5<z<15$ (represented here by a single effective redshift, $z=13$). We use only the binned GLIMPSE measurements \citep[Table~1 of][]{Chemerynska26}, excluding the $M_{\rm UV}=-12.5$ bins because of their low completeness, and exclude the HST measurements at $z\simeq10$ \citep{Oesch18} to avoid counting galaxies twice in overlapping fields. We restrict the analysis to $-20\leq M_{\rm UV}\leq-10$, an approximate range in which the underlying \texttt{21cmFAST} galaxy model is expected to be reliable. At the bright end, the model includes neither dust attenuation nor AGN emission and feedback, which become important for $M_{\rm UV}\lesssim-20$; previous analyses with the same galaxy model therefore used only fainter galaxies \citep{Park19}. At the faint end, $M_{\rm UV}\simeq-10$ corresponds, for the fiducial parameters of Table~\ref{tab:mock_params}, to haloes of $\sim$(1--3)$\times10^8\,M_\odot$ at $z=6$--13, close to the atomic-cooling threshold. Fainter galaxies are increasingly hosted by MCGs and by haloes in which star formation is suppressed, so their predicted abundance depends on the least constrained parts of the model. This cut removes the brightest galaxies, where the reported excess is most pronounced \citep[e.g.][]{Harikane23,Finkelstein23}, making our test conservative.

We split the measurements into a lower-redshift subset, $d_A$ ($z\leq10$: HST at $z=6$, 7, and 8; Donnan and GLIMPSE at $z=9$ and 10), and a higher-redshift subset, $d_B$ ($z>10$: Donnan at $z=11$ and 12.5; GLIMPSE at $z=11$ and 13). The likelihood is Gaussian in $\phi$, with the uncertainty of each bin taken as the mean of its upper and lower errors. \texttt{21cmEMUv3} predicts the UV LF at $z=6$, 7, 8, 9, 10, 12, and 15.\footnote{We perform this test only with \texttt{21cmEMUv3}, since \texttt{21cmEMUv1} predicts the UV LF only up to $z=10$.} We interpolate $\log_{10}\phi$ linearly in $z$ between the two nearest outputs, without extrapolation. The joint ($\mathcal{L}_A\mathcal{L}_B$) and separate ($\mathcal{L}_A$, $\mathcal{L}_B$) fits use the same model and prior. We perform two tests:
\begin{enumerate}[Test 1.]
    \item \textit{Redshift consistency:} $d_A$ versus $d_B$, with the reported uncertainties;
    \item \textit{Likelihood stress test:} as Test~1, but with every $z>10$ uncertainty doubled.
\end{enumerate}
Table~\ref{tab:jwst_inferences} lists the inferences: the subset-$A$ inference is shared, so the two tests require five. Unlike the HERA tests, these tests use no conditioning observations: including the $\tau_e$ and dark-fraction constraints as $d_C$ does not significantly affect $\log S$ in Test~1.

\begin{table}[tbp]
\centering
\begin{tabular}{ll}
\hline
Likelihood & Role \\
\hline
\multicolumn{2}{l}{\textit{Test 1: $z\leq10$ versus $z>10$}}\\
$\mathcal{L}_A\mathcal{L}_B$ & Joint \\
$\mathcal{L}_A$ & Subset $A$ \\
$\mathcal{L}_B$ & Subset $B$ \\
\hline
\multicolumn{2}{l}{\textit{Test 2: as Test 1, with the $z>10$ uncertainties doubled}}\\
$\mathcal{L}_A\mathcal{L}_B^{\times2}$ & Joint \\
$\mathcal{L}_A$ & Subset $A$ \\
$\mathcal{L}_B^{\times2}$ & Subset $B$ \\
\hline
\end{tabular}
\caption{
Inferences performed for the JWST consistency tests, all with \texttt{21cmEMUv3} and the same prior. The subset-$A$ inference is shared by both tests, so five inferences are required. $\mathcal{L}_B^{\times2}$ is $\mathcal{L}_B$ with every uncertainty doubled. There are no conditioning observations. The tests of Appendix~\ref{sec:jwst_app} use the same configurations with GLIMPSE measurements removed.}
\label{tab:jwst_inferences}
\end{table}

\begin{figure}
    \centering
    \includegraphics[width=0.95\linewidth]{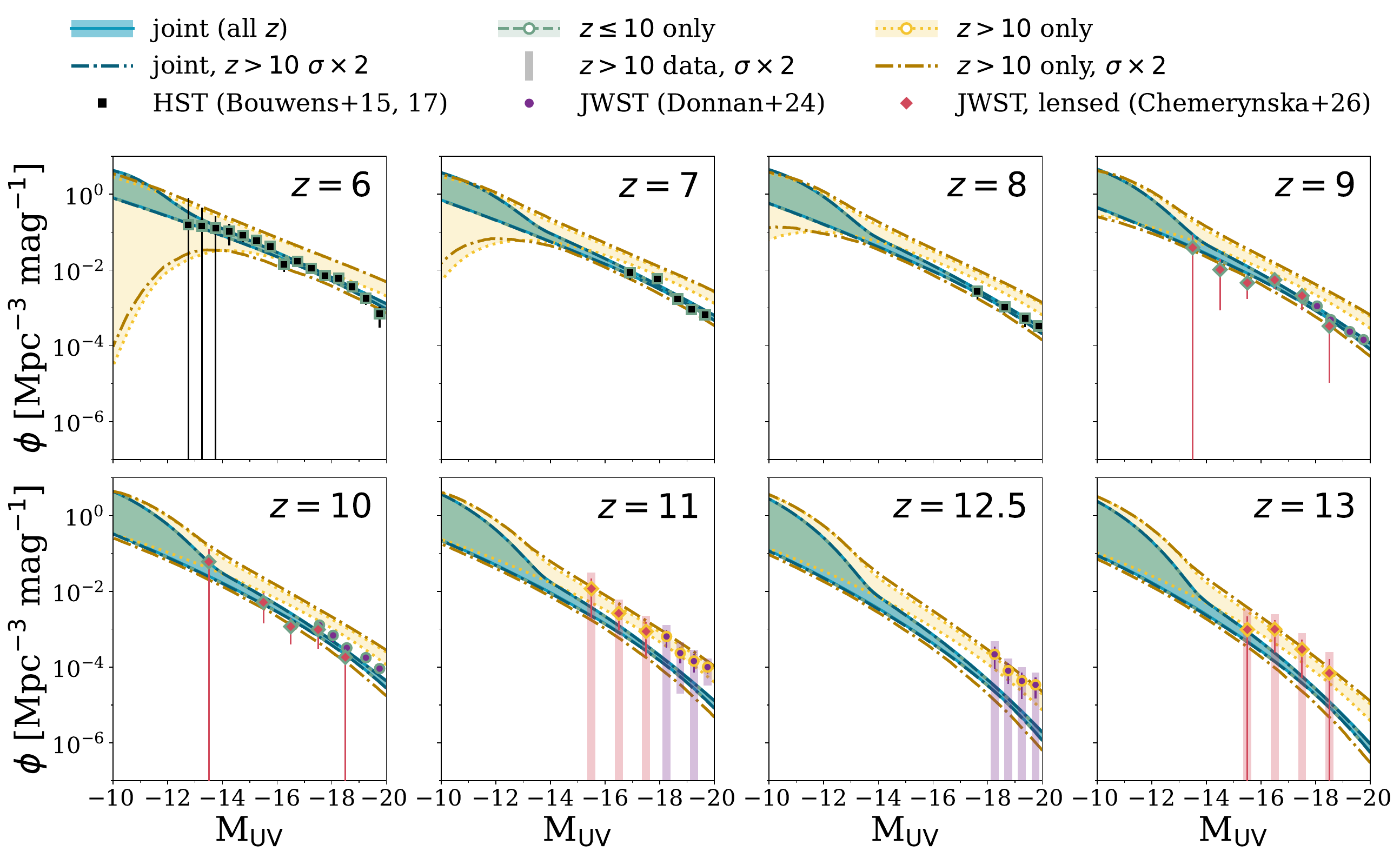}
    \caption{HST \citep{Bouwens15,Bouwens17} and JWST \citep{Donnan24, Chemerynska26} UV LFs used in the JWST tests (markers; error bars show the uncertainties used in the Gaussian likelihood) and the 95\% intervals of the posterior predicted UV LF from the joint (blue, solid), lower-redshift (green, dashed), and higher-redshift (yellow, dotted) inferences of Test~1. Dash-dotted outlines show the 95\% intervals of Test~2 (joint, dark blue; $z>10$ only, dark yellow), whose likelihood uses the doubled $z>10$ uncertainties shown as wide pale bars. Marker outlines indicate the subset containing each measurement: green for $d_A$ ($z\leq10$) and yellow for $d_B$ ($z>10$). The predictions are interpolated in redshift as in the likelihood. }
    \label{fig:jwst_uvlfs}
\end{figure}

\begin{figure}
    \centering
    \includegraphics[width=0.6\linewidth]{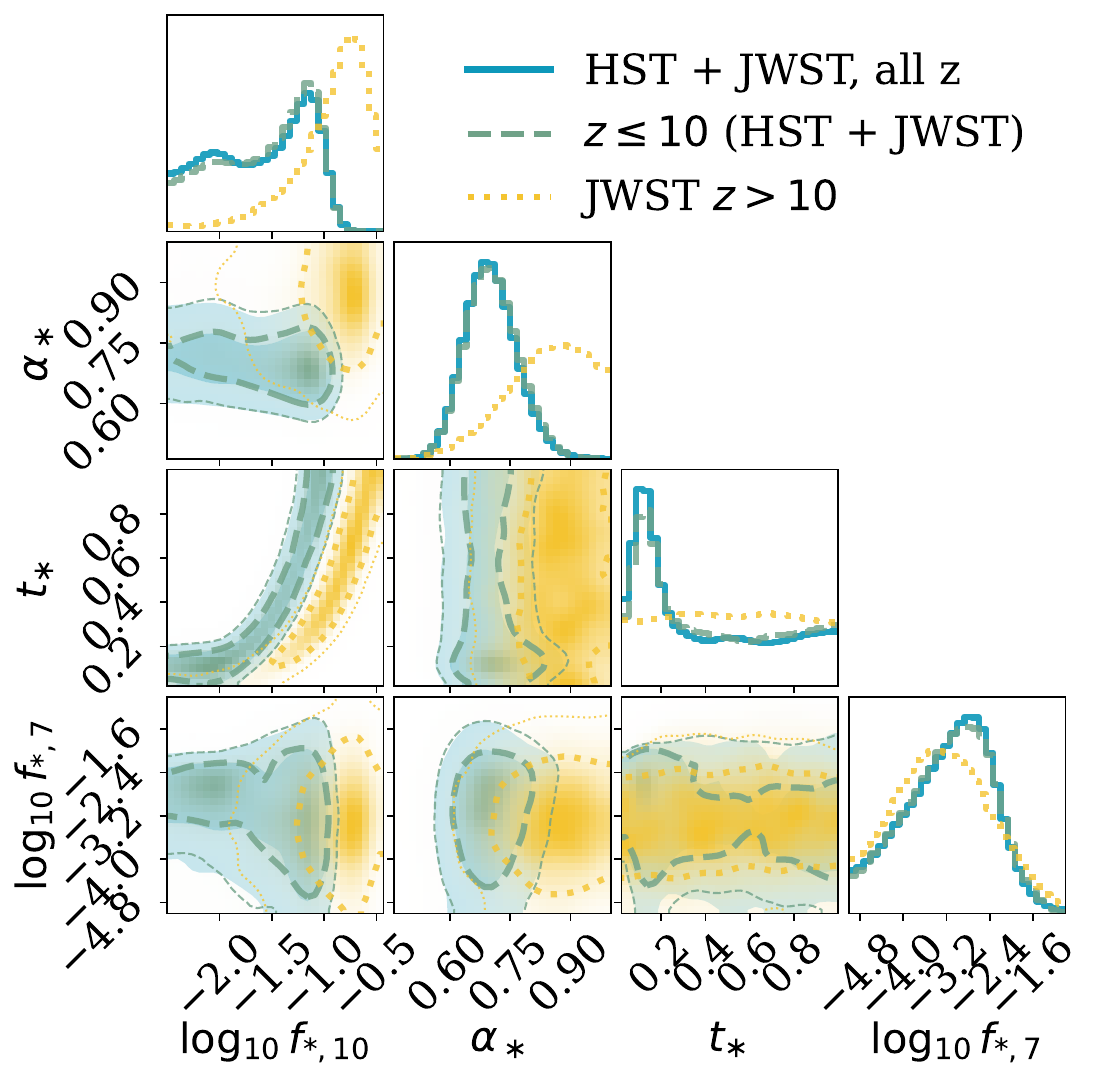}
    \caption{Posteriors of the SHMR parameters from the joint (blue), lower-redshift (green), and higher-redshift (yellow) inferences of Test~1, marginalised over all other parameters. Contours enclose 68\% and 95\% of the posterior mass. The $\log_{10}f_{\ast,7}$ posteriors follow the prior (Table~\ref{tab:priors}).}
    \label{fig:jwst_corner}
\end{figure}

Figure~\ref{fig:jwst_uvlfs} shows the posterior predictions for Test~1. The lower-redshift fit reproduces the UV LFs at $z\leq10$ but underpredicts the abundances at higher redshifts, most visibly towards the bright end at $z=12.5$ and 13. The higher-redshift fit favours larger abundances, matching the $z>10$ measurements more closely while also raising the predicted abundances at lower redshifts. The joint fit remains close to the lower-redshift solution and continues to underpredict the $z>10$ abundances: its tight constraints are a compromise that neither subset supports. The model fails to reproduce the redshift evolution of the UV LF: parameter values favoured by the lower-redshift population do not produce enough galaxies at earlier times. Figure~\ref{fig:jwst_corner} shows the corresponding SHMR parameters. Since $f_{\ast,10}$ and $t_\ast$ are degenerate, the UV LFs mainly constrain their ratio, which sets the normalisation of the SFR: the higher-redshift fit favours $\log_{10}(f_{\ast,10}/t_\ast)=-0.55^{+0.05}_{-0.06}$, about three times the lower-redshift value ($-1.02\pm0.03$), and a steeper slope, $\alpha_\ast=0.85^{+0.10}_{-0.11}$ ($0.70\pm0.06$; medians and 68\% credible intervals, CIs). The MCG normalisation is not constrained: all three posteriors of $\log_{10}f_{\ast,7}$ ($-3.2\pm0.8$) follow its prior ($-3.2\pm0.9$; Table~\ref{tab:priors}). The full posteriors are shown in Figure~\ref{fig:jwst_full_corner} (Appendix~\ref{sec:jwst_app}).

Suspiciousness quantifies this disagreement: $\log S_{AB}=-11.62$, an equivalent significance of $5.0\sigma$ under the Gaussian-limit calibration (Table~\ref{tab:jwst_results}). Doubling the $z>10$ uncertainties (Test~2) reduces this to $\log S_{AB}=-2.25$ ($2.1\sigma$). The significance of the tension therefore hinges on the precision of the $z>10$ measurements. We also test the contribution of the GLIMPSE measurements: removing them, at $z\simeq13$ or entirely, leaves $\log S_{AB}\leq-9.3$ (Appendix~\ref{sec:jwst_app}), so the tension does not depend on GLIMPSE alone.

\begin{table}[tbp]
\centering
\begin{tabular}{clcrrrrc}
\hline
Test & Partition, $d_A$ vs $d_B$ & Model & $\log R$ & $Q$ & $\log S$ & $d_G$ & $n_\sigma$ \\
\hline
1 & $z\leq10$ vs $z>10$ & v3 & $-9.16$ & $2.46$ & $-11.62$ & $0.76$ & $5.0$ \\
2 & as Test 1, $z>10$ uncertainties $\times2$ & v3 & $-0.89$ & $1.36$ & $-2.25$ & $1.83$ & $2.1$ \\
\hline
\end{tabular}
\caption{
Consistency of the lower- and higher-redshift UV LFs under \texttt{21cmEMUv3}, without conditioning observations. The equivalent Gaussian significance $n_\sigma$ (Equation~\ref{eq:tension_significance}) is quoted only for $d_G\geq0.5$. Test~2 doubles every $z>10$ uncertainty. The significance depends on the adopted $d_G$: for $0.5\leq d_G\leq2$, it spans $4.7$--$5.1\sigma$ for Test~1 and $2.1$--$2.6\sigma$ for Test~2. Tests without the GLIMPSE measurements are in Table~\ref{tab:jwst_glimpse}.}
\label{tab:jwst_results}
\end{table}

These results are a proof of concept for one combination of data and likelihood, not a definitive assessment of the high-redshift UV LFs, as the reported tension depends on the choice of both. The abundance of galaxies beyond $z\approx12$ remains debated: GLIMPSE indicates a slow evolution \citep{Chemerynska26}, other surveys report a steeper decline beyond $z\approx9$--12 \citep{Willott24,McLeod26}, and spectroscopic samples are still small: as of 2024, the compilation of Harikane et al.\ (2024)~\cite{Harikane24} contained 25 galaxies at $z_{\rm spec}=8.6$--13.2. Our Gaussian likelihood also neglects cosmic variance \citep[e.g.][]{Weibel25}, Eddington bias \citep{Eddington13}, and contamination by lower-redshift interlopers \citep[e.g.][]{ArrabalHaro23}, which matter for the sparse samples at $z>10$. Test~2 is only a crude proxy for these effects, but its much weaker tension shows that a robust conclusion requires modelling them. A complete assessment, with more observations and a likelihood that includes these effects, is beyond the scope of this paper.

Suspiciousness thus recovers the known UV LF tension as evidence that the galaxy model underlying \texttt{21cmEMUv3}, together with the adopted observational likelihoods, is misspecified. In this case, the misspecification is already visible in the posterior predictive distributions. An apparently satisfactory joint fit can, however, also hide disagreement between the parameter values favoured by different parts of the data, as we show in Section~\ref{sec:ska}.

\section{Application to mock SKA observations}
\label{sec:ska}

In the previous section, we showed that the suspiciousness detects a misspecified galaxy model when the UV LFs at lower and higher redshifts cannot be fitted with the same parameters. There, the misspecified model visibly fails to fit the data. We now consider a more challenging case, in which the fit appears satisfactory but the inferred astrophysics is misleading. We analyse the same mock SKA-Low 21-cm PS \citep{Mellema13,Koopmans15} with \texttt{21cmEMUv3}, whose physical model generated the mock (well specified), and with \texttt{21cmEMUv1}, which omits MCGs (misspecified). The observations and their uncertainties are identical in both analyses. The generating model is used only to assess the recovered astrophysics, never by the consistency diagnostic.

\subsection{Mock observation}
\label{sec:ska_mock}
We adopt the spherically averaged mock SKA PS of Breitman et al.\ (2026)~\cite{Breitman26}: the mean of 29 independent \texttt{21cmFAST} realisations at the fiducial parameters of Table~\ref{tab:mock_params}, taken from the \texttt{21cmEMUv3} test database. These parameters are consistent with the HST UV LFs used as conditioning data \citep{Bouwens15,Bouwens17,Oesch18} and representative of values commonly adopted in the literature \citep[e.g.][]{Park19,Qin20a,Munoz22}. The mock is therefore derived from simulations, not from an emulator. We do not add a thermal-noise realisation (but see Appendix~\ref{sec:calibration}). Averaging suppresses, but does not eliminate, sample variance in the mock, while the likelihood includes the thermal noise and sample variance expected for the assumed observation.

\begin{table}[ht]
\centering
\begin{tabular}{ll}
Parameter            & Values \\ \hline
$\log_{10} f_{*,10}$       & -1.2         \\
$\alpha_{\ast}$   & 0.5        \\
$t_{\ast}$         & 0.55  \\
$\log_{10} f_{\rm esc, 10}$     & -1.3    \\
$\alpha_{\rm esc}$     & 0.0     \\
$\log_{10} f_{\ast,7}$ &  -2.5  \\
$\log_{10} f_{\rm esc, 7}$ &-1.5\\
$\log_{10}L_{\rm X<2keV}^{\textsc{acg}}/{\rm SFR}$ & 40.5\\
$\log_{10}L_{\rm X<2keV}^{\textsc{mcg}}/{\rm SFR}$ & 41.5 \\
$E_{\rm 0}$ & 500 \\
$\sigma_8$ & 0.8118\\
\end{tabular}
\caption{Mock SKA observation fiducial parameters.}
\label{tab:mock_params}
\end{table}

Following Breitman et al.\ (2026)~\cite{Breitman26}, we assume $1080\,\mathrm{h}$ of observation with the SKA-Low AA$^\ast$ layout \citep{SKAO23}, computing sensitivities with {\tt 21cmSense} \citep{Pober13,Pober14,Murray24}. Of the ten frequency bands between 50 and $215\,\mathrm{MHz}$ ($z\simeq5.6$--27.4), we use the seven within the \texttt{21cmEMUv1} redshift range ($z\simeq6.1$, 6.9, 7.9, 9.1, 10.4, 10.8, and 16.8) in both analyses, each with 18 logarithmically spaced wavemode bins over $k=0.024$--$1.1\,\mathrm{Mpc}^{-1}$. Of these 126 bins, the 69 with non-zero sensitivity enter the likelihood: the size of the SKA-Low stations sets the shortest baseline, leaving little or no sensitivity at $k\lesssim0.1\,\mathrm{Mpc}^{-1}$. Figure~\ref{fig:ska_summaries} shows the mock PS and histories (red) and the PS measurements entering the likelihood (black squares).

As in Section~\ref{sec:HERA}, we include the UV LF, optical-depth, and dark-fraction constraints, treating all observations as conditionally independent given the model parameters:
\begin{equation}
    \mathcal{L} = \mathcal{L}_{\rm PS}\mathcal{L}_{\rm UV}\mathcal{L}_{\tau_e}\mathcal{L}_{\rm HI}.
    \label{eq:ska_total_likelihood}
\end{equation}
For the mock PS, we use a Gaussian likelihood,
\begin{equation}
    \log\mathcal{L}_{\rm PS}
    =
    -\frac{1}{2}
    \boldsymbol{r}^{\mathsf T}
    \mathbf{C}_{\rm SKA}^{-1}\boldsymbol{r}
    -\frac{1}{2}
    \log\det\!\left(2\pi\mathbf{C}_{\rm SKA}\right),
    \label{eq:ska_ps_likelihood}
\end{equation}
where $\boldsymbol{r} =d_{\rm PS}-\boldsymbol{\mu}_{\rm PS}(\boldsymbol{\theta})$ is the residual between the mock PS and the emulator prediction, and the diagonal covariance $\mathbf{C}_{\rm SKA}$ contains thermal noise and sample variance. Both are identical in the v3 and v1 analyses.

\subsection{Apparently satisfactory fits, yet misleading astrophysics}
\label{sec:ska_results}
\begin{figure}[tbp]
    \centering
    \includegraphics[width=0.49\linewidth]{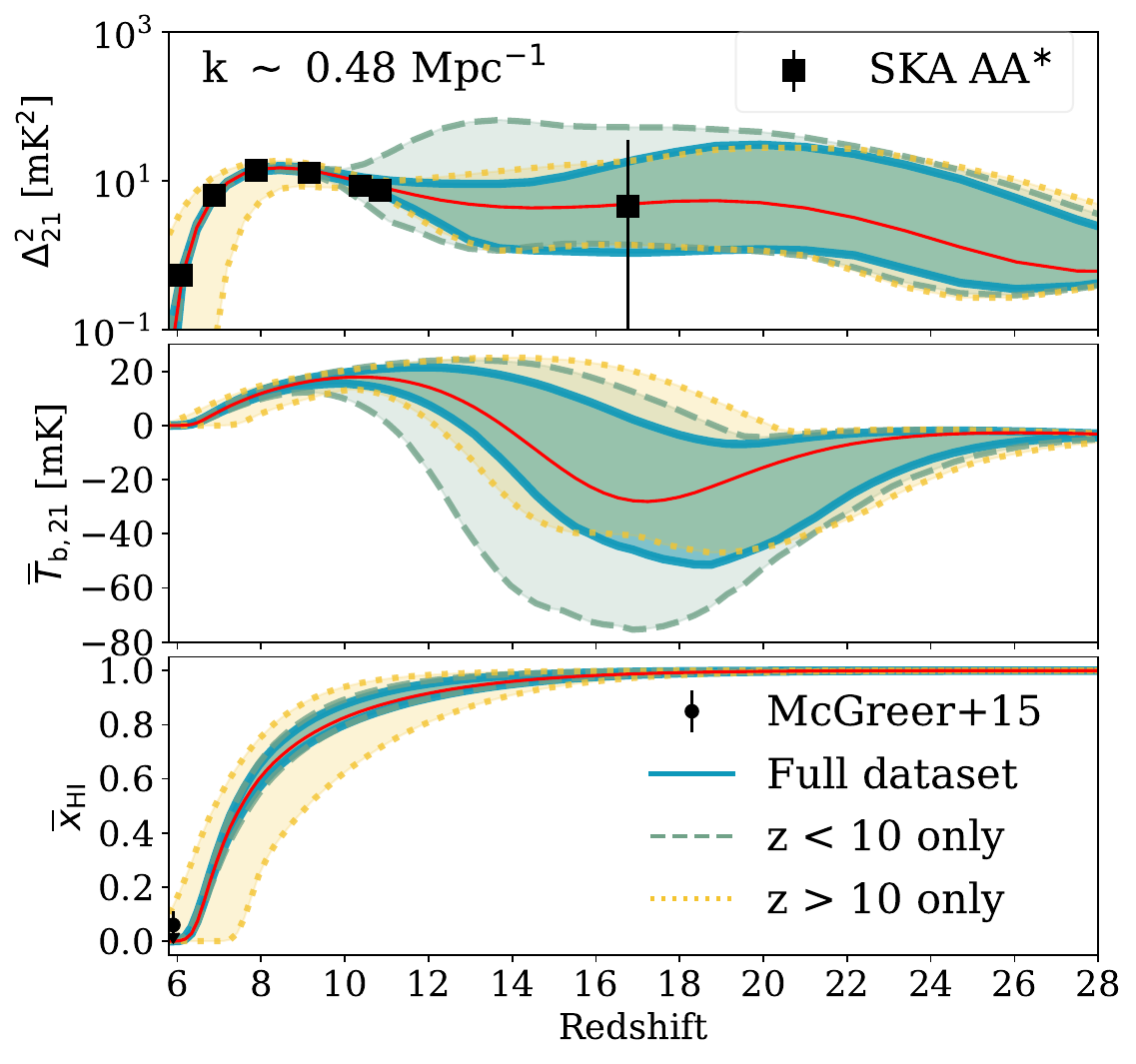}\hfill
    \includegraphics[width=0.49\linewidth]{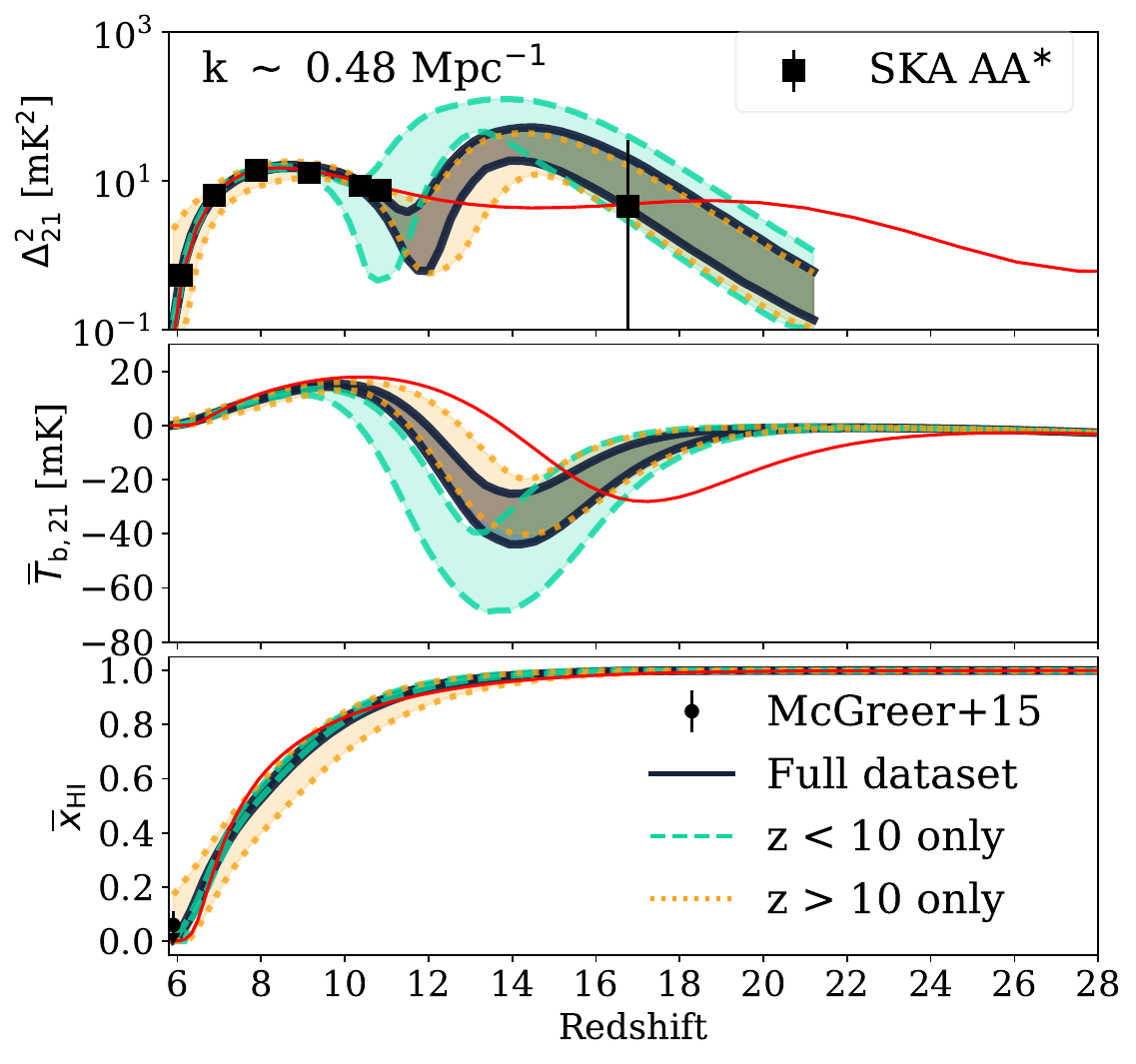}
    \caption{Posterior predictions for the well-specified \texttt{21cmEMUv3} analysis (left) and the misspecified \texttt{21cmEMUv1} analysis (right). From top to bottom, the panels show the 21-cm PS at $k\simeq0.48\,\mathrm{Mpc}^{-1}$, the global 21-cm brightness temperature, and the volume-averaged neutral hydrogen fraction. Red curves show the mock signal and histories. Black squares mark the mock SKA PS measurements \citep{Breitman26} in the top panel, and the dark-fraction upper bound on $\overline{x}_{\rm HI}$ \citep{McGreer15} is shown in the bottom panel. Solid, dashed, and dotted 95\% CI boundaries distinguish posterior predictions from the full PS dataset, the $z<10$ subset, and the $z>10$ subset, respectively. All three analyses include the UV LF, optical-depth, and dark-fraction constraints. Although both models describe the PS measurements (black squares), v1 clearly deviates from the mock PS at $11\lesssim z\lesssim16$ and from the mock global-signal history, and its inferred astrophysical parameters are biased (Figure~\ref{fig:ska_corner}).}
    \label{fig:ska_summaries}
\end{figure}

\begin{figure}[tbp]
    \centering
    \includegraphics[width=0.49\linewidth]{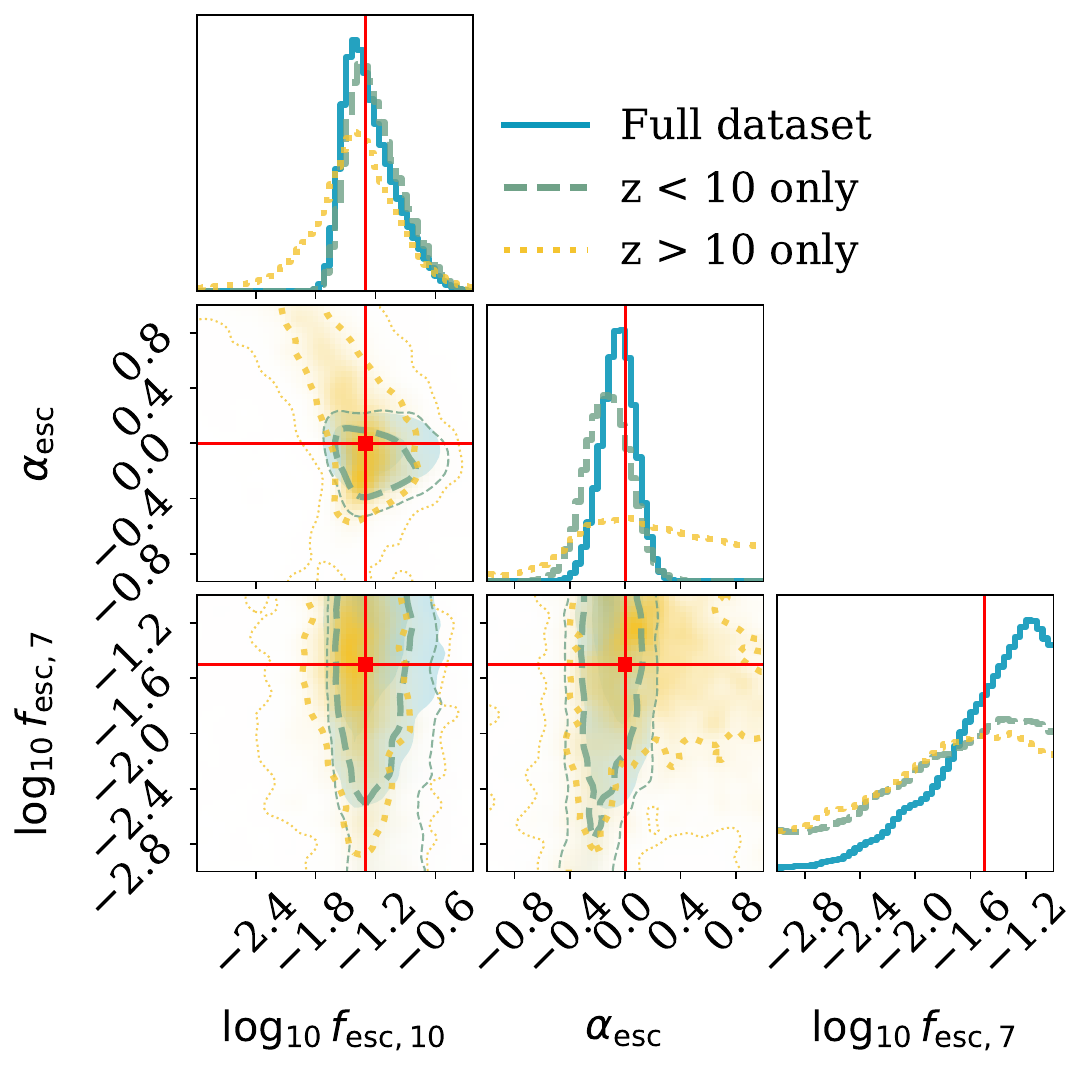}\hfill
    \includegraphics[width=0.49\linewidth]{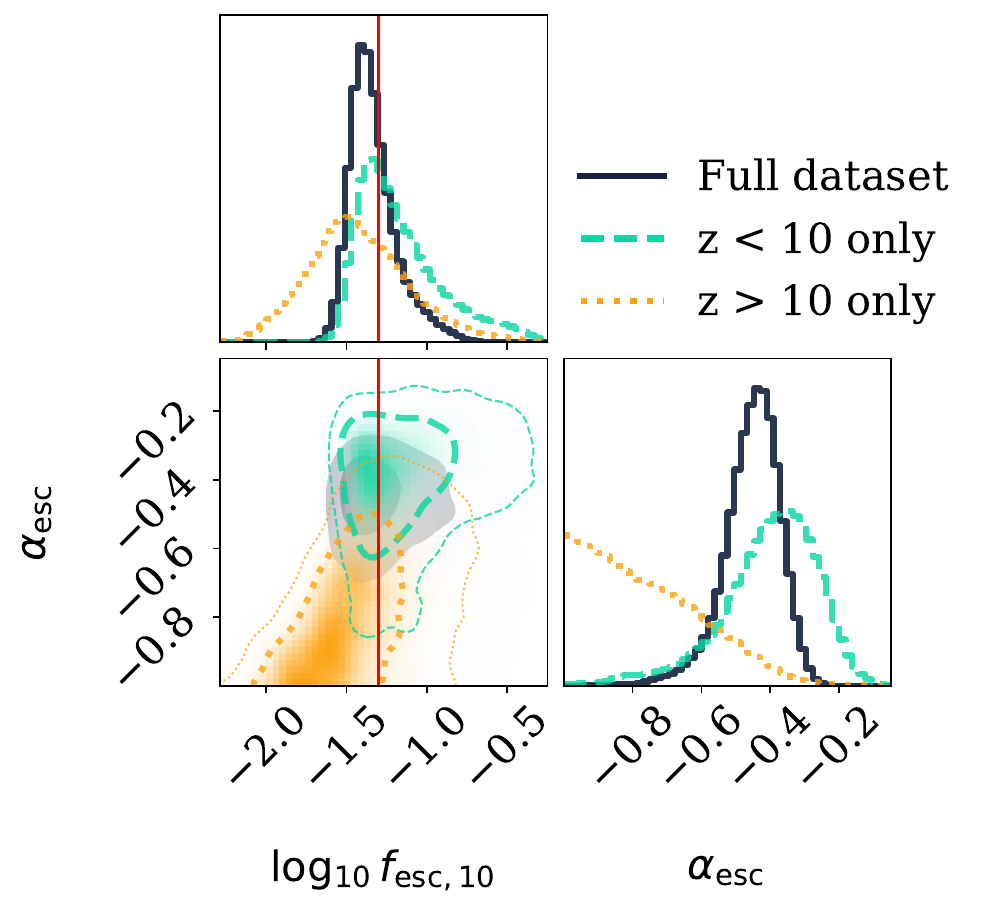}
    \caption{Escape-fraction constraints from the well-specified \texttt{21cmEMUv3} analysis (left) and the misspecified \texttt{21cmEMUv1} analysis (right). The diagonal panels show marginal posterior distributions, and the off-diagonal panels show two-dimensional projections. Filled, dashed, and dotted contours enclose 68\% and 95\% of the posterior mass for the full PS dataset, the $z<10$ subset, and the $z>10$ subset, respectively. Red crosshairs indicate the mock parameter values where shown. The MCG escape-fraction normalisation, $f_{\rm esc,7}$, appears only in the v3 model. The v1 joint posterior tightly constrains a negative escape-fraction slope, whereas the mock has $\alpha_{\rm esc}=0$.}
    \label{fig:ska_corner}
\end{figure}

Figure~\ref{fig:ska_summaries} compares the posterior predictions with the mock (red). In the left panel, the predictions of the well-specified v3 analysis of the full PS (solid blue lines) fit the mock PS and remain consistent with the mock histories. The reionisation history is tightly constrained, whereas a broader range of global signals and high-redshift PS remains allowed. Figure~\ref{fig:ska_noMM_full_corner} (Appendix~\ref{sec:ska_app}) shows the corresponding posteriors.

The misspecified v1 analysis (right panel) also fits the PS measurements: the predictions from the full PS (solid black lines) overlap with all the displayed PS measurements (black squares). While its EoR history remains mostly well recovered, its physical interpretation has changed significantly. The inferred global absorption trough occurs later and is typically deeper than in the mock, whose history lies outside the full-data 95\% CI over most of cosmic dawn. Between the sampled redshifts, the inferred PS develops a minimum and a peak that are absent from the mock. A good fit at the measured points therefore does not guarantee a faithful history. Nor would a goodness-of-fit test reliably reveal the problem: for noisy realisations of the mock, only 15 of 50 v1 fits would fail such a test at $p<0.05$ (Appendix~\ref{sec:calibration}).

These differences follow from \texttt{21cmEMUv1} emulating only one galaxy population, ACGs, whereas \texttt{21cmEMUv3}, which generated the mock, emulates both ACGs and MCGs. In the fiducial model, ACGs drive most of reionisation, so an ACG-only model can still reproduce the neutral-fraction history \citep{Qin20a,Munoz22}. MCGs, however, contribute substantially to the early Ly$\alpha$ and X-ray backgrounds, which set the timing and depth of the absorption trough and shape the 21-cm PS \citep[e.g.][]{Fialkov13,Mirocha18,Qin20a,Munoz22,Ventura25}. Without them, the v1 inference adjusts the ACG properties to reproduce the measured PS, attributing the signal to the wrong sources: for example, it raises the ACG X-ray luminosity to $\log_{10}(L_{\rm X<2keV}^{\textsc{acg}}/{\rm SFR})=41.5^{+0.3}_{-0.2}$, an order of magnitude above the mock value of 40.5.

Figure~\ref{fig:ska_corner} shows that these compensating adjustments differ between epochs. In the v1 analysis, the PS at $z>10$ favours a more negative escape-fraction slope, $\alpha_{\rm esc}$, than the PS at $z<10$, i.e. a different halo-mass dependence of the ionising escape fraction. The joint posterior is nonetheless narrow, centred on $\alpha_{\rm esc}\simeq-0.5$ rather than the mock value, $\alpha_{\rm esc}=0$, so its precision gives no indication of either the bias or the disagreement between the subsets. Some v1 parameters also move to the edges of their prior: in the full-PS fit, $\log_{10}(M_{\rm turn}/M_\odot)=8.11^{+0.15}_{-0.08}$ lies against its lower bound of 8, and the $z>10$ fit pushes $\alpha_{\rm esc}$ towards $-1$ (Figure~\ref{fig:ska_MM_full_corner}). The diagnostic therefore tests the v1 model together with its prior, which is part of the assumed model family. The subsets also imply different global signals (Figure~\ref{fig:ska_summaries}, right panel): the $z>10$ fit recovers the depth of the absorption trough but not its timing, whereas the $z<10$ fit misplaces both. None of these failures is apparent from the fit to the PS. They are visible here only because the generating histories are known; the global signal, in particular, is inferred rather than measured.

\subsection{Diagnosing the inconsistency}
\label{sec:ska_consistency}

We now test whether the observations alone can reveal the problem. Our primary test (Test~1; Table~\ref{tab:ska_inferences}) divides the mock PS at $z=10$ ($z\simeq6.1$--9.1 versus 10.4--16.8), conditional on the UV LF, optical-depth, and dark-fraction constraints, asking whether the physical description that accounts for the PS at one epoch remains consistent with the PS at another. Similar splits can expose biased inferences in other contexts: in a CMB lensing forecast, Fronenberg et al.\ (2024)~\cite[fig.~11]{Fronenberg23} show that when the true expansion history changes abruptly but the analysis assumes $\Lambda$CDM, the CMB lensing constraints, which integrate over all redshifts, are biased and disagree with the unbiased high-redshift constraints from line-intensity-mapping nulling. The separate and joint fits use the same model and prior. For v3, $\log S=0.80$, with no indication of tension. For v1, $\log S=-7.45$ (Table~\ref{tab:ska_results}): splitting the data reveals a loss of fit that is not apparent from the full-data fit. The negative $d_G=-1.50$ prevents a Gaussian-limit significance, so we calibrate this result with noisy mock realisations (Appendix~\ref{sec:calibration} and Figure~\ref{fig:ska_calibration}). None of the 150 realisations of a well-specified analysis is as discrepant, which bounds the significance at $n_\sigma\geq2.72$; extrapolating their distribution with the Gaussian-limit form gives $\sim3.4\sigma$, a value that depends on the assumed shape of the tail.

The PS redshift split is only one possible partition of the likelihood. The UV LFs and the reionisation constraints enter every fit, so a misspecified model may also have to compromise between different observables. Comparing the PS, the UV LFs, and the reionisation constraints with each other (Tests~2--4) asks whether misspecification can be revealed by these other datasets, rather than by the PS alone; together with Test~1, the four tests require nine inferences per model (Table~\ref{tab:ska_inferences}). These comparisons are presented in Appendix~\ref{sec:ska_obs}; of these, only the PS--UV LF comparison also flags v1. Internal consistency thus detects the misspecification of v1 without using the mock histories or comparing the evidences of v1 and v3.

Here, a direct model comparison would also disfavour v1: with the full likelihood, $\log\mathcal{Z}$ is higher for v3 than for v1 by 14.3. The preference is weaker for each half of the PS, by 2.8 and 3.8 for the $z<10$ and $z>10$ fits, respectively (Table~\ref{tab:ska_results}), so with less constraining observations, the evidence could distinguish the two models much less clearly. Unlike the suspiciousness, such a comparison also requires a better model to compare against, and for real observations, such a model is not always readily available.

\begin{table}[tbp]
\centering
\begin{tabular}{ll}
\hline
Likelihood & Role \\
\hline
\multicolumn{2}{l}{\textit{Test 1: PS at $z<10$ versus $z>10$, conditional on UV, $\tau_e$, $\overline{x}_{\rm HI}$}}\\
$\mathcal{L}_{\rm PS}\mathcal{L}_{\rm UV}\mathcal{L}_{\rm EoR}$ & Joint \\
$\mathcal{L}_{{\rm PS},\,z<10}\mathcal{L}_{\rm UV}\mathcal{L}_{\rm EoR}$ & Subset $A$ \\
$\mathcal{L}_{{\rm PS},\,z>10}\mathcal{L}_{\rm UV}\mathcal{L}_{\rm EoR}$ & Subset $B$ \\
$\mathcal{L}_{\rm UV}\mathcal{L}_{\rm EoR}$ & Conditioning data only \\
\hline
\multicolumn{2}{l}{\textit{Test 2: PS versus UV, conditional on $\tau_e$, $\overline{x}_{\rm HI}$}}\\
$\mathcal{L}_{\rm PS}\mathcal{L}_{\rm UV}\mathcal{L}_{\rm EoR}$ & Joint \\
$\mathcal{L}_{\rm PS}\mathcal{L}_{\rm EoR}$ & Subset $A$ \\
$\mathcal{L}_{\rm UV}\mathcal{L}_{\rm EoR}$ & Subset $B$ \\
$\mathcal{L}_{\rm EoR}$ & Conditioning data only \\
\hline
\multicolumn{2}{l}{\textit{Test 3: PS versus $\tau_e$, $\overline{x}_{\rm HI}$, conditional on UV}}\\
$\mathcal{L}_{\rm PS}\mathcal{L}_{\rm UV}\mathcal{L}_{\rm EoR}$ & Joint \\
$\mathcal{L}_{\rm PS}\mathcal{L}_{\rm UV}$ & Subset $A$ \\
$\mathcal{L}_{\rm UV}\mathcal{L}_{\rm EoR}$ & Subset $B$ \\
$\mathcal{L}_{\rm UV}$ & Conditioning data only \\
\hline
\multicolumn{2}{l}{\textit{Test 4: UV versus $\tau_e$, $\overline{x}_{\rm HI}$, conditional on PS}}\\
$\mathcal{L}_{\rm PS}\mathcal{L}_{\rm UV}\mathcal{L}_{\rm EoR}$ & Joint \\
$\mathcal{L}_{\rm PS}\mathcal{L}_{\rm UV}$ & Subset $A$ \\
$\mathcal{L}_{\rm PS}\mathcal{L}_{\rm EoR}$ & Subset $B$ \\
$\mathcal{L}_{\rm PS}$ & Conditioning data only \\
\hline
\end{tabular}
\caption{
Inferences performed for the mock SKA consistency tests, each run with \texttt{21cmEMUv3} (well specified) and \texttt{21cmEMUv1} (misspecified), using the same parameter prior within each model. Several inferences are shared between tests, so each model requires nine. $\mathcal{L}_{\rm EoR}\equiv\mathcal{L}_{\tau_e}\mathcal{L}_{\rm HI}$; ``UV'' denotes the UV LFs. Test~1 is discussed in Section~\ref{sec:ska_consistency} and Tests~2--4 in Appendix~\ref{sec:ska_obs}.}
\label{tab:ska_inferences}
\end{table}

\begin{table}[tbp]
\centering
\begin{tabular}{clcrrrrc}
\hline
Test & $d_A$ vs $d_B$ $\mid$ $d_C$ & Model & $\log R$ & $Q$ & $\log S$ & $d_G$ & $n_\sigma$ \\
\hline
1 & PS at $z<10$ vs $z>10$ $\mid$ UV, $\tau_e$, $\overline{x}_{\rm HI}$ & v3 & $3.38$ & $2.59$ & $0.80$ & $1.36$ & $0.0$ \\
  &                                                        & v1 & $-4.22$ & $3.23$ & $-7.45$ & $-1.50$ & -- \\
\hline
\end{tabular}
\caption{
Consistency of the mock SKA PS below and above $z=10$ (Test~1); Tests~2--4 are in Table~\ref{tab:ska_obs_results}. For comparison, $\log\mathcal{Z}$ is higher for v3 than for v1 by 14.3 for the joint fit and by 2.8 and 3.8 for the $z<10$ and $z>10$ fits, and nearly equal ($|\Delta\log\mathcal{Z}|=0.04$) for the conditioning data alone. The mock is generated with the \texttt{21cmEMUv3} physical model, so the v3 analyses are well specified and the v1 analyses misspecified. ``UV'' denotes the UV LFs and ``PS'' the mock SKA PS. $\log R$, $Q$, $\log S$, and $d_G$ are the conditional quantities of Equations~\ref{eq:conditional_logR}--\ref{eq:shared_dimensionality}; $n_\sigma$ is quoted only for $d_G\geq0.5$.}
\label{tab:ska_results}
\end{table}

\section{Discussion and conclusion}
\label{sec:conclusion}

In this work, we introduced the suspiciousness \citep{HandleyLemos19}, developed to quantify tensions between datasets, to CD/EoR inference as a model criticism tool, complementing parameter estimation and model comparison with an assessment of model adequacy. We asked whether different parts of the observations, such as different redshift ranges or observables, can be described by the same model parameters, and quantified their agreement with the suspiciousness conditioned on complementary observations. It compares the evidence ratio of the two subsets with the information they share, and its Gaussian-limit calibration accounts for the number of parameter directions that both constrain.

Each test requires three or four inferences: fits to each subset and to their combination and, for conditional tests, a fit to the conditioning observations alone, which can be shared between tests. Such repeated analyses are feasible with the \texttt{21cmEMU} emulators \citep{Breitman24,Breitman26}: all inferences in this work were performed serially on a single laptop, each taking from a few to tens of minutes.

We applied the diagnostic to three cases: (i) the HERA 21-cm PS upper limits; (ii) the HST and JWST UV LFs; and (iii) a mock SKA-Low AA$^\ast$ 1080\,h 21-cm PS detection. As expected, the HERA upper limits have little diagnostic power. Upper limits exclude models rather than favouring particular parameter values, so the two subsets share almost no constrained parameter directions ($d_G\simeq0$) and are unlikely to be in significant tension.

The UV LFs provide a test in which the model is expected to fail. Under the \texttt{21cmEMUv3} galaxy model, the suspiciousness recovers the expected tension between the measurements at $z\leq10$ and $z>10$, at $5.0\sigma$. This tension, however, hinges on the precision of the $z>10$ measurements: doubling their uncertainties reduces it to $2.1\sigma$. In this case, the failure is also visible by eye in the posterior predictive UV LFs (Figure~\ref{fig:jwst_uvlfs}).

The mock SKA analysis tests a more dangerous case, in which the misspecified model fits the observations. The \texttt{21cmEMUv1} analysis, which omits MCGs, reproduces the mock PS but attributes the signal to the wrong sources, misplacing the global absorption trough and biasing the halo-mass dependence of the escape fraction (Section~\ref{sec:ska_results}). Its posterior predictive PS gives no indication of this failure. Splitting the PS at $z=10$, the suspiciousness flags v1 with $\log S=-7.45$: lower than in all 150 noisy realisations of a well-specified analysis ($n_\sigma\geq2.72$), and $\sim3.4\sigma$ when their distribution is extrapolated (Appendix~\ref{sec:calibration}). The well-specified v3 analysis shows no tension ($0.0\sigma$), and 48 of 50 noisy realisations of the misspecified case lie below the 5th percentile of the well-specified ones, so the detection does not rely on a particular noise realisation. Bayesian model comparison also disfavours v1, by $\Delta\log\mathcal{Z}=14.3$, but only because the better model is available to compare against, which the suspiciousness does not require.

These results also illustrate what the diagnostic cannot establish. Not every misspecification creates an internal tension: a sufficiently flexible model can describe all parts of the observations with the same biased parameters, so finding no tension does not establish that a model is adequate. The sensitivity also depends on the chosen partition. In the SKA analysis, only the comparisons between PS redshift ranges and between the PS and UV LFs flag v1 (Appendix~\ref{sec:ska_obs}), and testing several partitions requires accounting for a trials factor. A detected tension, moreover, does not identify its cause. The diagnostic tests the whole assumed model family, including the likelihood and the prior: the JWST tension may reflect the galaxy model, the adopted $z>10$ uncertainties, or both, and the v1 SKA result is conditional on a prior whose bounds several of its parameters reach. Finally, when the Gaussian-limit calibration is unavailable, as for the v1 SKA result ($d_G<0$), calibrating with repeated mocks is costly: an empirical significance of $3\sigma$ requires about 370 noisy realisations, each with three inferences. Such a calibration also depends on how the realisations are generated: ours are centred on a single parameter vector and contain neither emulation error nor residual sample variance (Appendix~\ref{sec:calibration}).

The distinction between precise and adequate inference becomes increasingly consequential as CD/EoR observations improve. More precise measurements both constrain parameters more tightly within the assumed model family and increase the statistical power of model-criticism diagnostics. If the model is misspecified, the former can produce increasingly precise yet systematically biased scientific conclusions, while the latter can make discrepancies between the model and the observations detectable. Combining complementary observables can further increase this diagnostic power by requiring the same model to reproduce multiple aspects of the underlying physics simultaneously.

The importance of assessing model adequacy is already becoming apparent in community-wide inference exercises. The recent SKA Science Data Challenge 3b (SDC3b; Bonaldi et al., in prep.) tasked participants with recovering the EoR history from mock 21-cm PS observations generated using a partially unknown forward-modelling pipeline. Although the challenge was not designed as a model criticism exercise, the publicly available data-challenge results already demonstrate that different analysis pipelines can produce substantially different inferred EoR histories\footnote{SDC3b results are available at \url{https://sdc3.skao.int/challenges/inference/results}.}. While these differences cannot be uniquely attributed to astrophysical model misspecification, they illustrate how approximations throughout the forward-modelling pipeline can propagate into scientific inference. Related biases from incomplete or approximate models have been reported in other 21-cm analyses \citep[e.g.][]{Prelogovic24,Lazare23,Greig24a,Nikolic24}, including biases from the choice of numerical algorithm \citep{Berklas25} and from mismatches between simulators \citep[e.g.][]{Krishna26,Pietschke26}. Machine-learning analyses in cosmology face a related problem, robustness across simulation suites \citep[e.g.][]{deSanti23,Breitman25}, which has motivated training on several simulators while discarding simulator-specific features \citep[e.g.][]{Jo25}, and tests for misspecification have been developed for simulation-based inference \citep[e.g.][]{AnauMontel24} and cosmology \citep[e.g.][]{Akhmetzhanova25}.

Internal consistency tests need not wait for the SKA. Current UV LFs are already precise enough to reveal a tension (Section~\ref{sec:UVLFs}), and the other probes discussed in Section~\ref{sec:intro}, from quasar damping wings and the Ly$\alpha$ forest to CMB polarisation, can be partitioned in the same way to test whether a single model describes them all. For the 21-cm signal, a first PS detection by current interferometers would already provide the two-sided constraints that upper limits lack. As the community moves toward increasingly realistic end-to-end analyses and CD/EoR observations become more precise, model criticism must become a routine step of Bayesian inference, alongside parameter estimation and model comparison.

\acknowledgments

DB thanks Andrei Mesinger, Steven Murray, Hannah Fronenberg, and Roberto Trotta for helpful discussions and feedback during the preparation of the manuscript.

DB used Claude (Anthropic), accessed through the Claude Team plan for scientists, for code development (e.g.\ debugging and ensuring reproducibility), bookkeeping of the analysis runs, and text editing and proofreading. DB reviewed, validated and critically assessed the entire code and manuscript and takes full responsibility for their content.

\clearpage
\appendix
\section{Parameter priors}
\label{sec:priors}

\begin{table}[h]
\centering
\begin{tabular}{lll}
\hline
Parameter & \texttt{21cmEMUv3} & \texttt{21cmEMUv1} \\
\hline
$\log_{10} f_{\ast,10}$ & $\mathcal{N}(-1.0,\,1.1)$ on $[-2.5,\,0.5]$ & $\mathcal{U}[-2,\,-0.5]$ \\
$\alpha_{\ast}$ & $\mathcal{N}(0.8,\,0.2)$ on $[0,\,1]$ & $\mathcal{U}[0,\,1]$ \\
$t_{\ast}$ & $\mathcal{U}[0.01,\,1]$ & $\mathcal{U}[0.01,\,1]$ \\
$\log_{10} f_{\rm esc,10}$ & $\mathcal{U}[-3,\,0]$ & $\mathcal{U}[-3,\,0]$ \\
$\alpha_{\rm esc}$ & $\mathcal{U}[-1,\,1]$ & $\mathcal{U}[-1,\,1]$ \\
$\log_{10} f_{\ast,7}$ & $\mathcal{N}(-3.3,\,1.0)$ on $[-5,\,-1]$ & -- \\
$\log_{10} f_{\rm esc,7}$ & $\mathcal{U}[-3,\,-1]$ & -- \\
$\log_{10}\!\left(L_{\rm X<2keV}^{\textsc{acg}}/{\rm SFR}\right)$ & $\mathcal{U}[38,\,42]$ & $\mathcal{U}[38,\,42]$ \\
$\log_{10}\!\left(L_{\rm X<2keV}^{\textsc{mcg}}/{\rm SFR}\right)$ & $\mathcal{U}[39,\,43]$ & -- \\
$E_0$ [eV] & $\mathcal{U}[100,\,1500]$ & $\mathcal{U}[100,\,1500]$ \\
$\log_{10}\left(M_{\rm turn}/M_\odot\right)$ & -- & $\mathcal{U}[8,\,10]$ \\
$\alpha_X$ & -- & $\mathcal{U}[-1,\,3]$ \\
$\sigma_8$ & $0.8118$ (fixed) & -- \\
\hline
\end{tabular}
\caption{Parameter priors of every inference in this work (HERA, JWST, and mock SKA), for each emulator. $\mathcal{U}[a,b]$ is uniform on $[a,b]$; $\mathcal{N}(\mu,\sigma)$ on $[a,b]$ is a normal distribution with mean $\mu$ and standard deviation $\sigma$, truncated to $[a,b]$. The truncated normal priors of \texttt{21cmEMUv3} are the informed prior of Breitman et al.\ (2026)~\cite[Section~5.1]{Breitman26}, based on the Renaissance simulations \citep{Xu16}. In the HERA analyses, $L_{\rm X<2keV}^{\textsc{mcg}}/{\rm SFR}=L_{\rm X<2keV}^{\textsc{acg}}/{\rm SFR}$ (Section~\ref{sec:HERA}).}
\label{tab:priors}
\end{table}

\section{Conditional statistics from runs with the original prior}
\label{sec:conditional}

We sample every inference with the original prior, $p(\boldsymbol{\theta})$, rather than with $\mathcal{P}_C$, and obtain the conditional statistics of Section~\ref{sec:sus} from these runs. Assuming the likelihood contributions factorise at fixed $\boldsymbol{\theta}$, so that $\mathcal{L}_{XC}=\mathcal{L}_X\mathcal{L}_C$, the evidence of $d_X$ with $\mathcal{P}_C$ as the prior is $\mathcal{Z}_{X\mid C}=\mathcal{Z}_{XC}/\mathcal{Z}_C$. Here, $X\in\{A,B,AB\}$, and the evidences on the right are computed using the original prior. The conditional evidence ratio is consequently
\begin{equation}
    \log R_{AB\mid C}
    =
    \log\mathcal{Z}_{ABC}
    +\log\mathcal{Z}_C
    -\log\mathcal{Z}_{AC}
    -\log\mathcal{Z}_{BC}.
    \label{eq:conditional_logR}
\end{equation}
Where $\mathcal{L}_C$ confines the posteriors, widening the original prior rescales $\mathcal{Z}_{XC}$ and $\mathcal{Z}_C$ by the same factor, so $\log R_{AB\mid C}$ is then also insensitive to it. The Kullback--Leibler divergences relative to $\mathcal{P}_C$ are
\begin{align}
    \mathcal{D}_{X\mid C}
    &=
    \int
    \mathcal{P}_{XC}(\boldsymbol{\theta})
    \log\!\left[
        \frac{\mathcal{P}_{XC}(\boldsymbol{\theta})}
             {\mathcal{P}_C(\boldsymbol{\theta})}
    \right]
    \mathrm{d}\boldsymbol{\theta}
    \nonumber\\
    &=
    \left\langle
        \log\mathcal{L}_X
    \right\rangle_{\mathcal{P}_{XC}}
    -\log\mathcal{Z}_{X\mid C}.
    \label{eq:conditional_KL}
\end{align}
This gives
\begin{align}
    Q_{AB\mid C}
    &=
    \mathcal{D}_{A\mid C}
    +\mathcal{D}_{B\mid C}
    -\mathcal{D}_{AB\mid C},
    \label{eq:conditional_Q}\\
    \log S_{AB\mid C}
    &=
    \log R_{AB\mid C}-Q_{AB\mid C},
    \label{eq:conditional_logS}
\end{align}
or, equivalently,
\begin{align}
    \log S_{AB\mid C}
    &=
    \left\langle
        \log\mathcal{L}_A+\log\mathcal{L}_B
    \right\rangle_{\mathcal{P}_{ABC}}
    \nonumber\\
    &\quad-
    \left\langle
        \log\mathcal{L}_A
    \right\rangle_{\mathcal{P}_{AC}}
    -
    \left\langle
        \log\mathcal{L}_B
    \right\rangle_{\mathcal{P}_{BC}}.
    \label{eq:conditional_logS_likelihood}
\end{align}
The Bayesian model dimensionalities and their combination are
\begin{equation}
    d_{X\mid C}
    =
    2\,\operatorname{Var}_{\mathcal{P}_{XC}}
    \!\left(\log\mathcal{L}_X\right),
    \qquad X\in\{A,B,AB\},
    \label{eq:conditional_dimensionality}
\end{equation}
and
\begin{equation}
    d_G
    =
    d_{A\mid C}+d_{B\mid C}-d_{AB\mid C}.
    \label{eq:shared_dimensionality}
\end{equation}
We evaluate every posterior average and variance in these expressions from the weighted samples of the corresponding nested-sampling run, using the likelihood of the tested data alone, $\log\mathcal{L}_X=\log\mathcal{L}_{XC}-\log\mathcal{L}_C$, at each sample.

\section{HERA}
\label{sec:hera_app}

\begin{figure}[tbp]
    \centering
    \includegraphics[width=0.49\linewidth]{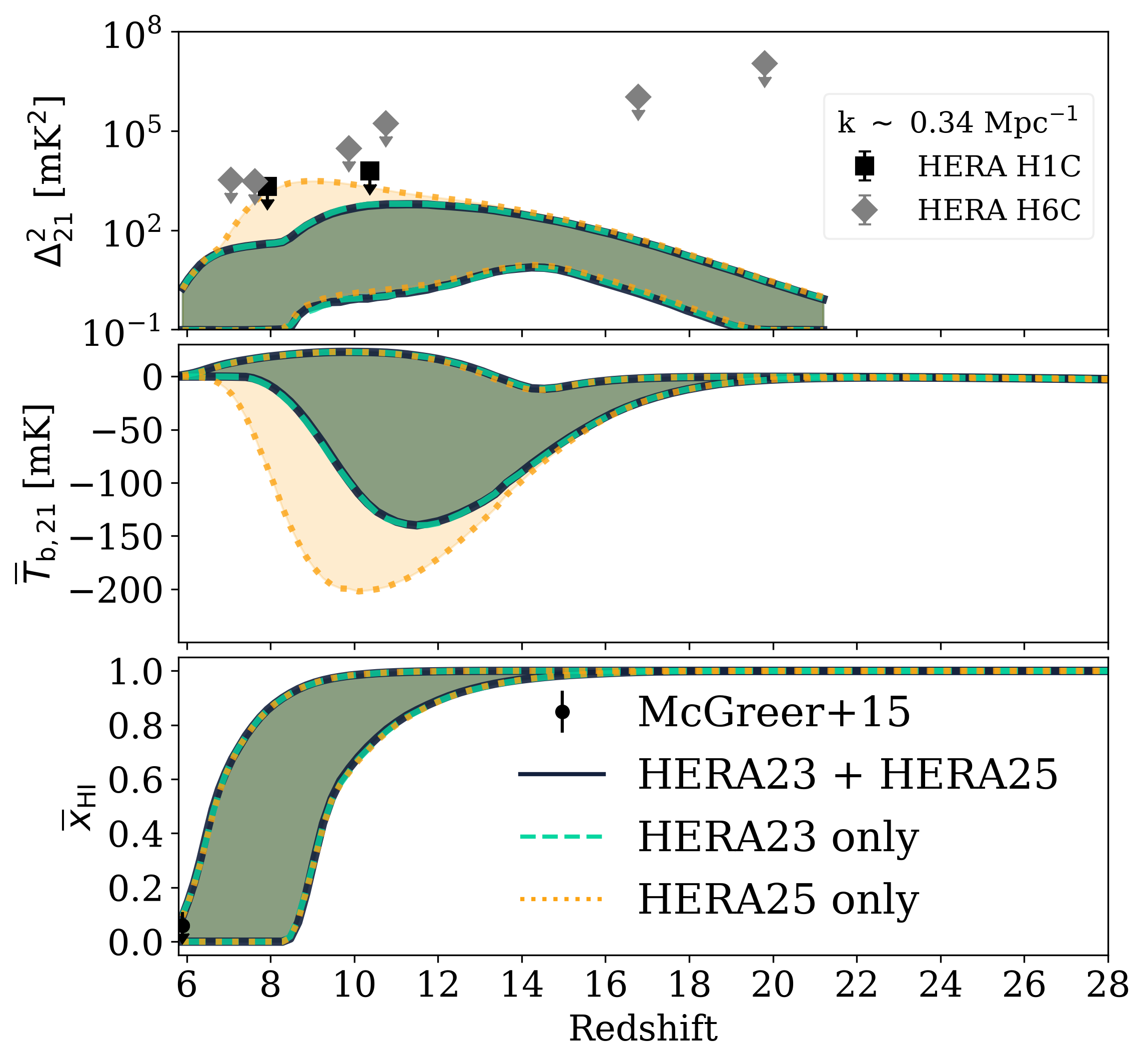}\hfill
    \includegraphics[width=0.49\linewidth]{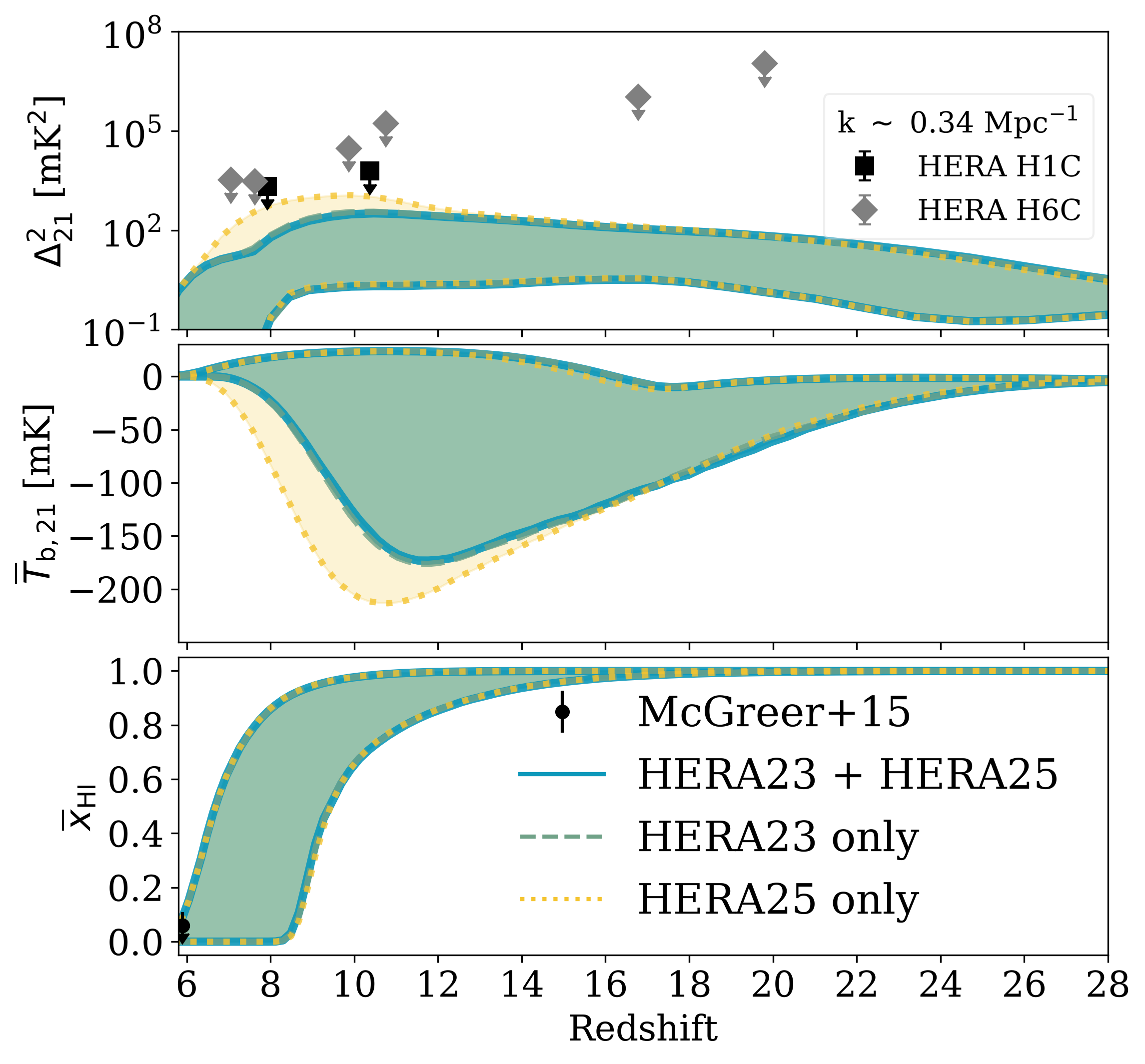}\\[0.5em]
    \includegraphics[width=0.49\linewidth]{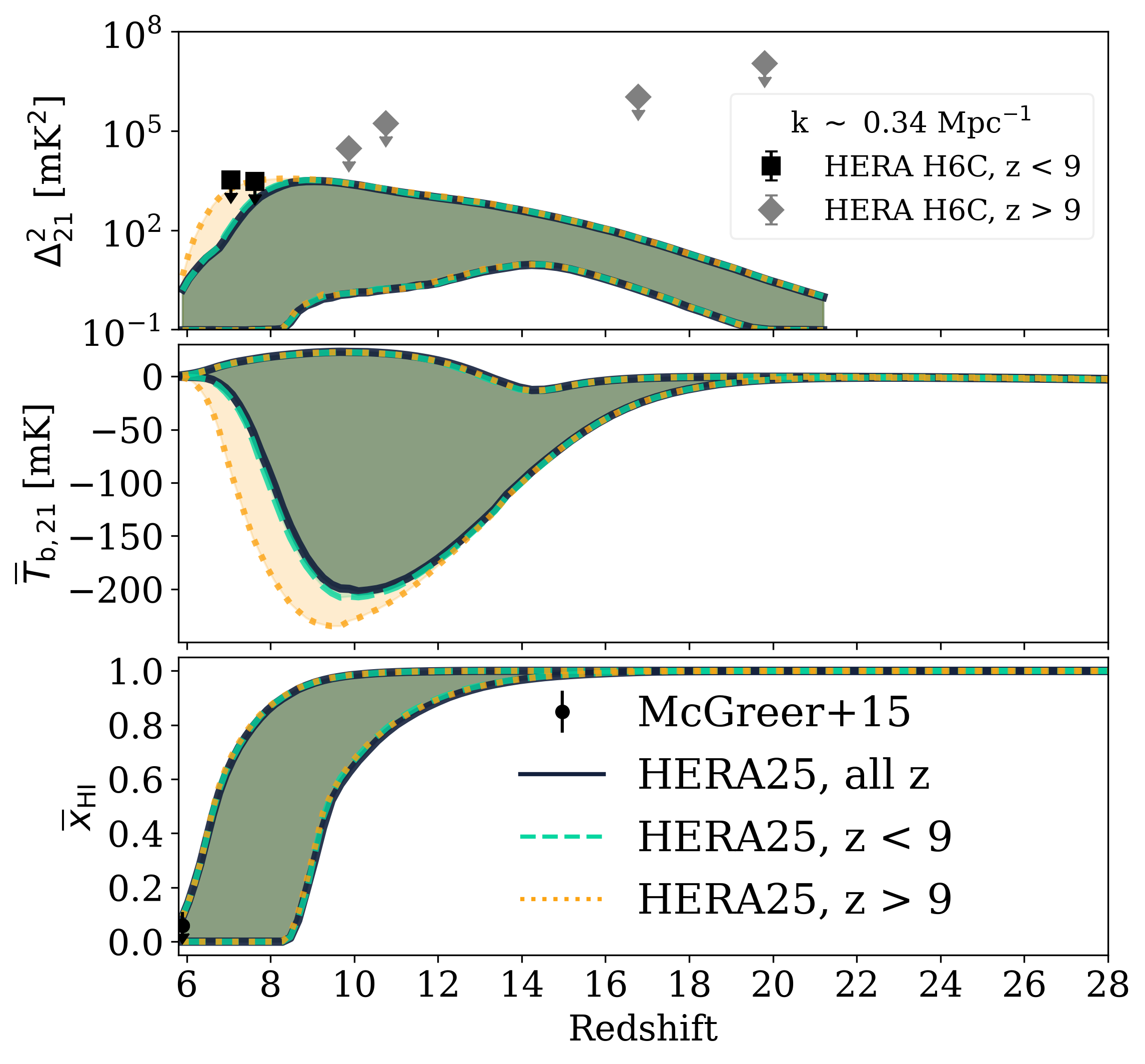}\hfill
    \includegraphics[width=0.49\linewidth]{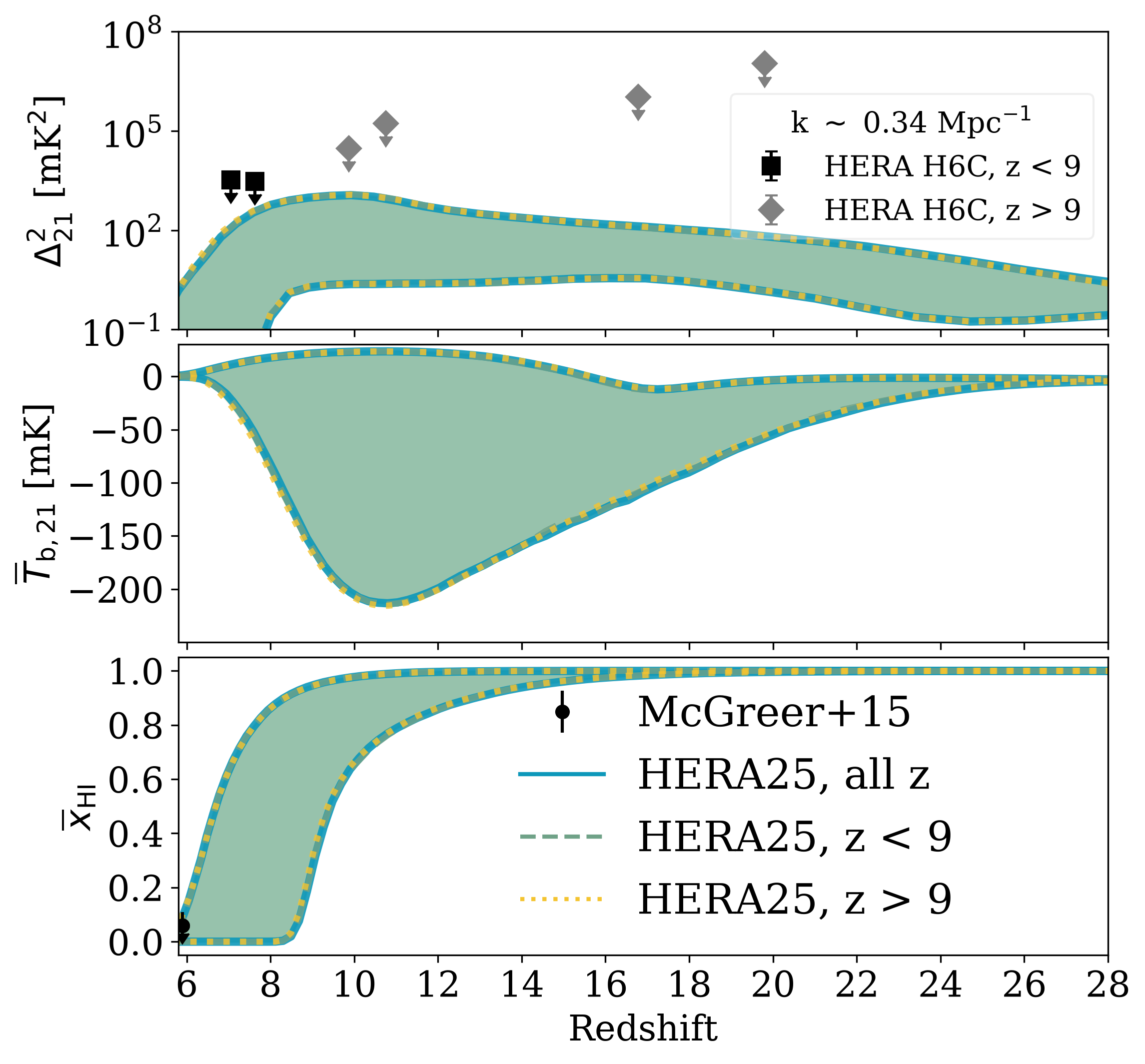}
    \caption{Posterior predictions of the HERA consistency tests with \texttt{21cmEMUv1} (left) and \texttt{21cmEMUv3} (right), for Test~1 (top: H23 and H25 jointly, H23 only, and H25 only) and Test~2 (bottom: H25 at all redshifts, $z<9$, and $z>9$). From top to bottom, each plot shows the 21-cm PS at $k\simeq0.34\,\mathrm{Mpc}^{-1}$ with the HERA upper limits (H23 and H25 are labelled H1C and H6C), the global 21-cm signal, and the neutral fraction with the dark-fraction bound \citep{McGreer15}. Solid, dashed, and dotted lines bound the 95\% intervals of the joint, $d_A$-only, and $d_B$-only inferences, all of which include $d_C$. With \texttt{21cmEMUv3}, ACGs and MCGs have the same X-ray luminosity per unit SFR.}
    \label{fig:hera_summaries}
\end{figure}

\begin{figure}[tbp]
    \centering
    \includegraphics[width=0.49\linewidth]{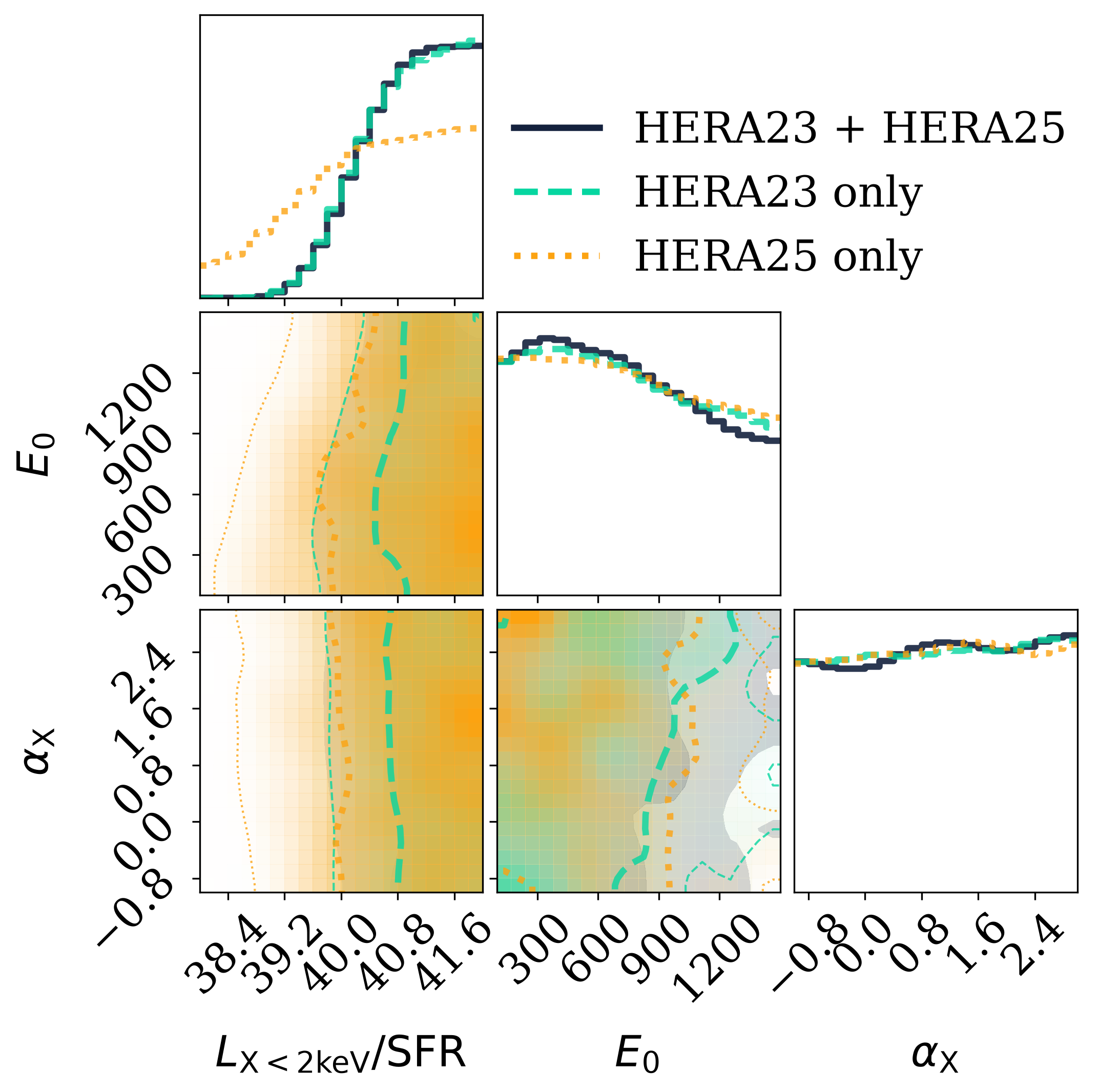}\hfill
    \includegraphics[width=0.49\linewidth]{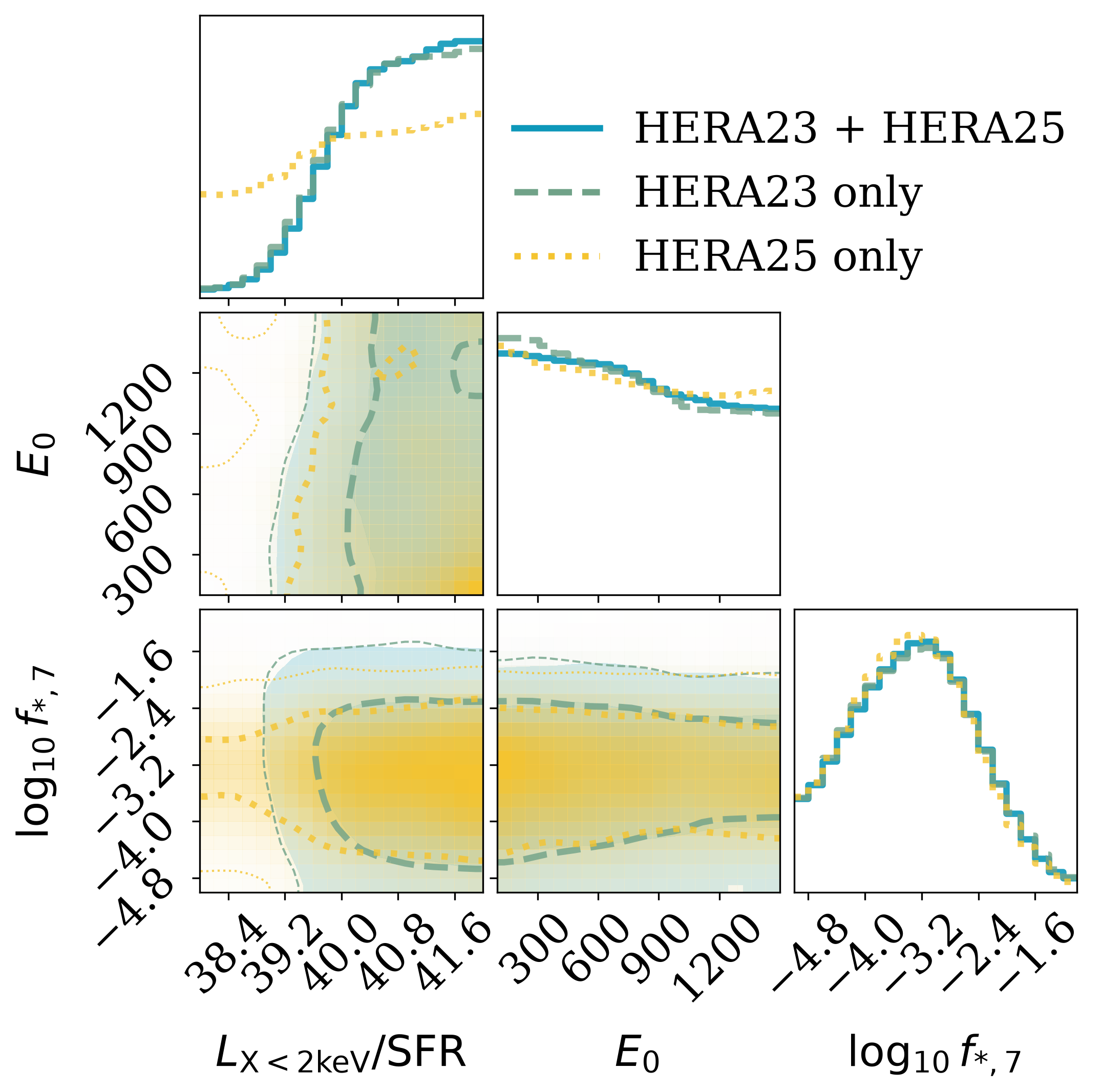}\\[0.5em]
    \includegraphics[width=0.49\linewidth]{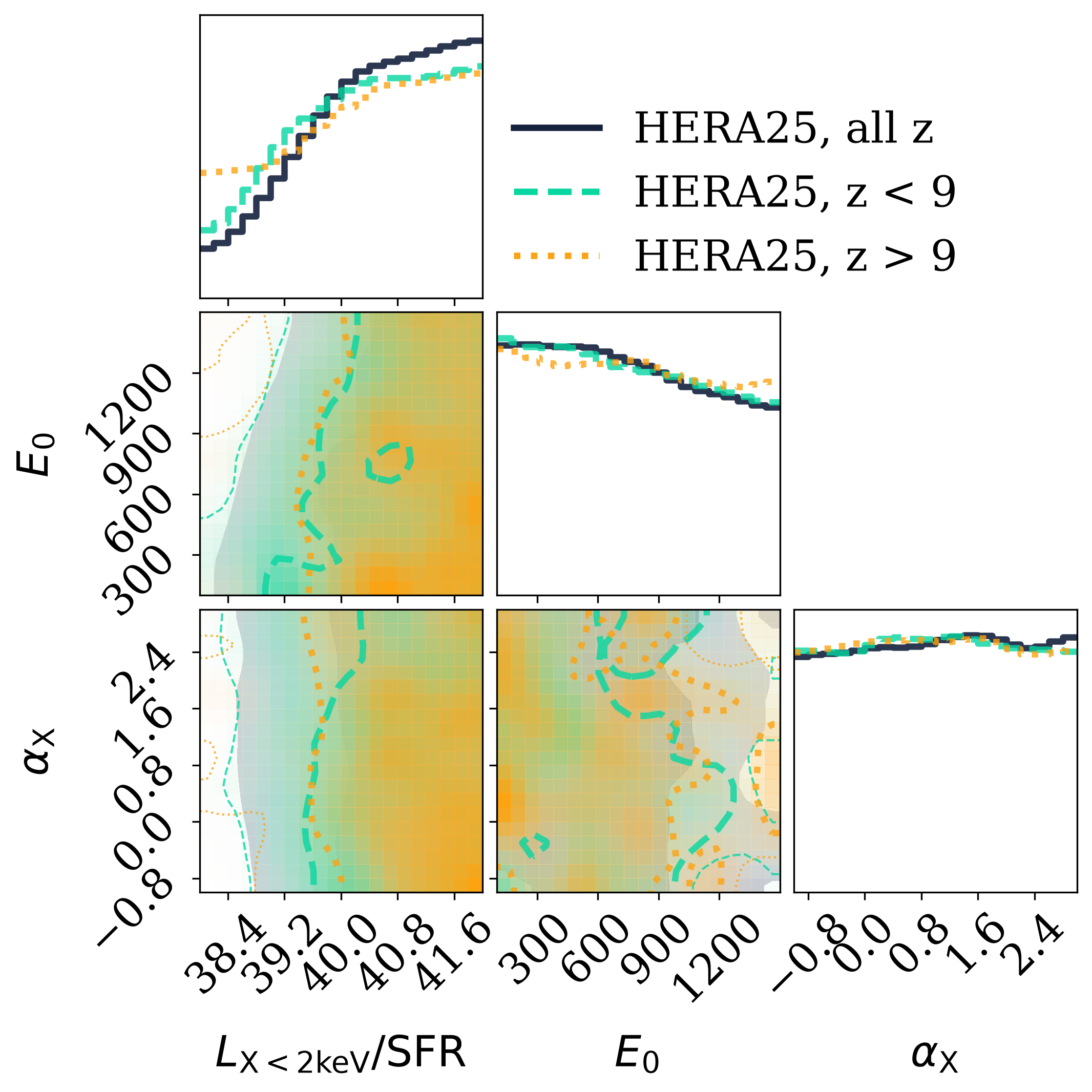}\hfill
    \includegraphics[width=0.49\linewidth]{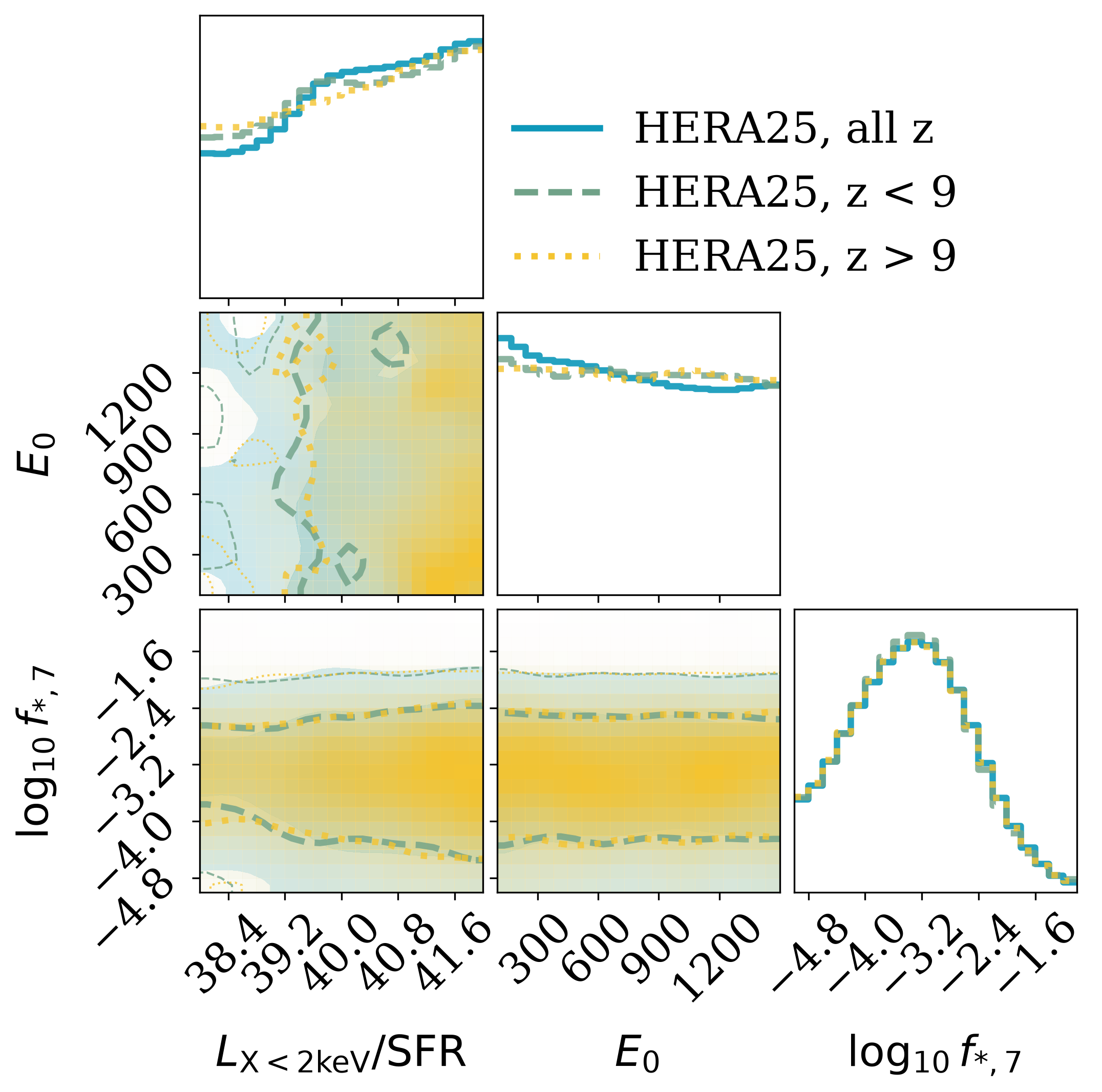}
    \caption{Posteriors of the X-ray parameters in the HERA consistency tests, marginalised over all other parameters, for Test~1 (top) and Test~2 (bottom). Left: \texttt{21cmEMUv1}, with $\log_{10}(L_{\rm X<2keV}/{\rm SFR})$, $E_0$, and $\alpha_X$. Right: \texttt{21cmEMUv3}, with $\log_{10}(L_{\rm X<2keV}/{\rm SFR})$, shared by ACGs and MCGs, $E_0$, and the MCG SHMR normalisation, $\log_{10}f_{\ast,7}$. Filled, dashed, and dotted contours enclose 68\% and 95\% of the posterior mass for the joint, $d_A$-only, and $d_B$-only inferences, respectively. Only $L_{\rm X<2keV}/{\rm SFR}$ is constrained beyond the prior and $d_C$.}
    \label{fig:hera_xray}
\end{figure}

Figures~\ref{fig:hera_summaries} and~\ref{fig:hera_xray} show the posterior predictions and the posteriors of the X-ray parameters of the HERA tests (Section~\ref{sec:HERA}). The remaining parameters have nearly identical posteriors in all inferences and are not shown.

\section{UV luminosity functions}
\label{sec:jwst_app}

\begin{figure}
    \centering
    \includegraphics[width=0.99\linewidth]{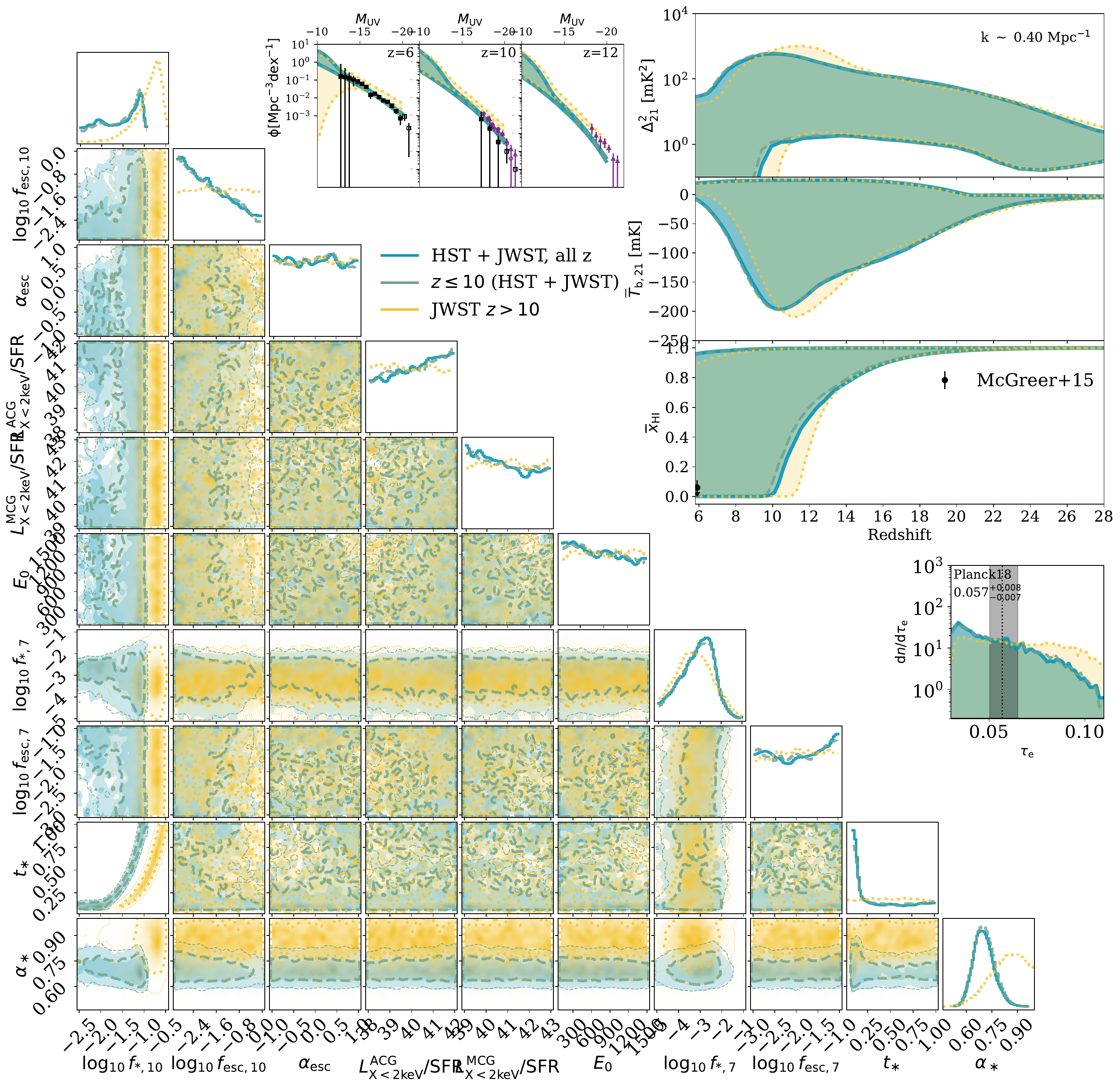}
    \caption{Full posteriors of Test~1 of the UV LFs under \texttt{21cmEMUv3}: the joint inference (filled, solid), the $z\leq10$ subset (dashed), and the $z>10$ subset (dotted). Contours enclose 68\% and 95\% of the posterior mass. The insets show the 95\% intervals of the posterior predictions for the UV LFs at $z=6$, 10, and 12, with the measurements, and for the 21-cm PS at $k\simeq0.40\,\mathrm{Mpc}^{-1}$, the global 21-cm signal, and the neutral fraction, and the predicted distribution of $\tau_e$; these inferences do not constrain the last four. The $\tau_e$ \citep{Planck18} and dark-fraction \citep{McGreer15} constraints are shown for reference only.}
    \label{fig:jwst_full_corner}
\end{figure}

Table~\ref{tab:jwst_glimpse} repeats Test~1 without the highest-redshift GLIMPSE interval ($z\simeq13$) and without any GLIMPSE measurements. The tension persists in both cases, with $\log S_{AB}=-10.77$ ($4.9\sigma$) and $-9.34$, respectively, although the negative $d_G$ of the latter prevents a calibrated significance. Including the $\tau_e$ and dark-fraction constraints as $d_C$ does not significantly affect $\log S$ without the $z\simeq13$ interval. Figure~\ref{fig:jwst_full_corner} shows the full posteriors of Test~1 (Section~\ref{sec:UVLFs}).

\begin{table}[h]
\centering
\begin{tabular}{clcrrrrc}
\hline
Test & Partition, $d_A$ vs $d_B$ & Model & $\log R$ & $Q$ & $\log S$ & $d_G$ & $n_\sigma$ \\
\hline
1 & $z\leq10$ vs $z>10$ & v3 & $-9.16$ & $2.46$ & $-11.62$ & $0.76$ & $5.0$ \\
-- & as Test 1, without GLIMPSE $z=13$ & v3 & $-8.50$ & $2.28$ & $-10.77$ & $0.64$ & $4.9$ \\
-- & as Test 1, without GLIMPSE & v3 & $-7.43$ & $1.92$ & $-9.34$ & $-0.46$ & -- \\
\hline
\end{tabular}
\caption{
As Table~\ref{tab:jwst_results} (Test~1, first row), without the GLIMPSE measurements at $z\simeq13$ (second row) or without any GLIMPSE measurements (third row). The equivalent Gaussian significance $n_\sigma$ (Equation~\ref{eq:tension_significance}) is quoted only for $d_G\geq0.5$. For the second row, it spans $4.5$--$5.0\sigma$ for $0.5\leq d_G\leq2$; the negative $d_G$ of the third row prevents a calibrated significance.}
\label{tab:jwst_glimpse}
\end{table}

\section{SKA}
\label{sec:ska_app}

\begin{figure}
    \centering
    \includegraphics[width=0.99\linewidth]{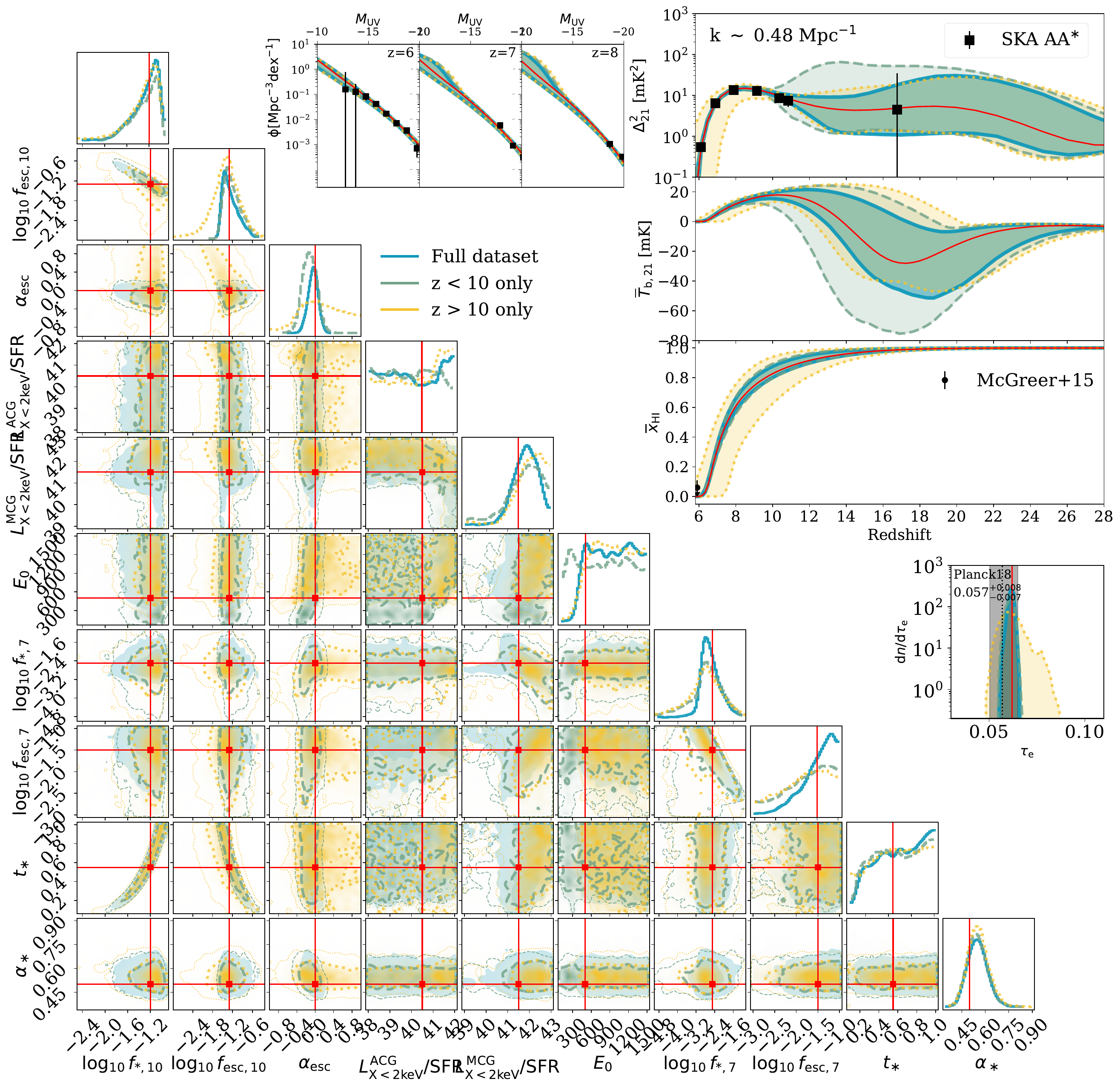}
    \caption{Full posteriors of the well-specified \texttt{21cmEMUv3} analysis of the mock SKA PS, for the full PS (filled, solid), the $z<10$ subset (dashed), and the $z>10$ subset (dotted), each including the UV LF, optical-depth, and dark-fraction constraints. Contours enclose 68\% and 95\% of the posterior mass. Red lines mark the mock parameter values (Table~\ref{tab:mock_params}). The insets show the 95\% intervals of the posterior predictions for the UV LFs at $z=6$, 7, and 8, the 21-cm PS at $k\simeq0.48\,\mathrm{Mpc}^{-1}$, the global 21-cm signal, and the neutral fraction, and the predicted distribution of $\tau_e$, with the mock in red and the measurements in black.}
    \label{fig:ska_noMM_full_corner}
\end{figure}

\begin{figure}
    \centering
    \includegraphics[width=0.99\linewidth]{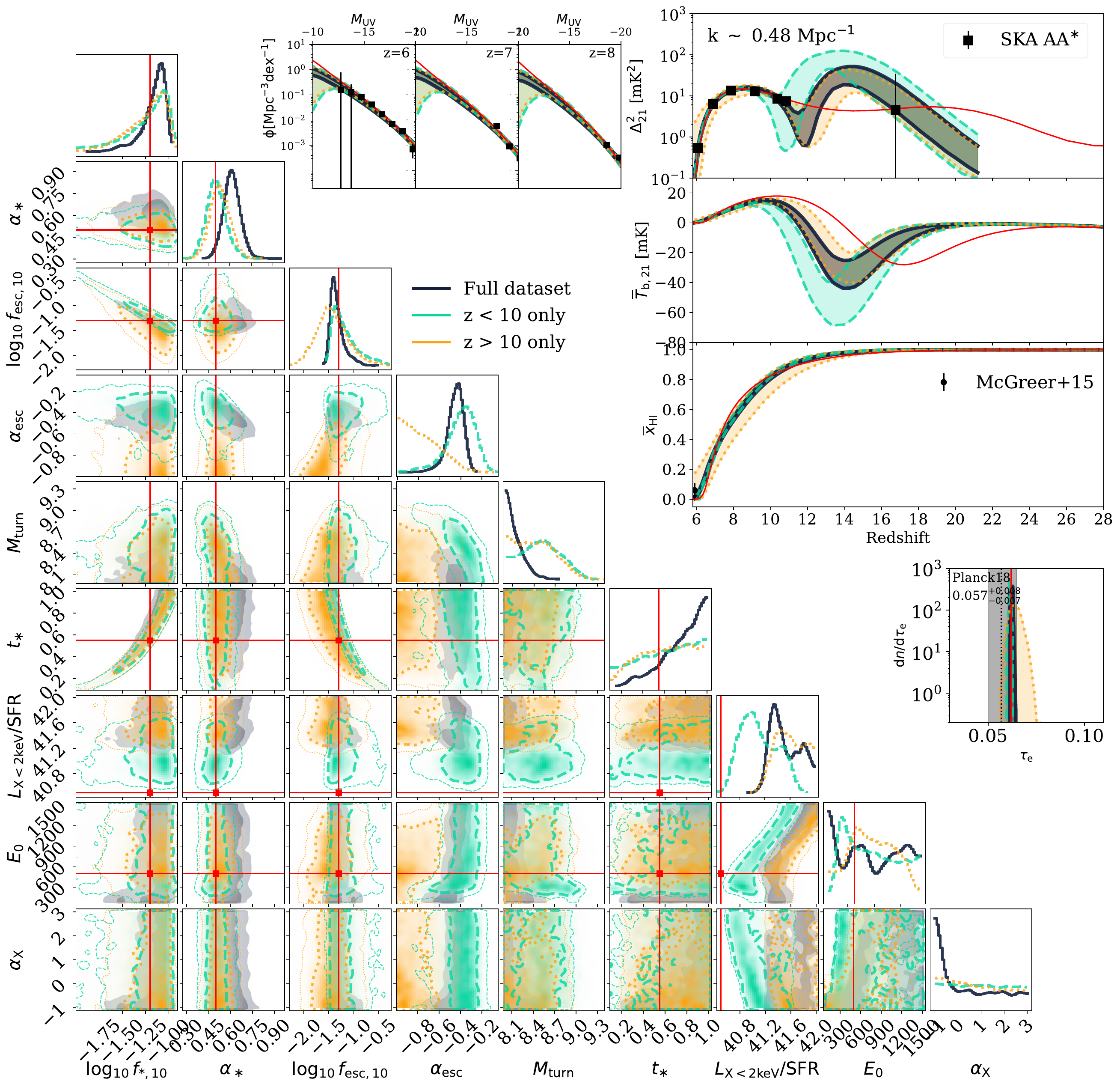}
    \caption{As Figure~\ref{fig:ska_noMM_full_corner}, for the misspecified \texttt{21cmEMUv1} analysis. Red lines mark the mock values of the parameters shared with \texttt{21cmEMUv3} where they lie within the plotted range (the mock $\alpha_{\rm esc}=0$ does not); $M_{\rm turn}$ and $\alpha_X$ have no mock counterpart.}
    \label{fig:ska_MM_full_corner}
\end{figure}

Figures~\ref{fig:ska_noMM_full_corner} and~\ref{fig:ska_MM_full_corner} show the full posteriors of Test~1 for the well-specified and misspecified analyses of Section~\ref{sec:ska}.

\subsection{Consistency between observables}
\label{sec:ska_obs}
Tests~2--4 use the same full likelihood as Test~1 but compare different observables (Table~\ref{tab:ska_inferences}): the PS versus the UV LFs, conditional on $\tau_e$ and $\overline{x}_{\rm HI}$ (Test~2); the PS versus $\tau_e$ and $\overline{x}_{\rm HI}$, conditional on the UV LFs (Test~3); and the UV LFs versus $\tau_e$ and $\overline{x}_{\rm HI}$, conditional on the PS (Test~4). Only the PS--UV LF comparison flags v1, with $\log S=-5.93$ ($3.5\sigma$ in the Gaussian limit), compared with $\log S=0.20$ for v3 (Table~\ref{tab:ska_obs_results}). The diagnostic therefore responds where the misspecified model must compromise, between PS redshift ranges and between the PS and UV LFs, but not in every partition of the likelihood. Because several partitions are tested, the significance of any single comparison should be read with a trials factor in mind \citep[e.g.][]{GrossVitells10}.

\begin{table}[h]
\centering
\begin{tabular}{clcrrrrc}
\hline
Test & $d_A$ vs $d_B$ $\mid$ $d_C$ & Model & $\log R$ & $Q$ & $\log S$ & $d_G$ & $n_\sigma$ \\
\hline
2 & PS vs UV $\mid$ $\tau_e$, $\overline{x}_{\rm HI}$ & v3 & $0.13$ & $-0.07$ & $0.20$ & $0.17$ & -- \\
  &                                                  & v1 & $-4.04$ & $1.89$ & $-5.93$ & $1.16$ & $3.5$ \\
3 & PS vs $\tau_e$, $\overline{x}_{\rm HI}$ $\mid$ UV & v3 & $1.51$ & $1.23$ & $0.27$ & $0.92$ & $0.7$ \\
  &                                                  & v1 & $1.50$ & $1.20$ & $0.30$ & $0.87$ & $0.6$ \\
4 & UV vs $\tau_e$, $\overline{x}_{\rm HI}$ $\mid$ PS & v3 & $-0.05$ & $-0.03$ & $-0.02$ & $0.13$ & -- \\
  &                                                  & v1 & $-0.10$ & $-0.02$ & $-0.08$ & $0.00$ & -- \\
\hline
\end{tabular}
\caption{
Consistency between observables in the mock SKA analysis (Tests~2--4). The mock is generated with the \texttt{21cmEMUv3} physical model, so the v3 analyses are well specified and the v1 analyses misspecified. ``UV'' denotes the UV LFs and ``PS'' the mock SKA PS. $\log R$, $Q$, $\log S$, and $d_G$ are the conditional quantities of Equations~\ref{eq:conditional_logR}--\ref{eq:shared_dimensionality}; $n_\sigma$ is quoted only for $d_G\geq0.5$.}
\label{tab:ska_obs_results}
\end{table}

\subsection{Calibration and robustness}
\label{sec:calibration}
\begin{figure}
    \centering
    \includegraphics[width=0.75\linewidth]{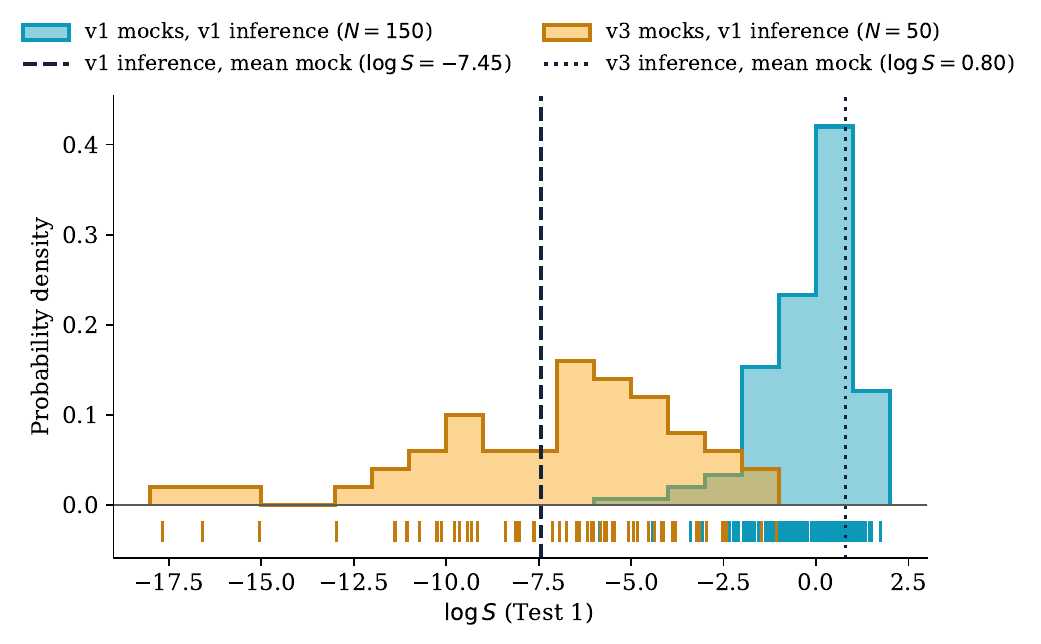}
    \caption{Test~1 suspiciousness for noisy realisations of the mock SKA PS, all analysed with \texttt{21cmEMUv1}: the null ensemble (blue), generated with v1 and hence well specified, and the misspecified ensemble (orange), generated with v3 (Section~\ref{sec:ska_mock}). Ticks below the axis mark individual realisations. The vertical lines mark the mean-mock values of Table~\ref{tab:ska_results} under v1 (dashed) and v3 (dotted).}
    \label{fig:ska_calibration}
\end{figure}

The mean mock of Section~\ref{sec:ska_mock} contains no noise realisation, and for v1 its negative $d_G$ prevents a Gaussian-limit significance. We therefore calibrate Test~1 with two ensembles of noisy mocks, both analysed with \texttt{21cmEMUv1} (Figure~\ref{fig:ska_calibration}). Each realisation adds independent Gaussian noise to every PS bin, with the variances of $\mathbf{C}_{\rm SKA}$, and is analysed as the mean mock: one fit to the full PS and one to the PS on each side of $z=10$, all including the UV LF, optical-depth, and dark-fraction constraints. These constraints are the same in every realisation, so the fit to them alone, $\mathcal{P}_C$, is shared. In the null ensemble, the mean is the \texttt{21cmEMUv1} prediction at the fiducial values of the parameters it shares with v3 (Table~\ref{tab:mock_params}), with $\log_{10}(M_{\rm turn}/M_\odot)=8$, the lower edge of its prior, and $\alpha_X=1$; the v1 analysis is then well specified. In the misspecified ensemble, the mean is the mock of Section~\ref{sec:ska_mock}. Both ensembles use the binning, covariance, and priors of the main analysis, and we include every completed realisation, without selection.

The $N=150$ null realisations have a median $\log S=0.14$, with 16th and 84th percentiles of $-1.34$ and $0.89$, and none lies below $-5.84$. All but three have $d_G>0.9$: two have $0<d_G<0.5$, and one, the lowest $\log S$ of the ensemble, has $d_G=-1.10$. For the 147 with $d_G\geq0.5$, the Gaussian-limit tail probabilities (Equation~\ref{eq:tension_probability}) are consistent with a uniform distribution in the tail relevant for a detection: 6 fall below 0.05 and 15 below 0.1. A Kolmogorov--Smirnov test nonetheless rejects uniformity at the 5\% level ($p=0.03$), owing to an excess of intermediate values (59\% lie below 0.5). For a well-specified model, the analytic calibration of Section~\ref{sec:sus} is therefore approximately, but not exactly, correct in this setting.

The $N=50$ misspecified realisations have a median $\log S=-6.13$, with 16th and 84th percentiles of $-10.15$ and $-3.87$. As for the mean mock, most of them (34 of 50) have $d_G<0$. The mean-mock value, $\log S=-7.45$, lies at the 38th percentile of this ensemble, so it is typical of a noisy observation rather than a favourable case. Of the 50 realisations, 48 (96\%; 86--99.5\% at 95\% confidence) lie below the 5th percentile of the null ensemble, and 28 (56\%; 41--70\%) below every null realisation, so a single SKA observation would reveal the inconsistency in most cases; the remaining two, with $\log S=-1.47$ and $-1.08$, overlap with the lower part of the null distribution. A goodness-of-fit test is much less sensitive. Over the 69 PS bins, the best-fitting v1 PS of the misspecified realisations has a median $\chi^2=74.7$ (16th--84th percentiles: 68.0--87.7), against 63.0 (50.6--73.3) for the null realisations, consistent with $69-9=60$ degrees of freedom. Only 15 of the 50 misspecified realisations exceed the 95th percentile of the $\chi^2_{60}$ distribution, as do 8 of the 150 null ones, whereas 48 fall below the 5th percentile of the null $\log S$ values, a threshold with the same 5\% false-alarm rate.

No null realisation is as discrepant as the v1 mean-mock value. When none of $N$ null realisations lies below the observed value, the empirical tail probability is $p=1/(N+1)$, the smallest that $N$ realisations can resolve, and the equivalent significance (Equation~\ref{eq:tension_significance}) is a lower bound: $n_\sigma\geq2.72$ for $N=150$. Larger values require extrapolating the tail of the null distribution. A distribution of the Gaussian-limit form, $\log S=a-X/2$ with $X\sim\chi^2_d$, matched to the mean and variance of the null ensemble, gives $3.4\sigma$ (3.3--$3.6\sigma$ for the 16th--84th percentiles of bootstrap resamplings of the ensemble); a normal distribution with the same moments, whose lower tail is lighter, gives $6.0\sigma$. For v3, $\log S=0.80$ exceeds 121 of the 150 null values ($0.2\sigma$), consistent with its Gaussian-limit value of $0.0\sigma$, although the null ensemble is built with v1 rather than v3.

Two limitations of the null ensemble affect the extrapolated values. First, it is centred on a single v1 parameter vector, with $M_{\rm turn}$ at the lower edge of its prior, rather than on the v1 fit to the mock, and only one of its realisations has $d_G<0$, whereas the mean mock and 34 of the 50 misspecified realisations do. A null ensemble centred on the v1 fit to the mock would remove this difference. Second, its mocks are generated with the emulator used to analyse them, so they contain neither emulation error nor the residual sample variance of the 29-simulation mean (Section~\ref{sec:ska_mock}), whereas the misspecified analyses test the v1 emulator and its physical model together. Both effects are small compared with the thermal noise: the best \texttt{21cmEMUv3} fit to the noise-free mean mock has $\chi^2=0.4$ over the 69 PS bins.

\bibliographystyle{JHEP}
\bibliography{refs}

@ARTICLE{Eddington13,
       author = {{Eddington}, A.~S.},
        title = "{On a formula for correcting statistics for the effects of a known error of observation}",
      journal = {\mnras},
         year = 1913,
       volume = {73},
        pages = {359},
          doi = {10.1093/mnras/73.5.359}
}

@ARTICLE{Box80,
       author = {{Box}, George E.~P.},
        title = "{Sampling and Bayes' Inference in Scientific Modelling and Robustness}",
      journal = {Journal of the Royal Statistical Society. Series A (General)},
         year = 1980,
       volume = {143},
       number = {4},
        pages = {383--430},
          doi = {10.2307/2982063}
}

@ARTICLE{Oke83,
       author = {{Oke}, J.~B. and {Gunn}, J.~E.},
        title = "{Secondary standard stars for absolute spectrophotometry.}",
      journal = {\apj},
         year = 1983,
        month = mar,
       volume = {266},
        pages = {713-717},
          doi = {10.1086/160817},
       adsurl = {https://ui.adsabs.harvard.edu/abs/1983ApJ...266..713O}
}

@BOOK{BernardoSmith94,
       author = {{Bernardo}, Jos{\'e} M. and {Smith}, Adrian F.~M.},
        title = "{Bayesian Theory}",
    publisher = {John Wiley \& Sons},
      address = {New York},
       series = {Wiley Series in Probability and Statistics},
         year = 1994,
          doi = {10.1002/9780470316870}
}

@ARTICLE{Madau97,
       author = {{Madau}, Piero and {Meiksin}, Avery and {Rees}, Martin J.},
        title = "{21 Centimeter Tomography of the Intergalactic Medium at High Redshift}",
      journal = {\apj},
         year = 1997,
        month = feb,
       volume = {475},
       number = {2},
        pages = {429-444},
          doi = {10.1086/303549},
archivePrefix = {arXiv},
       eprint = {astro-ph/9608010},
 primaryClass = {astro-ph},
       adsurl = {https://ui.adsabs.harvard.edu/abs/1997ApJ...475..429M}
}

@ARTICLE{Scoccimarro98,
       author = {{Scoccimarro}, Roman},
        title = "{Transients from initial conditions: a perturbative analysis}",
      journal = {\mnras},
         year = 1998,
        month = oct,
       volume = {299},
       number = {4},
        pages = {1097-1118},
          doi = {10.1046/j.1365-8711.1998.01845.x},
archivePrefix = {arXiv},
       eprint = {astro-ph/9711187},
 primaryClass = {astro-ph},
       adsurl = {https://ui.adsabs.harvard.edu/abs/1998MNRAS.299.1097S}
}

@ARTICLE{Furlanetto06,
       author = {{Furlanetto}, Steven R.},
        title = "{The global 21-centimeter background from high redshifts}",
      journal = {\mnras},
         year = 2006,
        month = sep,
       volume = {371},
       number = {2},
        pages = {867-878},
          doi = {10.1111/j.1365-2966.2006.10725.x},
archivePrefix = {arXiv},
       eprint = {astro-ph/0604040},
 primaryClass = {astro-ph},
       adsurl = {https://ui.adsabs.harvard.edu/abs/2006MNRAS.371..867F}
}

@ARTICLE{Mesinger07,
       author = {{Mesinger}, Andrei and {Furlanetto}, Steven},
        title = "{Efficient Simulations of Early Structure Formation and Reionization}",
      journal = {\apj},
         year = 2007,
        month = nov,
       volume = {669},
       number = {2},
        pages = {663-675},
          doi = {10.1086/521806},
archivePrefix = {arXiv},
       eprint = {0704.0946},
 primaryClass = {astro-ph},
       adsurl = {https://ui.adsabs.harvard.edu/abs/2007ApJ...669..663M}
}

@ARTICLE{GrossVitells10,
       author = {{Gross}, Eilam and {Vitells}, Ofer},
        title = "{Trial factors for the look elsewhere effect in high energy physics}",
      journal = {European Physical Journal C},
         year = 2010,
       volume = {70},
        pages = {525},
          doi = {10.1140/epjc/s10052-010-1470-8},
archivePrefix = {arXiv},
       eprint = {1005.1891},
 primaryClass = {physics.data-an}
}

@ARTICLE{Mesinger10,
       author = {{Mesinger}, Andrei},
        title = "{Was reionization complete by z \raisebox{-0.5ex}\textasciitilde 5-6?}",
      journal = {\mnras},
         year = 2010,
        month = sep,
       volume = {407},
       number = {2},
        pages = {1328-1337},
          doi = {10.1111/j.1365-2966.2010.16995.x},
archivePrefix = {arXiv},
       eprint = {0910.4161},
 primaryClass = {astro-ph.CO},
       adsurl = {https://ui.adsabs.harvard.edu/abs/2010MNRAS.407.1328M}
}

@ARTICLE{Parsons10,
       author = {{Parsons}, Aaron R. and {Backer}, Donald C. and others},
        title = "{The Precision Array for Probing the Epoch of Re-ionization: Eight Station Results}",
      journal = {\aj},
         year = 2010,
        month = apr,
       volume = {139},
       number = {4},
        pages = {1468},
          doi = {10.1088/0004-6256/139/4/1468},
archivePrefix = {arXiv},
       eprint = {0904.2334},
 primaryClass = {astro-ph.CO},
       adsurl = {https://ui.adsabs.harvard.edu/abs/2010AJ....139.1468P}
}

@ARTICLE{Mesinger11,
       author = {{Mesinger}, Andrei and {Furlanetto}, Steven and {Cen}, Renyue},
        title = "{21CMFAST: a fast, seminumerical simulation of the high-redshift 21-cm signal}",
      journal = {\mnras},
         year = 2011,
        month = feb,
       volume = {411},
       number = {2},
        pages = {955-972},
          doi = {10.1111/j.1365-2966.2010.17731.x},
archivePrefix = {arXiv},
       eprint = {1003.3878},
 primaryClass = {astro-ph.CO},
       adsurl = {https://ui.adsabs.harvard.edu/abs/2011MNRAS.411..955M}
}

@ARTICLE{Kleijn12,
       author = {{Kleijn}, B.~J.~K. and {van der Vaart}, A.~W.},
        title = "{The Bernstein-Von-Mises theorem under misspecification}",
      journal = {Electronic Journal of Statistics},
         year = 2012,
       volume = {6},
        pages = {354--381},
          doi = {10.1214/12-EJS675}
}

@ARTICLE{Pritchard12,
       author = {{Pritchard}, Jonathan R. and {Loeb}, Abraham},
        title = "{21 cm cosmology in the 21st century}",
      journal = {Reports on Progress in Physics},
         year = 2012,
        month = aug,
       volume = {75},
       number = {8},
          eid = {086901},
        pages = {086901},
          doi = {10.1088/0034-4885/75/8/086901},
archivePrefix = {arXiv},
       eprint = {1109.6012},
 primaryClass = {astro-ph.CO},
       adsurl = {https://ui.adsabs.harvard.edu/abs/2012RPPh...75h6901P}
}

@ARTICLE{VehtariOjanen12,
       author = {{Vehtari}, Aki and {Ojanen}, Janne},
        title = "{A survey of Bayesian predictive methods for model assessment, selection and comparison}",
      journal = {Statistics Surveys},
         year = 2012,
       volume = {6},
        pages = {142--228},
          doi = {10.1214/12-SS102}
}

@ARTICLE{Fialkov13,
       author = {{Fialkov}, Anastasia and {Barkana}, Rennan and {Visbal}, Eli and {Tseliakhovich}, Dmitriy and {Hirata}, Christopher M.},
        title = "{The 21-cm signature of the first stars during the Lyman-Werner feedback era}",
      journal = {\mnras},
         year = 2013,
        month = jul,
       volume = {432},
       number = {4},
        pages = {2909-2916},
          doi = {10.1093/mnras/stt650},
archivePrefix = {arXiv},
       eprint = {1212.0513},
 primaryClass = {astro-ph.CO},
       adsurl = {https://ui.adsabs.harvard.edu/abs/2013MNRAS.432.2909F}
}

@ARTICLE{GelmanShalizi13,
       author = {{Gelman}, Andrew and {Shalizi}, Cosma Rohilla},
        title = "{Philosophy and the practice of Bayesian statistics}",
      journal = {British Journal of Mathematical and Statistical Psychology},
         year = 2013,
       volume = {66},
       number = {1},
        pages = {8--38},
archivePrefix = {arXiv},
       eprint = {1006.3868},
 primaryClass = {stat.ME}
}

@ARTICLE{Haarlem13,
       author = {{van Haarlem}, M.~P. and {Wise}, M.~W. and {Gunst}, A.~W. and {Heald}, G. and {McKean}, J.~P. and {Hessels}, J.~W.~T. and {de Bruyn}, A.~G. and {Nijboer}, R. and {Swinbank}, J. and {Fallows}, R. and {Brentjens}, M. and {Nelles}, A. and {Beck}, R. and {Falcke}, H. and {Fender}, R. and {H{\"o}randel}, J. and {Koopmans}, L.~V.~E. and {Mann}, G. and {Miley}, G. and {R{\"o}ttgering}, H. and {Stappers}, B.~W. and {Wijers}, R.~A.~M.~J. and {Zaroubi}, S. and {van den Akker}, M. and {Alexov}, A. and {Anderson}, J. and {Anderson}, K. and {van Ardenne}, A. and {Arts}, M. and {Asgekar}, A. and {Avruch}, I.~M. and {Batejat}, F. and {B{\"a}hren}, L. and {Bell}, M.~E. and {Bell}, M.~R. and {van Bemmel}, I. and {Bennema}, P. and {Bentum}, M.~J. and {Bernardi}, G. and {Best}, P. and {B{\^\i}rzan}, L. and {Bonafede}, A. and {Boonstra}, A. -J. and {Braun}, R. and {Bregman}, J. and {Breitling}, F. and {van de Brink}, R.~H. and {Broderick}, J. and {Broekema}, P.~C. and {Brouw}, W.~N. and {Br{\"u}ggen}, M. and {Butcher}, H.~R. and {van Cappellen}, W. and {Ciardi}, B. and {Coenen}, T. and {Conway}, J. and {Coolen}, A. and {Corstanje}, A. and {Damstra}, S. and {Davies}, O. and {Deller}, A.~T. and {Dettmar}, R. -J. and {van Diepen}, G. and {Dijkstra}, K. and {Donker}, P. and {Doorduin}, A. and {Dromer}, J. and {Drost}, M. and {van Duin}, A. and {Eisl{\"o}ffel}, J. and {van Enst}, J. and {Ferrari}, C. and {Frieswijk}, W. and {Gankema}, H. and {Garrett}, M.~A. and {de Gasperin}, F. and {Gerbers}, M. and {de Geus}, E. and {Grie{\ss}meier}, J. -M. and {Grit}, T. and {Gruppen}, P. and {Hamaker}, J.~P. and {Hassall}, T. and {Hoeft}, M. and {Holties}, H.~A. and {Horneffer}, A. and {van der Horst}, A. and {van Houwelingen}, A. and {Huijgen}, A. and {Iacobelli}, M. and {Intema}, H. and {Jackson}, N. and {Jelic}, V. and {de Jong}, A. and {Juette}, E. and {Kant}, D. and {Karastergiou}, A. and {Koers}, A. and {Kollen}, H. and {Kondratiev}, V.~I. and {Kooistra}, E. and {Koopman}, Y. and {Koster}, A. and {Kuniyoshi}, M. and {Kramer}, M. and {Kuper}, G. and {Lambropoulos}, P. and {Law}, C. and {van Leeuwen}, J. and {Lemaitre}, J. and {Loose}, M. and {Maat}, P. and {Macario}, G. and {Markoff}, S. and {Masters}, J. and {McFadden}, R.~A. and {McKay-Bukowski}, D. and {Meijering}, H. and {Meulman}, H. and {Mevius}, M. and {Middelberg}, E. and {Millenaar}, R. and {Miller-Jones}, J.~C.~A. and {Mohan}, R.~N. and {Mol}, J.~D. and {Morawietz}, J. and {Morganti}, R. and {Mulcahy}, D.~D. and {Mulder}, E. and {Munk}, H. and {Nieuwenhuis}, L. and {van Nieuwpoort}, R. and {Noordam}, J.~E. and {Norden}, M. and {Noutsos}, A. and {Offringa}, A.~R. and {Olofsson}, H. and {Omar}, A. and {Orr{\'u}}, E. and {Overeem}, R. and {Paas}, H. and {Pandey-Pommier}, M. and {Pandey}, V.~N. and {Pizzo}, R. and {Polatidis}, A. and {Rafferty}, D. and {Rawlings}, S. and {Reich}, W. and {de Reijer}, J. -P. and {Reitsma}, J. and {Renting}, G.~A. and {Riemers}, P. and {Rol}, E. and {Romein}, J.~W. and {Roosjen}, J. and {Ruiter}, M. and {Scaife}, A. and {van der Schaaf}, K. and {Scheers}, B. and {Schellart}, P. and {Schoenmakers}, A. and {Schoonderbeek}, G. and {Serylak}, M. and {Shulevski}, A. and {Sluman}, J. and {Smirnov}, O. and {Sobey}, C. and {Spreeuw}, H. and {Steinmetz}, M. and {Sterks}, C.~G.~M. and {Stiepel}, H. -J. and {Stuurwold}, K. and {Tagger}, M. and {Tang}, Y. and {Tasse}, C. and {Thomas}, I. and {Thoudam}, S. and {Toribio}, M.~C. and {van der Tol}, B. and {Usov}, O. and {van Veelen}, M. and {van der Veen}, A. -J. and {ter Veen}, S. and {Verbiest}, J.~P.~W. and {Vermeulen}, R. and {Vermaas}, N. and {Vocks}, C. and {Vogt}, C. and {de Vos}, M. and {van der Wal}, E. and {van Weeren}, R. and {Weggemans}, H. and {Weltevrede}, P. and {White}, S. and {Wijnholds}, S.~J. and {Wilhelmsson}, T. and {Wucknitz}, O. and {Yatawatta}, S. and {Zarka}, P. and {Zensus}, A. and {van Zwieten}, J.},
        title = "{LOFAR: The LOw-Frequency ARray}",
      journal = {\aap},
         year = 2013,
        month = aug,
       volume = {556},
          eid = {A2},
        pages = {A2},
          doi = {10.1051/0004-6361/201220873},
archivePrefix = {arXiv},
       eprint = {1305.3550},
 primaryClass = {astro-ph.IM},
       adsurl = {https://ui.adsabs.harvard.edu/abs/2013A&A...556A...2V}
}

@ARTICLE{Mellema13,
       author = {{Mellema}, Garrelt and {Koopmans}, L{\'e}on V.~E. and {Abdalla}, Filipe A. and {Bernardi}, Gianni and {Ciardi}, Benedetta and {Daiboo}, Soobash and {de Bruyn}, A.~G. and {Datta}, Kanan K. and {Falcke}, Heino and {Ferrara}, Andrea and {Iliev}, Ilian T. and {Iocco}, Fabio and {Jeli{\'c}}, Vibor and {Jensen}, Hannes and {Joseph}, Ronniy and {Labroupoulos}, Panos and {Meiksin}, Avery and {Mesinger}, Andrei and {Offringa}, Andr{\'e} R. and {Pandey}, V.~N. and {Pritchard}, Jonathan R. and {Santos}, Mario G. and {Schwarz}, Dominik J. and {Semelin}, Benoit and {Vedantham}, Harish and {Yatawatta}, Sarod and {Zaroubi}, Saleem},
        title = "{Reionization and the Cosmic Dawn with the Square Kilometre Array}",
      journal = {Experimental Astronomy},
         year = 2013,
        month = aug,
       volume = {36},
       number = {1-2},
        pages = {235-318},
          doi = {10.1007/s10686-013-9334-5},
archivePrefix = {arXiv},
       eprint = {1210.0197},
 primaryClass = {astro-ph.CO},
       adsurl = {https://ui.adsabs.harvard.edu/abs/2013ExA....36..235M}
}

@ARTICLE{Pober13,
       author = {{Pober}, Jonathan C. and {Parsons}, Aaron R. and {DeBoer}, David R. and {McDonald}, Patrick and {McQuinn}, Matthew and {Aguirre}, James E. and {Ali}, Zaki and {Bradley}, Richard F. and {Chang}, Tzu-Ching and {Morales}, Miguel F.},
        title = "{The Baryon Acoustic Oscillation Broadband and Broad-beam Array: Design Overview and Sensitivity Forecasts}",
      journal = {\aj},
         year = 2013,
        month = mar,
       volume = {145},
       number = {3},
          eid = {65},
        pages = {65},
          doi = {10.1088/0004-6256/145/3/65},
archivePrefix = {arXiv},
       eprint = {1210.2413},
 primaryClass = {astro-ph.CO},
       adsurl = {https://ui.adsabs.harvard.edu/abs/2013AJ....145...65P}
}

@ARTICLE{Tingay13,
       author = {{Tingay}, S.~J. and {Goeke}, R. and {Bowman}, J.~D. and {Emrich}, D. and {Ord}, S.~M. and {Mitchell}, D.~A. and {Morales}, M.~F. and {Booler}, T. and {Crosse}, B. and {Wayth}, R.~B. and {Lonsdale}, C.~J. and {Tremblay}, S. and {Pallot}, D. and {Colegate}, T. and {Wicenec}, A. and {Kudryavtseva}, N. and {Arcus}, W. and {Barnes}, D. and {Bernardi}, G. and {Briggs}, F. and {Burns}, S. and {Bunton}, J.~D. and {Cappallo}, R.~J. and {Corey}, B.~E. and {Deshpande}, A. and {Desouza}, L. and {Gaensler}, B.~M. and {Greenhill}, L.~J. and {Hall}, P.~J. and {Hazelton}, B.~J. and {Herne}, D. and {Hewitt}, J.~N. and {Johnston-Hollitt}, M. and {Kaplan}, D.~L. and {Kasper}, J.~C. and {Kincaid}, B.~B. and {Koenig}, R. and {Kratzenberg}, E. and {Lynch}, M.~J. and {Mckinley}, B. and {Mcwhirter}, S.~R. and {Morgan}, E. and {Oberoi}, D. and {Pathikulangara}, J. and {Prabu}, T. and {Remillard}, R.~A. and {Rogers}, A.~E.~E. and {Roshi}, A. and {Salah}, J.~E. and {Sault}, R.~J. and {Udaya-Shankar}, N. and {Schlagenhaufer}, F. and {Srivani}, K.~S. and {Stevens}, J. and {Subrahmanyan}, R. and {Waterson}, M. and {Webster}, R.~L. and {Whitney}, A.~R. and {Williams}, A. and {Williams}, C.~L. and {Wyithe}, J.~S.~B.},
        title = "{The Murchison Widefield Array: The Square Kilometre Array Precursor at Low Radio Frequencies}",
      journal = {\pasa},
         year = 2013,
        month = jan,
       volume = {30},
          eid = {e007},
        pages = {e007},
          doi = {10.1017/pasa.2012.007},
archivePrefix = {arXiv},
       eprint = {1206.6945},
 primaryClass = {astro-ph.IM},
       adsurl = {https://ui.adsabs.harvard.edu/abs/2013PASA...30....7T}
}

@ARTICLE{Walker13,
       author = {{Walker}, Stephen G.},
        title = "{Bayesian inference with misspecified models}",
      journal = {Journal of Statistical Planning and Inference},
         year = 2013,
       volume = {143},
       number = {10},
        pages = {1621--1633},
          doi = {10.1016/j.jspi.2013.05.013}
}

@ARTICLE{Madau14,
       author = {{Madau}, Piero and {Dickinson}, Mark},
        title = "{Cosmic Star-Formation History}",
      journal = {\araa},
         year = 2014,
        month = aug,
       volume = {52},
        pages = {415-486},
          doi = {10.1146/annurev-astro-081811-125615},
archivePrefix = {arXiv},
       eprint = {1403.0007},
 primaryClass = {astro-ph.CO},
       adsurl = {https://ui.adsabs.harvard.edu/abs/2014ARA&A..52..415M}
}

@ARTICLE{Pober14,
       author = {{Pober}, Jonathan C. and {Liu}, Adrian and {Dillon}, Joshua S. and {Aguirre}, James E. and {Bowman}, Judd D. and {Bradley}, Richard F. and {Carilli}, Chris L. and {DeBoer}, David R. and {Hewitt}, Jacqueline N. and {Jacobs}, Daniel C. and {McQuinn}, Matthew and {Morales}, Miguel F. and {Parsons}, Aaron R. and {Tegmark}, Max and {Werthimer}, Dan J.},
        title = "{What Next-generation 21 cm Power Spectrum Measurements can Teach us About the Epoch of Reionization}",
      journal = {\apj},
         year = 2014,
        month = feb,
       volume = {782},
       number = {2},
          eid = {66},
        pages = {66},
          doi = {10.1088/0004-637X/782/2/66},
archivePrefix = {arXiv},
       eprint = {1310.7031},
 primaryClass = {astro-ph.CO},
       adsurl = {https://ui.adsabs.harvard.edu/abs/2014ApJ...782...66P}
}

@ARTICLE{Bouwens15,
       author = {{Bouwens}, R.~J. and {Illingworth}, G.~D. and {Oesch}, P.~A. and {Trenti}, M. and {Labb{\'e}}, I. and {Bradley}, L. and {Carollo}, M. and {van Dokkum}, P.~G. and {Gonzalez}, V. and {Holwerda}, B. and {Franx}, M. and {Spitler}, L. and {Smit}, R. and {Magee}, D.},
        title = "{UV Luminosity Functions at Redshifts z {\ensuremath{\sim}} 4 to z {\ensuremath{\sim}} 10: 10,000 Galaxies from HST Legacy Fields}",
      journal = {\apj},
         year = 2015,
        month = apr,
       volume = {803},
       number = {1},
          eid = {34},
        pages = {34},
          doi = {10.1088/0004-637X/803/1/34},
archivePrefix = {arXiv},
       eprint = {1403.4295},
 primaryClass = {astro-ph.CO},
       adsurl = {https://ui.adsabs.harvard.edu/abs/2015ApJ...803...34B}
}

@ARTICLE{Greig15,
       author = {{Greig}, Bradley and {Mesinger}, Andrei},
        title = "{21CMMC: an MCMC analysis tool enabling astrophysical parameter studies of the cosmic 21 cm signal}",
      journal = {\mnras},
         year = 2015,
        month = jun,
       volume = {449},
       number = {4},
        pages = {4246-4263},
          doi = {10.1093/mnras/stv571},
archivePrefix = {arXiv},
       eprint = {1501.06576},
 primaryClass = {astro-ph.CO},
       adsurl = {https://ui.adsabs.harvard.edu/abs/2015MNRAS.449.4246G}
}

@INPROCEEDINGS{Koopmans15,
       author = {{Koopmans}, L. and {Pritchard}, J. and {Mellema}, G. and {Aguirre}, J. and {Ahn}, K. and {Barkana}, R. and {van Bemmel}, I. and {Bernardi}, G. and {Bonaldi}, A. and {Briggs}, F. and {de Bruyn}, A.~G. and {Chang}, T.~C. and {Chapman}, E. and {Chen}, X. and {Ciardi}, B. and {Dayal}, P. and {Ferrara}, A. and {Fialkov}, A. and {Fiore}, F. and {Ichiki}, K. and {Illiev}, I.~T. and {Inoue}, S. and {Jelic}, V. and {Jones}, M. and {Lazio}, J. and {Maio}, U. and {Majumdar}, S. and {Mack}, K.~J. and {Mesinger}, A. and {Morales}, M.~F. and {Parsons}, A. and {Pen}, U.~L. and {Santos}, M. and {Schneider}, R. and {Semelin}, B. and {de Souza}, R.~S. and {Subrahmanyan}, R. and {Takeuchi}, T. and {Vedantham}, H. and {Wagg}, J. and {Webster}, R. and {Wyithe}, S. and {Datta}, K.~K. and {Trott}, C.},
        title = "{The Cosmic Dawn and Epoch of Reionisation with SKA}",
    booktitle = {Advancing Astrophysics with the Square Kilometre Array (AASKA14)},
         year = 2015,
        month = apr,
          eid = {1},
        pages = {1},
          doi = {10.22323/1.215.0001},
archivePrefix = {arXiv},
       eprint = {1505.07568},
 primaryClass = {astro-ph.CO},
       adsurl = {https://ui.adsabs.harvard.edu/abs/2015aska.confE...1K}
}

@ARTICLE{Mason15,
       author = {{Mason}, Charlotte A. and {Trenti}, Michele and {Treu}, Tommaso},
        title = "{The Galaxy UV Luminosity Function before the Epoch of Reionization}",
      journal = {\apj},
         year = 2015,
        month = nov,
       volume = {813},
       number = {1},
          eid = {21},
        pages = {21},
          doi = {10.1088/0004-637X/813/1/21},
archivePrefix = {arXiv},
       eprint = {1508.01204},
 primaryClass = {astro-ph.GA},
       adsurl = {https://ui.adsabs.harvard.edu/abs/2015ApJ...813...21M}
}

@ARTICLE{McGreer15,
       author = {{McGreer}, Ian D. and {Mesinger}, Andrei and {D'Odorico}, Valentina},
        title = "{Model-independent evidence in favour of an end to reionization by z {\ensuremath{\approx}} 6}",
      journal = {\mnras},
         year = 2015,
        month = feb,
       volume = {447},
       number = {1},
        pages = {499-505},
          doi = {10.1093/mnras/stu2449},
archivePrefix = {arXiv},
       eprint = {1411.5375},
 primaryClass = {astro-ph.CO},
       adsurl = {https://ui.adsabs.harvard.edu/abs/2015MNRAS.447..499M}
}

@ARTICLE{Sun16,
       author = {{Sun}, G. and {Furlanetto}, S.~R.},
        title = "{Constraints on the star formation efficiency of galaxies during the epoch of reionization}",
      journal = {\mnras},
         year = 2016,
        month = jul,
       volume = {460},
       number = {1},
        pages = {417-433},
          doi = {10.1093/mnras/stw980},
archivePrefix = {arXiv},
       eprint = {1512.06219},
 primaryClass = {astro-ph.GA},
       adsurl = {https://ui.adsabs.harvard.edu/abs/2016MNRAS.460..417S}
}

@ARTICLE{Xu16,
       author = {{Xu}, Hao and {Wise}, John H. and {Norman}, Michael L. and {Ahn}, Kyungjin and {O'Shea}, Brian W.},
        title = "{Galaxy Properties and UV Escape Fractions during the Epoch of Reionization: Results from the Renaissance Simulations}",
      journal = {\apj},
         year = 2016,
        month = dec,
       volume = {833},
       number = {1},
          eid = {84},
        pages = {84},
          doi = {10.3847/1538-4357/833/1/84},
archivePrefix = {arXiv},
       eprint = {1604.07842},
 primaryClass = {astro-ph.GA},
       adsurl = {https://ui.adsabs.harvard.edu/abs/2016ApJ...833...84X}
}

@ARTICLE{Bouwens17,
       author = {{Bouwens}, R.~J. and {Oesch}, P.~A. and {Illingworth}, G.~D. and {Ellis}, R.~S. and {Stefanon}, M.},
        title = "{The z {\ensuremath{\sim}} 6 Luminosity Function Fainter than -15 mag from the Hubble Frontier Fields: The Impact of Magnification Uncertainties}",
      journal = {\apj},
         year = 2017,
       volume = {843},
       number = {2},
          eid = {129},
        pages = {129},
          doi = {10.3847/1538-4357/aa70a4},
archivePrefix = {arXiv},
       eprint = {1610.00283},
 primaryClass = {astro-ph.GA}
}

@ARTICLE{DeBoer17,
       author = {{DeBoer}, David R. and {Parsons}, Aaron R. and {Aguirre}, James E. and {Alexander}, Paul and {Ali}, Zaki S. and {Beardsley}, Adam P. and {Bernardi}, Gianni and {Bowman}, Judd D. and {Bradley}, Richard F. and {Carilli}, Chris L. and {Cheng}, Carina and {de Lera Acedo}, Eloy and {Dillon}, Joshua S. and {Ewall-Wice}, Aaron and {Fadana}, Gcobisa and {Fagnoni}, Nicolas and {Fritz}, Randall and {Furlanetto}, Steve R. and {Glendenning}, Brian and {Greig}, Bradley and {Grobbelaar}, Jasper and {Hazelton}, Bryna J. and {Hewitt}, Jacqueline N. and {Hickish}, Jack and {Jacobs}, Daniel C. and {Julius}, Austin and {Kariseb}, MacCalvin and {Kohn}, Saul A. and {Lekalake}, Telalo and {Liu}, Adrian and {Loots}, Anita and {MacMahon}, David and {Malan}, Lourence and {Malgas}, Cresshim and {Maree}, Matthys and {Martinot}, Zachary and {Mathison}, Nathan and {Matsetela}, Eunice and {Mesinger}, Andrei and {Morales}, Miguel F. and {Neben}, Abraham R. and {Patra}, Nipanjana and {Pieterse}, Samantha and {Pober}, Jonathan C. and {Razavi-Ghods}, Nima and {Ringuette}, Jon and {Robnett}, James and {Rosie}, Kathryn and {Sell}, Raddwine and {Smith}, Craig and {Syce}, Angelo and {Tegmark}, Max and {Thyagarajan}, Nithyanandan and {Williams}, Peter K.~G. and {Zheng}, Haoxuan},
        title = "{Hydrogen Epoch of Reionization Array (HERA)}",
      journal = {\pasp},
         year = 2017,
        month = apr,
       volume = {129},
       number = {974},
        pages = {045001},
          doi = {10.1088/1538-3873/129/974/045001},
archivePrefix = {arXiv},
       eprint = {1606.07473},
 primaryClass = {astro-ph.IM},
       adsurl = {https://ui.adsabs.harvard.edu/abs/2017PASP..129d5001D}
}

@ARTICLE{Mirocha18,
       author = {{Mirocha}, Jordan and {Mebane}, Richard H. and {Furlanetto}, Steven R. and {Singal}, Krishma and {Trinh}, Donald},
        title = "{Unique signatures of Population III stars in the global 21-cm signal}",
      journal = {\mnras},
         year = 2018,
        month = aug,
       volume = {478},
       number = {4},
        pages = {5591-5606},
          doi = {10.1093/mnras/sty1388},
archivePrefix = {arXiv},
       eprint = {1710.02530},
 primaryClass = {astro-ph.GA},
       adsurl = {https://ui.adsabs.harvard.edu/abs/2018MNRAS.478.5591M}
}

@ARTICLE{Oesch18,
       author = {{Oesch}, P.~A. and {Bouwens}, R.~J. and {Illingworth}, G.~D. and {Labb{\'e}}, I. and {Stefanon}, M.},
        title = "{The Dearth of z {\ensuremath{\sim}} 10 Galaxies in All HST Legacy Fields{\textemdash}The Rapid Evolution of the Galaxy Population in the First 500 Myr}",
      journal = {\apj},
         year = 2018,
        month = mar,
       volume = {855},
       number = {2},
          eid = {105},
        pages = {105},
          doi = {10.3847/1538-4357/aab03f},
archivePrefix = {arXiv},
       eprint = {1710.11131},
 primaryClass = {astro-ph.GA},
       adsurl = {https://ui.adsabs.harvard.edu/abs/2018ApJ...855..105O}
}

@ARTICLE{Planck18,
       author = {{Planck Collaboration} and {Aghanim}, N. and {Akrami}, Y. and {Ashdown}, M. and {Aumont}, J. and {Baccigalupi}, C. and {Ballardini}, M. and {Banday}, A.~J. and {Barreiro}, R.~B. and {Bartolo}, N. and {Basak}, S. and {Battye}, R. and {Benabed}, K. and {Bernard}, J. -P. and {Bersanelli}, M. and {Bielewicz}, P. and {Bock}, J.~J. and {Bond}, J.~R. and {Borrill}, J. and {Bouchet}, F.~R. and {Boulanger}, F. and {Bucher}, M. and {Burigana}, C. and {Butler}, R.~C. and {Calabrese}, E. and {Cardoso}, J. -F. and {Carron}, J. and {Challinor}, A. and {Chiang}, H.~C. and {Chluba}, J. and {Colombo}, L.~P.~L. and {Combet}, C. and {Contreras}, D. and {Crill}, B.~P. and {Cuttaia}, F. and {de Bernardis}, P. and {de Zotti}, G. and {Delabrouille}, J. and {Delouis}, J. -M. and {Di Valentino}, E. and {Diego}, J.~M. and {Dor{\'e}}, O. and {Douspis}, M. and {Ducout}, A. and {Dupac}, X. and {Dusini}, S. and {Efstathiou}, G. and {Elsner}, F. and {En{\ss}lin}, T.~A. and {Eriksen}, H.~K. and {Fantaye}, Y. and {Farhang}, M. and {Fergusson}, J. and {Fernandez-Cobos}, R. and {Finelli}, F. and {Forastieri}, F. and {Frailis}, M. and {Fraisse}, A.~A. and {Franceschi}, E. and {Frolov}, A. and {Galeotta}, S. and {Galli}, S. and {Ganga}, K. and {G{\'e}nova-Santos}, R.~T. and {Gerbino}, M. and {Ghosh}, T. and {Gonz{\'a}lez-Nuevo}, J. and {G{\'o}rski}, K.~M. and {Gratton}, S. and {Gruppuso}, A. and {Gudmundsson}, J.~E. and {Hamann}, J. and {Handley}, W. and {Hansen}, F.~K. and {Herranz}, D. and {Hildebrandt}, S.~R. and {Hivon}, E. and {Huang}, Z. and {Jaffe}, A.~H. and {Jones}, W.~C. and {Karakci}, A. and {Keih{\"a}nen}, E. and {Keskitalo}, R. and {Kiiveri}, K. and {Kim}, J. and {Kisner}, T.~S. and {Knox}, L. and {Krachmalnicoff}, N. and {Kunz}, M. and {Kurki-Suonio}, H. and {Lagache}, G. and {Lamarre}, J. -M. and {Lasenby}, A. and {Lattanzi}, M. and {Lawrence}, C.~R. and {Le Jeune}, M. and {Lemos}, P. and {Lesgourgues}, J. and {Levrier}, F. and {Lewis}, A. and {Liguori}, M. and {Lilje}, P.~B. and {Lilley}, M. and {Lindholm}, V. and {L{\'o}pez-Caniego}, M. and {Lubin}, P.~M. and {Ma}, Y. -Z. and {Mac{\'\i}as-P{\'e}rez}, J.~F. and {Maggio}, G. and {Maino}, D. and {Mandolesi}, N. and {Mangilli}, A. and {Marcos-Caballero}, A. and {Maris}, M. and {Martin}, P.~G. and {Martinelli}, M. and {Mart{\'\i}nez-Gonz{\'a}lez}, E. and {Matarrese}, S. and {Mauri}, N. and {McEwen}, J.~D. and {Meinhold}, P.~R. and {Melchiorri}, A. and {Mennella}, A. and {Migliaccio}, M. and {Millea}, M. and {Mitra}, S. and {Miville-Desch{\^e}nes}, M. -A. and {Molinari}, D. and {Montier}, L. and {Morgante}, G. and {Moss}, A. and {Natoli}, P. and {N{\o}rgaard-Nielsen}, H.~U. and {Pagano}, L. and {Paoletti}, D. and {Partridge}, B. and {Patanchon}, G. and {Peiris}, H.~V. and {Perrotta}, F. and {Pettorino}, V. and {Piacentini}, F. and {Polastri}, L. and {Polenta}, G. and {Puget}, J. -L. and {Rachen}, J.~P. and {Reinecke}, M. and {Remazeilles}, M. and {Renzi}, A. and {Rocha}, G. and {Rosset}, C. and {Roudier}, G. and {Rubi{\~n}o-Mart{\'\i}n}, J.~A. and {Ruiz-Granados}, B. and {Salvati}, L. and {Sandri}, M. and {Savelainen}, M. and {Scott}, D. and {Shellard}, E.~P.~S. and {Sirignano}, C. and {Sirri}, G. and {Spencer}, L.~D. and {Sunyaev}, R. and {Suur-Uski}, A. -S. and {Tauber}, J.~A. and {Tavagnacco}, D. and {Tenti}, M. and {Toffolatti}, L. and {Tomasi}, M. and {Trombetti}, T. and {Valenziano}, L. and {Valiviita}, J. and {Van Tent}, B. and {Vibert}, L. and {Vielva}, P. and {Villa}, F. and {Vittorio}, N. and {Wandelt}, B.~D. and {Wehus}, I.~K. and {White}, M. and {White}, S.~D.~M. and {Zacchei}, A. and {Zonca}, A.},
        title = "{Planck 2018 results. VI. Cosmological parameters}",
      journal = {\aap},
         year = 2020,
        month = sep,
       volume = {641},
          eid = {A6},
        pages = {A6},
          doi = {10.1051/0004-6361/201833910},
archivePrefix = {arXiv},
       eprint = {1807.06209},
 primaryClass = {astro-ph.CO},
       adsurl = {https://ui.adsabs.harvard.edu/abs/2020A&A...641A...6P}
}

@ARTICLE{Tacchella18,
       author = {{Tacchella}, Sandro and {Bose}, Sownak and {Conroy}, Charlie and {Eisenstein}, Daniel J. and {Johnson}, Benjamin D.},
        title = "{A Redshift-independent Efficiency Model: Star Formation and Stellar Masses in Dark Matter Halos at z {\ensuremath{\gtrsim}} 4}",
      journal = {\apj},
         year = 2018,
        month = dec,
       volume = {868},
       number = {2},
          eid = {92},
        pages = {92},
          doi = {10.3847/1538-4357/aae8e0},
archivePrefix = {arXiv},
       eprint = {1806.03299},
 primaryClass = {astro-ph.GA},
       adsurl = {https://ui.adsabs.harvard.edu/abs/2018ApJ...868...92T}
}

@ARTICLE{Cranmer19,
       author = {{Cranmer}, Kyle and {Brehmer}, Johann and {Louppe}, Gilles},
        title = "{The frontier of simulation-based inference}",
      journal = {Proceedings of the National Academy of Science},
         year = 2020,
        month = dec,
       volume = {117},
       number = {48},
        pages = {30055-30062},
          doi = {10.1073/pnas.1912789117},
archivePrefix = {arXiv},
       eprint = {1911.01429},
 primaryClass = {stat.ML},
       adsurl = {https://ui.adsabs.harvard.edu/abs/2020PNAS..11730055C}
}

@ARTICLE{Greig19,
       author = {{Greig}, Bradley and {Mesinger}, Andrei and {Ba{\~n}ados}, Eduardo},
        title = "{Constraints on reionization from the z = 7.5 QSO ULASJ1342+0928}",
      journal = {\mnras},
         year = 2019,
        month = apr,
       volume = {484},
       number = {4},
        pages = {5094-5101},
          doi = {10.1093/mnras/stz230},
       adsurl = {https://ui.adsabs.harvard.edu/abs/2019MNRAS.484.5094G}
}

@ARTICLE{HandleyLemos19,
       author = {{Handley}, Will and {Lemos}, Pablo},
        title = "{Quantifying tensions in cosmological parameters: Interpreting the DES evidence ratio}",
      journal = {\prd},
         year = 2019,
       volume = {100},
       number = {4},
          eid = {043504},
        pages = {043504},
          doi = {10.1103/PhysRevD.100.043504},
archivePrefix = {arXiv},
       eprint = {1902.04029},
 primaryClass = {astro-ph.CO}
}

@ARTICLE{HandleyLemos19b,
       author = {{Handley}, Will and {Lemos}, Pablo},
        title = "{Quantifying dimensionality: Bayesian cosmological model complexities}",
      journal = {\prd},
         year = 2019,
       volume = {100},
       number = {2},
          eid = {023512},
        pages = {023512},
          doi = {10.1103/PhysRevD.100.023512},
archivePrefix = {arXiv},
       eprint = {1903.06682},
 primaryClass = {astro-ph.CO}
}

@ARTICLE{Lemos20,
       author = {{Lemos}, Pablo and {K{\"o}hlinger}, Fabian and {Handley}, Will and {Joachimi}, Benjamin and {Whiteway}, Lorne and {Lahav}, Ofer},
        title = "{Quantifying Suspiciousness within correlated data sets}",
      journal = {\mnras},
         year = 2020,
       volume = {496},
       number = {4},
        pages = {4647-4653},
          doi = {10.1093/mnras/staa1836},
archivePrefix = {arXiv},
       eprint = {1910.07820},
 primaryClass = {astro-ph.CO}
}

@ARTICLE{Bevins24,
       author = {{Bevins}, Harry T.~J. and {Handley}, William J. and {Gessey-Jones}, Thomas},
        title = "{Calibrating Bayesian tension statistics using neural ratio estimators}",
      journal = {\prd},
         year = 2025,
        month = aug,
       volume = {112},
       number = {4},
          eid = {043520},
        pages = {043520},
          doi = {10.1103/47hn-h8kc},
archivePrefix = {arXiv},
       eprint = {2407.15478},
 primaryClass = {astro-ph.CO},
       adsurl = {https://ui.adsabs.harvard.edu/abs/2025PhRvD.112d3520B}
}

@ARTICLE{Li19,
       author = {{Li}, Wenyang and {Pober}, Jonathan C. and {Barry}, Nichole and {Hazelton}, Bryna J. and {Morales}, Miguel F. and {Trott}, Cathryn M. and {Lanman}, Adam and {Wilensky}, Michael and {Sullivan}, Ian and {Beardsley}, Adam P. and others},
        title = "{First Season MWA Phase II Epoch of Reionization Power Spectrum Results at Redshift 7}",
      journal = {\apj},
         year = 2019,
        month = dec,
       volume = {887},
       number = {2},
          eid = {141},
        pages = {141},
          doi = {10.3847/1538-4357/ab55e4},
archivePrefix = {arXiv},
       eprint = {1911.10216},
 primaryClass = {astro-ph.CO},
       adsurl = {https://ui.adsabs.harvard.edu/abs/2019ApJ...887..141L}
}

@ARTICLE{Park19,
       author = {{Park}, Jaehong and {Mesinger}, Andrei and {Greig}, Bradley and {Gillet}, Nicolas},
        title = "{Inferring the astrophysics of reionization and cosmic dawn from galaxy luminosity functions and the 21-cm signal}",
      journal = {\mnras},
         year = 2019,
        month = mar,
       volume = {484},
       number = {1},
        pages = {933-949},
          doi = {10.1093/mnras/stz032},
archivePrefix = {arXiv},
       eprint = {1809.08995},
 primaryClass = {astro-ph.GA},
       adsurl = {https://ui.adsabs.harvard.edu/abs/2019MNRAS.484..933P}
}

@ARTICLE{Yung19a,
       author = {{Yung}, L.~Y. Aaron and {Somerville}, Rachel S. and {Finkelstein}, Steven L. and {Popping}, Gerg{\"o} and {Dav{\'e}}, Romeel},
        title = "{Semi-analytic forecasts for JWST - I. UV luminosity functions at z = 4-10}",
      journal = {\mnras},
         year = 2019,
       volume = {483},
       number = {3},
        pages = {2983},
          doi = {10.1093/mnras/sty3241},
archivePrefix = {arXiv},
       eprint = {1803.09761},
 primaryClass = {astro-ph.GA}
}

@ARTICLE{Gelman20,
       author = {{Gelman}, Andrew and {Vehtari}, Aki and {Simpson}, Daniel and {Margossian}, Charles C. and {Carpenter}, Bob and {Yao}, Yuling and {Kennedy}, Lauren and {Gabry}, Jonah and {B{\"u}rkner}, Paul-Christian and {Modr{\'a}k}, Martin},
        title = "{Bayesian Workflow}",
      journal = {arXiv e-prints},
         year = 2020,
        month = nov,
          eid = {arXiv:2011.01808},
archivePrefix = {arXiv},
       eprint = {2011.01808},
 primaryClass = {stat.ME}
}

@ARTICLE{Ghara20,
       author = {{Ghara}, R. and {Giri}, S.~K. and {Mellema}, G. and {Ciardi}, B. and {Zaroubi}, S. and {Iliev}, I.~T. and {Koopmans}, L.~V.~E. and {Chapman}, E. and {Gazagnes}, S. and {Gehlot}, B.~K. and {Ghosh}, A. and {Jeli{\'c}}, V. and {Mertens}, F.~G. and {Mondal}, R. and {Schaye}, J. and {Silva}, M.~B. and {Asad}, K.~M.~B. and {Kooistra}, R. and {Mevius}, M. and {Offringa}, A.~R. and {Pandey}, V.~N. and {Yatawatta}, S.},
        title = "{Constraining the intergalactic medium at z = 9.1 using LOFAR Epoch of Reionization observations}",
      journal = {\mnras},
         year = 2020,
        month = feb,
       volume = {493},
       number = {4},
        pages = {4728-4747},
          doi = {10.1093/mnras/staa487},
archivePrefix = {arXiv},
       eprint = {2002.07195},
 primaryClass = {astro-ph.CO},
       adsurl = {https://ui.adsabs.harvard.edu/abs/2020MNRAS.493.4728G}
}

@ARTICLE{Murray20,
       author = {{Murray}, Steven and {Greig}, Bradley and {Mesinger}, Andrei and {Mu{\~n}oz}, Julian and {Qin}, Yuxiang and {Park}, Jaehong and {Watkinson}, Catherine},
        title = "{21cmFAST v3: A Python-integrated C code for generating 3D realizations of the cosmic 21cm signal.}",
      journal = {The Journal of Open Source Software},
         year = 2020,
        month = oct,
       volume = {5},
       number = {54},
          eid = {2582},
        pages = {2582},
          doi = {10.21105/joss.02582},
archivePrefix = {arXiv},
       eprint = {2010.15121},
 primaryClass = {astro-ph.IM},
       adsurl = {https://ui.adsabs.harvard.edu/abs/2020JOSS....5.2582M}
}

@ARTICLE{Qin20,
       author = {{Qin}, Yuxiang and {Poulin}, Vivian and {Mesinger}, Andrei and {Greig}, Bradley and {Murray}, Steven and {Park}, Jaehong},
        title = "{Reionization inference from the CMB optical depth and E-mode polarization power spectra}",
      journal = {\mnras},
         year = 2020,
        month = nov,
       volume = {499},
       number = {1},
        pages = {550-558},
          doi = {10.1093/mnras/staa2797},
archivePrefix = {arXiv},
       eprint = {2006.16828},
 primaryClass = {astro-ph.CO},
       adsurl = {https://ui.adsabs.harvard.edu/abs/2020MNRAS.499..550Q}
}

@ARTICLE{Qin20a,
       author = {{Qin}, Yuxiang and {Mesinger}, Andrei and {Park}, Jaehong and {Greig}, Bradley and {Mu{\~n}oz}, Julian B.},
        title = "{A tale of two sites - I. Inferring the properties of minihalo-hosted galaxies from current observations}",
      journal = {\mnras},
         year = 2020,
        month = jun,
       volume = {495},
       number = {1},
        pages = {123-140},
          doi = {10.1093/mnras/staa1131},
archivePrefix = {arXiv},
       eprint = {2003.04442},
 primaryClass = {astro-ph.CO},
       adsurl = {https://ui.adsabs.harvard.edu/abs/2020MNRAS.495..123Q}
}

@MISC{Dunlop21,
       author = {{Dunlop}, James S. and others},
        title = "{PRIMER: Public Release IMaging for Extragalactic Research}",
 howpublished = {JWST Proposal. Cycle 1, ID. \#1837},
         year = 2021
}

@ARTICLE{Fagnoni21a,
       author = {{Fagnoni}, Nicolas and {de Lera Acedo}, Eloy and {Drought}, Nick and {DeBoer}, David R. and {Riley}, Daniel and {Razavi-Ghods}, Nima and {Carey}, Steven and {Parsons}, Aaron R.},
        title = "{Design of the New Wideband Vivaldi Feed for the HERA Radio-Telescope Phase II}",
      journal = {IEEE Transactions on Antennas and Propagation},
         year = 2021,
        month = dec,
       volume = {69},
       number = {12},
        pages = {8143-8157},
          doi = {10.1109/TAP.2021.3083788},
archivePrefix = {arXiv},
       eprint = {2009.07939},
 primaryClass = {astro-ph.IM},
       adsurl = {https://ui.adsabs.harvard.edu/abs/arXiv:2009.07939}
}

@ARTICLE{Bosman22,
       author = {{Bosman}, Sarah E.~I. and {Davies}, Frederick B. and {Becker}, George D. and {Keating}, Laura C. and {Davies}, Rebecca L. and {Zhu}, Yongda and {Eilers}, Anna-Christina and {D'Odorico}, Valentina and {Bian}, Fuyan and {Bischetti}, Manuela and {Cristiani}, Stefano V. and {Fan}, Xiaohui and {Farina}, Emanuele P. and {Haehnelt}, Martin G. and {Hennawi}, Joseph F. and {Kulkarni}, Girish and {Mesinger}, Andrei and {Meyer}, Romain A. and {Onoue}, Masafusa and {Pallottini}, Andrea and {Qin}, Yuxiang and {Ryan-Weber}, Emma and {Schindler}, Jan-Torge and {Walter}, Fabian and {Wang}, Feige and {Yang}, Jinyi},
        title = "{Hydrogen reionization ends by z = 5.3: Lyman-{\ensuremath{\alpha}} optical depth measured by the XQR-30 sample}",
      journal = {\mnras},
         year = 2022,
        month = jul,
       volume = {514},
       number = {1},
        pages = {55-76},
          doi = {10.1093/mnras/stac1046},
archivePrefix = {arXiv},
       eprint = {2108.03699},
 primaryClass = {astro-ph.CO},
       adsurl = {https://ui.adsabs.harvard.edu/abs/2022MNRAS.514...55B}
}

@ARTICLE{Greig22,
       author = {{Greig}, Bradley and {Mesinger}, Andrei and {Davies}, Frederick B. and {Wang}, Feige and {Yang}, Jinyi and {Hennawi}, Joseph F.},
        title = "{IGM damping wing constraints on reionization from covariance reconstruction of two z {\ensuremath{\gtrsim}} 7 QSOs}",
      journal = {\mnras},
         year = 2022,
        month = jun,
       volume = {512},
       number = {4},
        pages = {5390-5403},
          doi = {10.1093/mnras/stac825},
archivePrefix = {arXiv},
       eprint = {2112.04091},
 primaryClass = {astro-ph.CO},
       adsurl = {https://ui.adsabs.harvard.edu/abs/2022MNRAS.512.5390G}
}

@ARTICLE{HERA22a,
       author = {{HERA Collaboration} and {Abdurashidova}, Zara and {Aguirre}, James E. and {Alexander}, Paul and {Ali}, Zaki S. and {Balfour}, Yanga and {Barkana}, Rennan and {Beardsley}, Adam P. and {Bernardi}, Gianni and {Billings}, Tashalee S. and {Bowman}, Judd D. and {Bradley}, Richard F. and {Bull}, Philip and {Burba}, Jacob and {Carey}, Steve and {Carilli}, Chris L. and {Cheng}, Carina and {DeBoer}, David R. and {Dexter}, Matt and {de Lera Acedo}, Eloy and {Dillon}, Joshua S. and {Ely}, John and {Ewall-Wice}, Aaron and {Fagnoni}, Nicolas and {Fialkov}, Anastasia and {Fritz}, Randall and {Furlanetto}, Steven R. and {Gale-Sides}, Kingsley and {Glendenning}, Brian and {Gorthi}, Deepthi and {Greig}, Bradley and {Grobbelaar}, Jasper and {Halday}, Ziyaad and {Hazelton}, Bryna J. and {Heimersheim}, Stefan and {Hewitt}, Jacqueline N. and {Hickish}, Jack and {Jacobs}, Daniel C. and {Julius}, Austin and {Kern}, Nicholas S. and {Kerrigan}, Joshua and {Kittiwisit}, Piyanat and {Kohn}, Saul A. and {Kolopanis}, Matthew and {Lanman}, Adam and {La Plante}, Paul and {Lekalake}, Telalo and {Lewis}, David and {Liu}, Adrian and {Ma}, Yin-Zhe and {MacMahon}, David and {Malan}, Lourence and {Malgas}, Cresshim and {Maree}, Matthys and {Martinot}, Zachary E. and {Matsetela}, Eunice and {Mesinger}, Andrei and {Mirocha}, Jordan and {Molewa}, Mathakane and {Morales}, Miguel F. and {Mosiane}, Tshegofalang and {Mu{\~n}oz}, Julian B. and {Murray}, Steven G. and {Neben}, Abraham R. and {Nikolic}, Bojan and {Nunhokee}, Chuneeta D. and {Parsons}, Aaron R. and {Patra}, Nipanjana and {Pieterse}, Samantha and {Pober}, Jonathan C. and {Qin}, Yuxiang and {Razavi-Ghods}, Nima and {Reis}, Itamar and {Ringuette}, Jon and {Robnett}, James and {Rosie}, Kathryn and {Santos}, Mario G. and {Sikder}, Sudipta and {Sims}, Peter and {Smith}, Craig and {Syce}, Angelo and {Thyagarajan}, Nithyanandan and {Williams}, Peter K.~G. and {Zheng}, Haoxuan},
        title = "{HERA Phase I Limits on the Cosmic 21 cm Signal: Constraints on Astrophysics and Cosmology during the Epoch of Reionization}",
      journal = {\apj},
         year = 2022,
        month = jan,
       volume = {924},
       number = {2},
          eid = {51},
        pages = {51},
          doi = {10.3847/1538-4357/ac2ffc},
archivePrefix = {arXiv},
       eprint = {2108.07282},
 primaryClass = {astro-ph.CO},
       adsurl = {https://ui.adsabs.harvard.edu/abs/2022ApJ...924...51A}
}

@ARTICLE{Munoz22,
       author = {{Mu{\~n}oz}, Julian B. and {Qin}, Yuxiang and {Mesinger}, Andrei and {Murray}, Steven G. and {Greig}, Bradley and {Mason}, Charlotte},
        title = "{The impact of the first galaxies on cosmic dawn and reionization}",
      journal = {\mnras},
         year = 2022,
        month = apr,
       volume = {511},
       number = {3},
        pages = {3657-3681},
          doi = {10.1093/mnras/stac185},
archivePrefix = {arXiv},
       eprint = {2110.13919},
 primaryClass = {astro-ph.CO},
       adsurl = {https://ui.adsabs.harvard.edu/abs/2022MNRAS.511.3657M}
}

@ARTICLE{Pacucci22,
       author = {{Pacucci}, Fabio and {Dayal}, Pratika and {Harikane}, Yuichi and {Inoue}, Akio K. and {Loeb}, Abraham},
        title = "{Are the newly-discovered z   13 drop-out sources starburst galaxies or quasars?}",
      journal = {\mnras},
         year = 2022,
        month = jul,
       volume = {514},
       number = {1},
        pages = {L6-L10},
          doi = {10.1093/mnrasl/slac035},
archivePrefix = {arXiv},
       eprint = {2201.00823},
 primaryClass = {astro-ph.GA},
       adsurl = {https://ui.adsabs.harvard.edu/abs/2022MNRAS.514L...6P}
}

@ARTICLE{ArrabalHaro23,
       author = {{Arrabal Haro}, Pablo and {Dickinson}, Mark and {Finkelstein}, Steven L. and others},
        title = "{Confirmation and refutation of very luminous galaxies in the early universe}",
      journal = {\nat},
         year = 2023,
        month = oct,
       volume = {622},
       number = {7984},
        pages = {707-711},
          doi = {10.1038/s41586-023-06521-7},
archivePrefix = {arXiv},
       eprint = {2303.15431},
 primaryClass = {astro-ph.GA},
       adsurl = {https://ui.adsabs.harvard.edu/abs/arXiv:2303.15431}
}

@ARTICLE{Dekel23,
       author = {{Dekel}, Avishai and {Sarkar}, Kartick C. and {Birnboim}, Yuval and {Mandelker}, Nir and {Li}, Zhaozhou},
        title = "{Efficient formation of massive galaxies at cosmic dawn by feedback-free starbursts}",
      journal = {\mnras},
         year = 2023,
        month = aug,
       volume = {523},
       number = {3},
        pages = {3201-3218},
          doi = {10.1093/mnras/stad1557},
archivePrefix = {arXiv},
       eprint = {2303.04827},
 primaryClass = {astro-ph.GA},
       adsurl = {https://ui.adsabs.harvard.edu/abs/2023MNRAS.523.3201D}
}

@ARTICLE{deSanti23,
       author = {{de Santi}, Natal{\'\i} S.~M. and {Shao}, Helen and {Villaescusa-Navarro}, Francisco and {Abramo}, L. Raul and {Teyssier}, Romain and {Villanueva-Domingo}, Pablo and {Ni}, Yueying and {Angl{\'e}s-Alc{\'a}zar}, Daniel and {Genel}, Shy and {Hern{\'a}ndez-Mart{\'\i}nez}, Elena and {Steinwandel}, Ulrich P. and {Lovell}, Christopher C. and {Dolag}, Klaus and {Castro}, Tiago and {Vogelsberger}, Mark},
        title = "{Robust Field-level Likelihood-free Inference with Galaxies}",
      journal = {\apj},
         year = 2023,
       volume = {952},
       number = {1},
          eid = {69},
        pages = {69},
          doi = {10.3847/1538-4357/acd1e2},
archivePrefix = {arXiv},
       eprint = {2302.14101},
 primaryClass = {astro-ph.CO}
}

@ARTICLE{Eisenstein23,
       author = {{Eisenstein}, Daniel J. and {Willott}, Chris and {Alberts}, Stacey and {Arribas}, Santiago and {Bonaventura}, Nina and {Bunker}, Andrew J. and others},
        title = "{Overview of the JWST Advanced Deep Extragalactic Survey (JADES)}",
      journal = {\apjs},
         year = 2026,
       volume = {283},
       number = {1},
        pages = {6},
          doi = {10.3847/1538-4365/ae3163},
archivePrefix = {arXiv},
       eprint = {2306.02465},
 primaryClass = {astro-ph.GA}
}

@ARTICLE{Ferrara23,
       author = {{Ferrara}, Andrea and {Pallottini}, Andrea and {Dayal}, Pratika},
        title = "{On the stunning abundance of super-early, luminous galaxies revealed by JWST}",
      journal = {\mnras},
         year = 2023,
        month = jul,
       volume = {522},
       number = {3},
        pages = {3986-3991},
          doi = {10.1093/mnras/stad1095},
archivePrefix = {arXiv},
       eprint = {2208.00720},
 primaryClass = {astro-ph.GA},
       adsurl = {https://ui.adsabs.harvard.edu/abs/2023MNRAS.522.3986F}
}

@ARTICLE{Finkelstein23,
       author = {{Finkelstein}, Steven L. and {Bagley}, Micaela B. and {Ferguson}, Henry C. and {Wilkins}, Stephen M. and {Kartaltepe}, Jeyhan S. and {Papovich}, Casey and {Yung}, L.~Y. Aaron and {Arrabal Haro}, Pablo and {Behroozi}, Peter and {Dickinson}, Mark and {Kocevski}, Dale D. and {Koekemoer}, Anton M. and {Larson}, Rebecca L. and {Le Bail}, Aur{\'e}lien and {Morales}, Alexa M. and {P{\'e}rez-Gonz{\'a}lez}, Pablo G. and {Burgarella}, Denis and {Dav{\'e}}, Romeel and {Hirschmann}, Michaela and {Somerville}, Rachel S. and {Wuyts}, Stijn and {Bromm}, Volker and {Casey}, Caitlin M. and {Fontana}, Adriano and {Fujimoto}, Seiji and {Gardner}, Jonathan P. and {Giavalisco}, Mauro and {Grazian}, Andrea and {Grogin}, Norman A. and {Hathi}, Nimish P. and {Hutchison}, Taylor A. and {Jha}, Saurabh W. and {Jogee}, Shardha and {Kewley}, Lisa J. and {Kirkpatrick}, Allison and {Long}, Arianna S. and {Lotz}, Jennifer M. and {Pentericci}, Laura and {Pierel}, Justin D.~R. and {Pirzkal}, Nor and {Ravindranath}, Swara and {Ryan}, Russell E. and {Trump}, Jonathan R. and {Yang}, Guang and {Bhatawdekar}, Rachana and {Bisigello}, Laura and {Buat}, V{\'e}ronique and {Calabr{\`o}}, Antonello and {Castellano}, Marco and {Cleri}, Nikko J. and {Cooper}, M.~C. and {Croton}, Darren and {Daddi}, Emanuele and {Dekel}, Avishai and {Elbaz}, David and {Franco}, Maximilien and {Gawiser}, Eric and {Holwerda}, Benne W. and {Huertas-Company}, Marc and {Jaskot}, Anne E. and {Leung}, Gene C.~K. and {Lucas}, Ray A. and {Mobasher}, Bahram and {Pandya}, Viraj and {Tacchella}, Sandro and {Weiner}, Benjamin J. and {Zavala}, Jorge A.},
        title = "{CEERS Key Paper. I. An Early Look into the First 500 Myr of Galaxy Formation with JWST}",
      journal = {\apjl},
         year = 2023,
        month = mar,
       volume = {946},
       number = {1},
          eid = {L13},
        pages = {L13},
          doi = {10.3847/2041-8213/acade4},
archivePrefix = {arXiv},
       eprint = {2211.05792},
 primaryClass = {astro-ph.GA},
       adsurl = {https://ui.adsabs.harvard.edu/abs/2023ApJ...946L..13F}
}

@ARTICLE{Fronenberg23,
       author = {{Fronenberg}, Hannah and {Maniyar}, Abhishek S. and {Pullen}, Anthony R. and {Liu}, Adrian},
        title = "{Constraining cosmology with the CMB {\texttimes} line intensity mapping-nulling convergence}",
      journal = {\prd},
         year = 2024,
        month = jun,
       volume = {109},
       number = {12},
          eid = {123518},
        pages = {123518},
          doi = {10.1103/PhysRevD.109.123518},
archivePrefix = {arXiv},
       eprint = {2309.06477},
 primaryClass = {astro-ph.CO},
       adsurl = {https://ui.adsabs.harvard.edu/abs/2024PhRvD.109l3518F}
}

@ARTICLE{Harikane23,
       author = {{Harikane}, Yuichi and {Ouchi}, Masami and {Oguri}, Masamune and {Ono}, Yoshiaki and {Nakajima}, Kimihiko and {Isobe}, Yuki and {Umeda}, Hiroya and {Mawatari}, Ken and {Zhang}, Yechi},
        title = "{A Comprehensive Study of Galaxies at z   9-16 Found in the Early JWST Data: Ultraviolet Luminosity Functions and Cosmic Star Formation History at the Pre-reionization Epoch}",
      journal = {\apjs},
         year = 2023,
        month = mar,
       volume = {265},
       number = {1},
          eid = {5},
        pages = {5},
          doi = {10.3847/1538-4365/acaaa9},
archivePrefix = {arXiv},
       eprint = {2208.01612},
 primaryClass = {astro-ph.GA},
       adsurl = {https://ui.adsabs.harvard.edu/abs/2023ApJS..265....5H}
}

@ARTICLE{HERA23,
       author = {{HERA Collaboration} and {Abdurashidova}, Zara and {Adams}, Tyrone and {Aguirre}, James E. and {Alexander}, Paul and {Ali}, Zaki S. and {Baartman}, Rushelle and {Balfour}, Yanga and {Barkana}, Rennan and {Beardsley}, Adam P. and {Bernardi}, Gianni and {Billings}, Tashalee S. and {Bowman}, Judd D. and {Bradley}, Richard F. and {Breitman}, Daniela and {Bull}, Philip and {Burba}, Jacob and {Carey}, Steve and {Carilli}, Chris L. and {Cheng}, Carina and {Choudhuri}, Samir and {DeBoer}, David R. and {de Lera Acedo}, Eloy and {Dexter}, Matt and {Dillon}, Joshua S. and {Ely}, John and {Ewall-Wice}, Aaron and {Fagnoni}, Nicolas and {Fialkov}, Anastasia and {Fritz}, Randall and {Furlanetto}, Steven R. and {Gale-Sides}, Kingsley and {Garsden}, Hugh and {Glendenning}, Brian and {Gorce}, Ad{\'e}lie and {Gorthi}, Deepthi and {Greig}, Bradley and {Grobbelaar}, Jasper and {Halday}, Ziyaad and {Hazelton}, Bryna J. and {Heimersheim}, Stefan and {Hewitt}, Jacqueline N. and {Hickish}, Jack and {Jacobs}, Daniel C. and {Julius}, Austin and {Kern}, Nicholas S. and {Kerrigan}, Joshua and {Kittiwisit}, Piyanat and {Kohn}, Saul A. and {Kolopanis}, Matthew and {Lanman}, Adam and {La Plante}, Paul and {Lewis}, David and {Liu}, Adrian and {Loots}, Anita and {Ma}, Yin-Zhe and {MacMahon}, David H.~E. and {Malan}, Lourence and {Malgas}, Keith and {Malgas}, Cresshim and {Maree}, Matthys and {Marero}, Bradley and {Martinot}, Zachary E. and {McBride}, Lisa and {Mesinger}, Andrei and {Mirocha}, Jordan and {Molewa}, Mathakane and {Morales}, Miguel F. and {Mosiane}, Tshegofalang and {Mu{\~n}oz}, Julian B. and {Murray}, Steven G. and {Nagpal}, Vighnesh and {Neben}, Abraham R. and {Nikolic}, Bojan and {Nunhokee}, Chuneeta D. and {Nuwegeld}, Hans and {Parsons}, Aaron R. and {Pascua}, Robert and {Patra}, Nipanjana and {Pieterse}, Samantha and {Qin}, Yuxiang and {Razavi-Ghods}, Nima and {Robnett}, James and {Rosie}, Kathryn and {Santos}, Mario G. and {Sims}, Peter and {Singh}, Saurabh and {Smith}, Craig and {Swarts}, Hilton and {Tan}, Jianrong and {Thyagarajan}, Nithyanandan and {Wilensky}, Michael J. and {Williams}, Peter K.~G. and {van Wyngaarden}, Pieter and {Zheng}, Haoxuan},
        title = "{Improved Constraints on the 21 cm EoR Power Spectrum and the X-Ray Heating of the IGM with HERA Phase I Observations}",
      journal = {\apj},
         year = 2023,
        month = mar,
       volume = {945},
       number = {2},
          eid = {124},
        pages = {124},
          doi = {10.3847/1538-4357/acaf50},
archivePrefix = {arXiv},
       eprint = {2210.04912},
 primaryClass = {astro-ph.CO},
       adsurl = {https://ui.adsabs.harvard.edu/abs/2023ApJ...945..124H}
}

@ARTICLE{Lange23,
       author = {{Lange}, Johannes U.},
        title = "{NAUTILUS: boosting Bayesian importance nested sampling with deep learning}",
      journal = {\mnras},
         year = 2023,
        month = oct,
       volume = {525},
       number = {2},
        pages = {3181-3194},
          doi = {10.1093/mnras/stad2441},
archivePrefix = {arXiv},
       eprint = {2306.16923},
 primaryClass = {astro-ph.IM},
       adsurl = {https://ui.adsabs.harvard.edu/abs/2023MNRAS.525.3181L}
}

@ARTICLE{Lazare23,
       author = {{Lazare}, Hovav and {Sarkar}, Debanjan and {Kovetz}, Ely D.},
        title = "{HERA bound on x-ray luminosity when accounting for population III stars}",
      journal = {\prd},
         year = 2024,
        month = feb,
       volume = {109},
       number = {4},
          eid = {043523},
        pages = {043523},
          doi = {10.1103/PhysRevD.109.043523},
archivePrefix = {arXiv},
       eprint = {2307.15577},
 primaryClass = {astro-ph.CO},
       adsurl = {https://ui.adsabs.harvard.edu/abs/2024PhRvD.109d3523L}
}

@ARTICLE{Mason23,
       author = {{Mason}, Charlotte A. and {Trenti}, Michele and {Treu}, Tommaso},
        title = "{The brightest galaxies at cosmic dawn}",
      journal = {\mnras},
         year = 2023,
        month = may,
       volume = {521},
       number = {1},
        pages = {497-503},
          doi = {10.1093/mnras/stad035},
archivePrefix = {arXiv},
       eprint = {2207.14808},
 primaryClass = {astro-ph.GA},
       adsurl = {https://ui.adsabs.harvard.edu/abs/2023MNRAS.521..497M}
}

@ARTICLE{Mirocha23,
       author = {{Mirocha}, Jordan and {Furlanetto}, Steven R.},
        title = "{Balancing the efficiency and stochasticity of star formation with dust extinction in z {\ensuremath{\gtrsim}} 10 galaxies observed by JWST}",
      journal = {\mnras},
         year = 2023,
        month = feb,
       volume = {519},
       number = {1},
        pages = {843-853},
          doi = {10.1093/mnras/stac3578},
archivePrefix = {arXiv},
       eprint = {2208.12826},
 primaryClass = {astro-ph.GA},
       adsurl = {https://ui.adsabs.harvard.edu/abs/2023MNRAS.519..843M}
}

@ARTICLE{Prelogovic23,
       author = {{Prelogovi{\'c}}, David and {Mesinger}, Andrei},
        title = "{Exploring the likelihood of the 21-cm power spectrum with simulation-based inference}",
      journal = {\mnras},
         year = 2023,
        month = sep,
       volume = {524},
       number = {3},
        pages = {4239-4255},
          doi = {10.1093/mnras/stad2027},
archivePrefix = {arXiv},
       eprint = {2305.03074},
 primaryClass = {astro-ph.CO},
       adsurl = {https://ui.adsabs.harvard.edu/abs/2023MNRAS.524.4239P}
}

@ARTICLE{Rieke23,
       author = {{Rieke}, Marcia J. and {Kelly}, Douglas M. and {Misselt}, Karl and others},
        title = "{Performance of NIRCam on JWST in Flight}",
      journal = {\pasp},
         year = 2023,
       volume = {135},
       number = {1044},
          eid = {028001},
        pages = {028001},
          doi = {10.1088/1538-3873/acac53},
archivePrefix = {arXiv},
       eprint = {2212.12069},
 primaryClass = {astro-ph.IM}
}

@ARTICLE{Rigby23,
       author = {{Rigby}, Jane and {Perrin}, Marshall and {McElwain}, Michael and {Kimble}, Randy and {Friedman}, Scott and {Lallo}, Matt and {Doyon}, Ren{\'e} and {Feinberg}, Lee and {Ferruit}, Pierre and {Glasse}, Alistair and {Rieke}, Marcia and {Rieke}, George and {Wright}, Gillian and {Willott}, Chris and {Colon}, Knicole and {Milam}, Stefanie and {Neff}, Susan and {Stark}, Christopher and {Valenti}, Jim and {others}},
        title = "{The Science Performance of JWST as Characterized in Commissioning}",
      journal = {\pasp},
         year = 2023,
        month = apr,
       volume = {135},
       number = {1046},
          eid = {048001},
        pages = {048001},
          doi = {10.1088/1538-3873/acb293},
archivePrefix = {arXiv},
       eprint = {2207.05632},
 primaryClass = {astro-ph.IM},
       adsurl = {https://ui.adsabs.harvard.edu/abs/2023PASP..135d8001R}
}

@ARTICLE{Shen23,
       author = {{Shen}, Xuejian and {Vogelsberger}, Mark and {Boylan-Kolchin}, Michael and {Tacchella}, Sandro and {Kannan}, Rahul},
        title = "{The impact of UV variability on the abundance of bright galaxies at z {\ensuremath{\geq}} 9}",
      journal = {\mnras},
         year = 2023,
        month = nov,
       volume = {525},
       number = {3},
        pages = {3254-3261},
          doi = {10.1093/mnras/stad2508},
archivePrefix = {arXiv},
       eprint = {2305.05679},
 primaryClass = {astro-ph.GA},
       adsurl = {https://ui.adsabs.harvard.edu/abs/2023MNRAS.525.3254S}
}

@MISC{SKAO23,
       author = {{SKA Observatory}},
        title = "{SKAO staged delivery, array assemblies and layouts}",
 howpublished = {SKAO technical document SKAO-TEL-0002299, revision 01},
         year = 2023,
        month = nov,
          url = {https://www.skao.int/sites/default/files/documents/SKAO-TEL-0002299_01_staged_delivery.pdf}
}

@ARTICLE{Sun23,
       author = {{Sun}, Guochao and {Faucher-Gigu{\`e}re}, Claude-Andr{\'e} and {Hayward}, Christopher C. and {Shen}, Xuejian},
        title = "{Seen and unseen: bursty star formation and its implications for observations of high-redshift galaxies with JWST}",
      journal = {\mnras},
         year = 2023,
        month = dec,
       volume = {526},
       number = {2},
        pages = {2665-2672},
          doi = {10.1093/mnras/stad2902},
archivePrefix = {arXiv},
       eprint = {2305.02713},
 primaryClass = {astro-ph.GA},
       adsurl = {https://ui.adsabs.harvard.edu/abs/2023MNRAS.526.2665S}
}

@ARTICLE{Adams24,
       author = {{Adams}, Nathan J. and {Conselice}, Christopher J. and {Austin}, Duncan and {Harvey}, Thomas and {Ferreira}, Leonardo and {Trussler}, James and {Juod{\v{z}}balis}, Ignas and {Li}, Qiong and {Windhorst}, Rogier and {Cohen}, Seth H. and {Jansen}, Rolf A. and {Summers}, Jake and {Tompkins}, Scott and {Driver}, Simon P. and {Robotham}, Aaron and {D'Silva}, Jordan C.~J. and {Yan}, Haojing and {Coe}, Dan and {Frye}, Brenda and {Grogin}, Norman A. and {Koekemoer}, Anton M. and {Marshall}, Madeline A. and {Pirzkal}, Nor and {Ryan}, Russell E. and {Maksym}, W. Peter and {Rutkowski}, Michael J. and {Willmer}, Christopher N.~A. and {Hammel}, Heidi B. and {Nonino}, Mario and {Bhatawdekar}, Rachana and {Wilkins}, Stephen M. and {Bradley}, Larry D. and {Broadhurst}, Tom and {Cheng}, Cheng and {Dole}, Herv{\'e} and {Hathi}, Nimish P. and {Zitrin}, Adi},
        title = "{EPOCHS. II. The Ultraviolet Luminosity Function from 7.5 < z < 13.5 Using 180 arcmin$^{2}$ of Deep, Blank Fields from the PEARLS Survey and Public JWST Data}",
      journal = {\apj},
         year = 2024,
        month = apr,
       volume = {965},
       number = {2},
          eid = {169},
        pages = {169},
          doi = {10.3847/1538-4357/ad2a7b},
archivePrefix = {arXiv},
       eprint = {2304.13721},
 primaryClass = {astro-ph.GA},
       adsurl = {https://ui.adsabs.harvard.edu/abs/2024ApJ...965..169A}
}

@ARTICLE{AnauMontel24,
       author = {{Anau Montel}, Noemi and {Alvey}, James and {Weniger}, Christoph},
        title = "{Tests for model misspecification in simulation-based inference: from local distortions to global model checks}",
      journal = {\prd},
         year = 2025,
       volume = {111},
       number = {8},
          eid = {083013},
        pages = {083013},
          doi = {10.1103/PhysRevD.111.083013},
archivePrefix = {arXiv},
       eprint = {2412.15100},
 primaryClass = {stat.ML}
}

@ARTICLE{Bagley24,
       author = {{Bagley}, Micaela B. and others},
        title = "{The Next Generation Deep Extragalactic Exploratory Public (NGDEEP) Survey}",
      journal = {\apjl},
         year = 2024,
       volume = {965},
       number = {1},
          eid = {L6},
        pages = {L6},
          doi = {10.3847/2041-8213/ad2f31},
archivePrefix = {arXiv},
       eprint = {2302.05466},
 primaryClass = {astro-ph.GA}
}

@ARTICLE{Berkhout24,
       author = {{Berkhout}, Lindsay M. and {Jacobs}, Daniel C. and {Abdurashidova}, Zuhra and {Adams}, Tyrone and {Aguirre}, James E. and {Alexander}, Paul and {Ali}, Zaki S. and {Baartman}, Rushelle and {Balfour}, Yanga and {Beardsley}, Adam P. and others},
        title = "{Hydrogen Epoch of Reionization Array (HERA) Phase II Deployment and Commissioning}",
      journal = {\pasp},
         year = 2024,
        month = apr,
       volume = {136},
       number = {4},
          eid = {045002},
        pages = {045002},
          doi = {10.1088/1538-3873/ad3122},
archivePrefix = {arXiv},
       eprint = {2401.04304},
 primaryClass = {astro-ph.IM},
       adsurl = {https://ui.adsabs.harvard.edu/abs/2024PASP..136d5002B}
}

@ARTICLE{Breitman24,
       author = {{Breitman}, Daniela and {Mesinger}, Andrei and {Murray}, Steven G. and {Prelogovi{\'c}}, David and {Qin}, Yuxiang and {Trotta}, Roberto},
        title = "{21CMEMU: an emulator of 21CMFAST summary observables}",
      journal = {\mnras},
         year = 2024,
        month = feb,
       volume = {527},
       number = {4},
        pages = {9833-9852},
          doi = {10.1093/mnras/stad3849},
archivePrefix = {arXiv},
       eprint = {2309.05697},
 primaryClass = {astro-ph.CO},
       adsurl = {https://ui.adsabs.harvard.edu/abs/2024MNRAS.527.9833B}
}

@ARTICLE{Carniani24,
       author = {{Carniani}, Stefano and {Hainline}, Kevin and {D'Eugenio}, Francesco and {Eisenstein}, Daniel J. and {Jakobsen}, Peter and {Witstok}, Joris and {Johnson}, Benjamin D. and {Chevallard}, Jacopo and {Maiolino}, Roberto and {Helton}, Jakob M. and {Willott}, Chris and {Robertson}, Brant and {Alberts}, Stacey and {Arribas}, Santiago and {Baker}, William M. and {Bhatawdekar}, Rachana and {Boyett}, Kristan and {Bunker}, Andrew J. and {Cameron}, Alex J. and {Cargile}, Phillip A. and {Charlot}, St{\'e}phane and {Curti}, Mirko and {Curtis-Lake}, Emma and {Egami}, Eiichi and {Giardino}, Giovanna and {Isaak}, Kate and {Ji}, Zhiyuan and {Jones}, Gareth C. and {Kumari}, Nimisha and {Maseda}, Michael V. and {Parlanti}, Eleonora and {P{\'e}rez-Gonz{\'a}lez}, Pablo G. and {Rawle}, Tim and {Rieke}, George and {Rieke}, Marcia and {Del Pino}, Bruno Rodr{\'\i}guez and {Saxena}, Aayush and {Scholtz}, Jan and {Smit}, Renske and {Sun}, Fengwu and {Tacchella}, Sandro and {{\"U}bler}, Hannah and {Venturi}, Giacomo and {Williams}, Christina C. and {Willmer}, Christopher N.~A.},
        title = "{Spectroscopic confirmation of two luminous galaxies at a redshift of 14}",
      journal = {\nat},
         year = 2024,
        month = sep,
       volume = {633},
       number = {8029},
        pages = {318-322},
          doi = {10.1038/s41586-024-07860-9},
archivePrefix = {arXiv},
       eprint = {2405.18485},
 primaryClass = {astro-ph.GA},
       adsurl = {https://ui.adsabs.harvard.edu/abs/2024Natur.633..318C}
}

@ARTICLE{Donnan24,
       author = {{Donnan}, C.~T. and {McLure}, R.~J. and {Dunlop}, J.~S. and {McLeod}, D.~J. and {Magee}, D. and {Arellano-C{\'o}rdova}, K.~Z. and {Barrufet}, L. and {Begley}, R. and {Bowler}, R.~A.~A. and {Carnall}, A.~C. and {Cullen}, F. and {Ellis}, R.~S. and {Fontana}, A. and {Illingworth}, G.~D. and {Grogin}, N.~A. and {Hamadouche}, M.~L. and {Koekemoer}, A.~M. and {Liu}, F.-Y. and {Mason}, C. and {Santini}, P. and {Stanton}, T.~M.},
        title = "{JWST PRIMER: a new multifield determination of the evolving galaxy UV luminosity function at redshifts z {\ensuremath{\simeq}} 9 - 15}",
      journal = {\mnras},
         year = 2024,
        month = sep,
       volume = {533},
       number = {3},
        pages = {3222-3237},
          doi = {10.1093/mnras/stae2037},
archivePrefix = {arXiv},
       eprint = {2403.03171},
 primaryClass = {astro-ph.GA},
       adsurl = {https://ui.adsabs.harvard.edu/abs/2024MNRAS.533.3222D}
}

@ARTICLE{Gelli24,
       author = {{Gelli}, Viola and {Mason}, Charlotte and {Hayward}, Christopher C.},
        title = "{The Impact of Mass-dependent Stochasticity at Cosmic Dawn}",
      journal = {\apj},
         year = 2024,
        month = nov,
       volume = {975},
       number = {2},
          eid = {192},
        pages = {192},
          doi = {10.3847/1538-4357/ad7b36},
archivePrefix = {arXiv},
       eprint = {2405.13108},
 primaryClass = {astro-ph.GA},
       adsurl = {https://ui.adsabs.harvard.edu/abs/2024ApJ...975..192G}
}

@ARTICLE{Greig24a,
       author = {{Greig}, Bradley and {Prelogovi{\'c}}, David and {Mirocha}, Jordan and {Qin}, Yuxiang and {Ting}, Yuan-Sen and {Mesinger}, Andrei},
        title = "{Exploring the role of the halo mass function for inferring astrophysical parameters during reionization}",
      journal = {\mnras},
         year = 2024,
        month = sep,
       volume = {533},
       number = {2},
        pages = {2502-2529},
          doi = {10.1093/mnras/stae1983},
archivePrefix = {arXiv},
       eprint = {2403.14061},
 primaryClass = {astro-ph.CO},
       adsurl = {https://ui.adsabs.harvard.edu/abs/2024MNRAS.533.2502G}
}

@article{Harikane24,
doi = {10.3847/1538-4357/ad0b7e},
url = {https://dx.doi.org/10.3847/1538-4357/ad0b7e},
year = {2024},
month = {jan},
publisher = {The American Astronomical Society},
volume = {960},
number = {1},
pages = {56},
author = {Harikane, Yuichi and Nakajima, Kimihiko and Ouchi, Masami and Umeda, Hiroya and Isobe, Yuki and Ono, Yoshiaki and Xu, Yi and Zhang, Yechi},
title = {Pure Spectroscopic Constraints on UV Luminosity Functions and Cosmic Star Formation History from 25 Galaxies at zspec = 8.61–13.20 Confirmed with JWST/NIRSpec},
journal = {The Astrophysical Journal}
}

@ARTICLE{Hegde24,
       author = {{Hegde}, Sahil and {Wyatt}, Michael M. and {Furlanetto}, Steven R.},
        title = "{A hidden population of active galactic nuclei can explain the overabundance of luminous z > 10 objects observed by JWST}",
      journal = {\jcap},
         year = 2024,
        month = aug,
       volume = {2024},
       number = {8},
          eid = {025},
        pages = {025},
          doi = {10.1088/1475-7516/2024/08/025},
archivePrefix = {arXiv},
       eprint = {2405.01629},
 primaryClass = {astro-ph.GA},
       adsurl = {https://ui.adsabs.harvard.edu/abs/2024JCAP...08..025H}
}

@ARTICLE{Murray24,
       author = {{Murray}, Steven and {Pober}, Jonathan and {Kolopanis}, Matthew},
        title = "{21cmSense v2: A modular, open-source 21 cm sensitivity calculator}",
      journal = {The Journal of Open Source Software},
         year = 2024,
        month = may,
       volume = {9},
       number = {97},
          eid = {6501},
        pages = {6501},
          doi = {10.21105/joss.06501},
archivePrefix = {arXiv},
       eprint = {2406.02415},
 primaryClass = {astro-ph.CO},
       adsurl = {https://ui.adsabs.harvard.edu/abs/2024JOSS....9.6501M}
}

@ARTICLE{Nikolic24,
       author = {{Nikoli{\'c}}, Ivan and {Mesinger}, Andrei and {Davies}, James E. and {Prelogovi{\'c}}, David},
        title = "{The importance of stochasticity in determining galaxy emissivities and UV LFs during cosmic dawn and reionization}",
      journal = {\aap},
         year = 2024,
        month = dec,
       volume = {692},
          eid = {A142},
        pages = {A142},
          doi = {10.1051/0004-6361/202451213},
archivePrefix = {arXiv},
       eprint = {2406.15237},
 primaryClass = {astro-ph.CO},
       adsurl = {https://ui.adsabs.harvard.edu/abs/2024A&A...692A.142N}
}

@ARTICLE{Prelogovic24,
       author = {{Prelogovi{\'c}}, David and {Mesinger}, Andrei},
        title = "{How informative are summaries of the cosmic 21 cm signal?}",
      journal = {\aap},
         year = 2024,
        month = aug,
       volume = {688},
          eid = {A199},
        pages = {A199},
          doi = {10.1051/0004-6361/202449309},
archivePrefix = {arXiv},
       eprint = {2401.12277},
 primaryClass = {astro-ph.CO},
       adsurl = {https://ui.adsabs.harvard.edu/abs/2024A&A...688A.199P}
}

@ARTICLE{Trinca24,
       author = {{Trinca}, Alessandro and {Schneider}, Raffaella and {Valiante}, Rosa and {Graziani}, Luca and {Ferrotti}, Arianna and {Omukai}, Kazuyuki and {Chon}, Sunmyon},
        title = "{Exploring the nature of UV-bright z {\ensuremath{\gtrsim}} 10 galaxies detected by JWST: star formation, black hole accretion, or a non-universal IMF?}",
      journal = {\mnras},
         year = 2024,
        month = apr,
       volume = {529},
       number = {4},
        pages = {3563-3581},
          doi = {10.1093/mnras/stae651},
archivePrefix = {arXiv},
       eprint = {2305.04944},
 primaryClass = {astro-ph.GA},
       adsurl = {https://ui.adsabs.harvard.edu/abs/2024MNRAS.529.3563T}
}

@ARTICLE{Willott24,
       author = {{Willott}, Chris J. and {Desprez}, Guillaume and {Asada}, Yoshihisa and {Sarrouh}, Ghassan T.~E. and {Abraham}, Roberto and {Brada{\v{c}}}, Maru{\v{s}}a and {Brammer}, Gabe and {Estrada-Carpenter}, Vince and {Iyer}, Kartheik G. and {Martis}, Nicholas S. and {Matharu}, Jasleen and {Mowla}, Lamiya and {Muzzin}, Adam and {Noirot}, Ga{\"e}l and {Sawicki}, Marcin and {Strait}, Victoria and {Rihtar{\v{s}}i{\v{c}}}, Gregor and {Withers}, Sunna},
        title = "{A Steep Decline in the Galaxy Space Density beyond Redshift 9 in the CANUCS UV Luminosity Function}",
      journal = {\apj},
         year = 2024,
        month = may,
       volume = {966},
       number = {1},
          eid = {74},
        pages = {74},
          doi = {10.3847/1538-4357/ad35bc},
archivePrefix = {arXiv},
       eprint = {2311.12234},
 primaryClass = {astro-ph.GA},
       adsurl = {https://ui.adsabs.harvard.edu/abs/2024ApJ...966...74W}
}

@ARTICLE{Yung24,
       author = {{Yung}, L.~Y. Aaron and {Somerville}, Rachel S. and {Finkelstein}, Steven L. and {Wilkins}, Stephen M. and {Gardner}, Jonathan P.},
        title = "{Are the ultra-high-redshift galaxies at z > 10 surprising in the context of standard galaxy formation models?}",
      journal = {\mnras},
         year = 2024,
        month = jan,
       volume = {527},
       number = {3},
        pages = {5929-5948},
          doi = {10.1093/mnras/stad3484},
archivePrefix = {arXiv},
       eprint = {2304.04348},
 primaryClass = {astro-ph.GA},
       adsurl = {https://ui.adsabs.harvard.edu/abs/2024MNRAS.527.5929Y}
}

@ARTICLE{Akhmetzhanova25,
       author = {{Akhmetzhanova}, Aizhan and {Cuesta-Lazaro}, Carolina and {Mishra-Sharma}, Siddharth},
        title = "{Detecting model misspecification in cosmology with scale-dependent normalizing flows}",
      journal = {Machine Learning: Science and Technology},
         year = 2025,
        month = dec,
       volume = {6},
       number = {4},
          eid = {045020},
        pages = {045020},
          doi = {10.1088/2632-2153/ae107d},
archivePrefix = {arXiv},
       eprint = {2508.05744},
 primaryClass = {astro-ph.CO},
       adsurl = {https://ui.adsabs.harvard.edu/abs/2025MLS&T...6d5020A}
}

@ARTICLE{Atek25,
       author = {{Atek}, Hakim and {Chisholm}, John and {Kokorev}, Vasily and {Endsley}, Ryan and {Pan}, Richard and {Furtak}, Lukas J. and others},
        title = "{JWST's GLIMPSE: an overview of the deepest probe of early galaxy formation and cosmic reionization}",
      journal = {arXiv e-prints},
         year = 2025,
        month = nov,
          eid = {arXiv:2511.07542},
archivePrefix = {arXiv},
       eprint = {2511.07542},
 primaryClass = {astro-ph.GA},
         note = {JWST program 3293}
}

@ARTICLE{Berklas25,
       author = {{Berklas}, A. and {Pober}, J.~C.},
        title = "{Exploring the Model Dependence of MCMC-based 21 cm Power Spectrum Parameter Constraints}",
      journal = {\apj},
         year = 2026,
       volume = {1001},
       number = {2},
        pages = {215},
          doi = {10.3847/1538-4357/ae5934},
archivePrefix = {arXiv},
       eprint = {2511.13854},
 primaryClass = {astro-ph.CO}
}

@ARTICLE{Breitman25,
	author = {{Breitman}, Daniela and {Mesinger}, Andrei and {Murray}, Steven G. and {Acharya}, Anshuman},
	title = {Sample variance denoising in cylindrical 21 cm power spectra},
	DOI= "10.1051/0004-6361/202556449",
	url= "https://doi.org/10.1051/0004-6361/202556449",
	journal = {A\&A},
	year = 2025,
	volume = 703,
	pages = "A130",
}

@ARTICLE{Cooper25,
       author = {{Cooper}, Nadia and {Norregaard}, Carina and {Meriot}, Romain and {Pritchard}, Jonathan R.},
        title = "{Simulation-based inference pipeline of the ionization history from the 2D 21 cm power spectrum}",
      journal = {\mnras},
         year = 2026,
        month = may,
       volume = {548},
       number = {2},
          eid = {stag577},
        pages = {stag577},
          doi = {10.1093/mnras/stag577},
archivePrefix = {arXiv},
       eprint = {2508.16329},
 primaryClass = {astro-ph.CO},
       adsurl = {https://ui.adsabs.harvard.edu/abs/2026MNRAS.548ag577C}
}

@ARTICLE{Davies25,
       author = {{Davies}, James E. and {Mesinger}, Andrei and {Murray}, Steven},
        title = "{Efficient simulation of discrete galaxy populations and associated radiation fields over the first billion years}",
      journal = {\aap},
         year = 2025,
       volume = {701},
        pages = {A236},
          doi = {10.1051/0004-6361/202554951},
archivePrefix = {arXiv},
       eprint = {2504.17254},
 primaryClass = {astro-ph.CO},
}

@ARTICLE{Dhandha25,
       author = {{Dhandha}, Jiten and {Fialkov}, Anastasia and {Gessey-Jones}, Thomas and {Bevins}, Harry T.~J. and {Tacchella}, Sandro and {Pochinda}, Simon and {Acedo}, Eloy de Lera and {Singh}, Saurabh and {Barkana}, Rennan},
        title = "{Narrowing the discovery space of the cosmological 21-cm signal using multi-wavelength constraints}",
      journal = {\mnras},
         year = 2025,
        month = dec,
       volume = {544},
       number = {2},
        pages = {1608-1626},
          doi = {10.1093/mnras/staf1736},
archivePrefix = {arXiv},
       eprint = {2508.13761},
 primaryClass = {astro-ph.CO},
       adsurl = {https://ui.adsabs.harvard.edu/abs/2025MNRAS.544.1608D}
}

@ARTICLE{HERA25,
       author = {{The HERA Collaboration} and {Abdurashidova}, Zuhra and {Adams}, Tyrone and {Aguirre}, James E. and {Baartman}, Rushelle and {Barkana}, Rennan and {Berkhout}, Lindsay M. and {Bernardi}, Gianni and {Billings}, Tashalee S. and {Bizarria}, Bruno B. and {Bowman}, Judd D. and {Breitman}, Daniela and {Bull}, Philip and {Burba}, Jacob and {Byrne}, Ruby and {Carey}, Steven and {Sushovan Chandra}, Rajorshi and {Chen}, Kai-Feng and {Choudhuri}, Samir and {Cox}, Tyler and {DeBoer}, David R. and {de Lera Acedo}, Eloy and {Dexter}, Matt and {Dhandha}, Jiten and {Dillon}, Joshua S. and {Dynes}, Scott and {Eksteen}, Nico and {Ely}, John and {Ewall-Wice}, Aaron and {Fagnoni}, Nicolas and {Fialkov}, Anastasia and {Furlanetto}, Steven R. and {Gale-Sides}, Kingsley and {Garsden}, Hugh and {Gorce}, Adelie and {Gorthi}, Deepthi and {Halday}, Ziyaad and {Hazelton}, Bryna J. and {Hewitt}, Jacqueline N. and {Hickish}, Jack and {Huang}, Tian and {Jacobs}, Daniel C. and {Josaitis}, Alec and {Kern}, Nicholas S. and {Kerrigan}, Joshua and {Kittiwisit}, Piyanat and {Kolopanis}, Matthew and {Lanman}, Adam and {La Plante}, Paul and {Liu}, Adrian and {Ma}, Yin-Zhe and {MacMahon}, David H.~E. and {Malan}, Lourence and {Malgas}, Cresshim and {Malgas}, Keith and {Marero}, Bradley and {Martinot}, Zachary E. and {McBride}, Lisa and {Mesinger}, Andrei and {Mirocha}, Jordan and {Mohamed-Hinds}, Nicel and {Molewa}, Mathakane and {Morales}, Miguel F. and {Mu{\~n}oz}, Julian B. and {Murray}, Steven G. and {Nikolic}, Bojan and {Nuwegeld}, Hans and {Parsons}, Aaron R. and {Pascua}, Robert and {Patra}, Nipanjana and {Pochinda}, Simon and {Qin}, Yuxiang and {Rath}, Eleanor and {Razavi-Ghods}, Nima and {Riley}, Daniel and {Rosie}, Kathryn and {Santos}, Mario G. and {Singh}, Saurabh and {Storer}, Dara and {Swarts}, Hilton and {Tan}, Jianrong and {Th{\'e}lie}, Emilie and {van Wyngaarden}, Pieter and {Wilensky}, Michael J. and {Williams}, Peter K.~G. and {Zheng}, Haoxuan},
        title = "{First Results from HERA Phase II}",
      journal = {\apj},
         year = 2026,
       volume = {998},
          eid = {33},
        pages = {33},
          doi = {10.3847/1538-4357/ae2d54},
archivePrefix = {arXiv},
       eprint = {2511.21289},
 primaryClass = {astro-ph.CO},
       adsurl = {https://ui.adsabs.harvard.edu/abs/2025arXiv251121289T}
}

@ARTICLE{Hutter25,
       author = {{Hutter}, Anne and {Cueto}, Elie R. and {Dayal}, Pratika and {Gottl{\"o}ber}, Stefan and {Trebitsch}, Maxime and {Yepes}, Gustavo},
        title = "{ASTRAEUS: X. Indications of a top-heavy initial mass function in highly star-forming galaxies from JWST observations at z > 10}",
      journal = {\aap},
         year = 2025,
        month = feb,
       volume = {694},
          eid = {A254},
        pages = {A254},
          doi = {10.1051/0004-6361/202452460},
archivePrefix = {arXiv},
       eprint = {2410.00730},
 primaryClass = {astro-ph.GA},
       adsurl = {https://ui.adsabs.harvard.edu/abs/2025A&A...694A.254H}
}

@ARTICLE{Jeong25,
       author = {{Jeong}, Tae Bong and {Jeon}, Myoungwon and {Song}, Hyunmi and {Bromm}, Volker},
        title = "{Simulating High-redshift Galaxies: Enhancing UV Luminosity with Star Formation Efficiency and a Top-heavy IMF}",
      journal = {\apj},
         year = 2025,
        month = feb,
       volume = {980},
       number = {1},
          eid = {10},
        pages = {10},
          doi = {10.3847/1538-4357/ada27d},
archivePrefix = {arXiv},
       eprint = {2411.17007},
 primaryClass = {astro-ph.GA},
       adsurl = {https://ui.adsabs.harvard.edu/abs/2025ApJ...980...10J}
}

@ARTICLE{Jo25,
       author = {{Jo}, Yongseok and {Genel}, Shy and {Sengupta}, Anirvan and {Wandelt}, Benjamin and {Somerville}, Rachel and {Villaescusa-Navarro}, Francisco},
        title = "{Toward Robustness across Cosmological Simulation Models IllustrisTNG, SIMBA, Astrid, and Swift-Eagle}",
      journal = {\apj},
         year = 2025,
       volume = {991},
       number = {1},
        pages = {120},
          doi = {10.3847/1538-4357/adec78},
archivePrefix = {arXiv},
       eprint = {2502.13239},
 primaryClass = {astro-ph.CO}
}

@ARTICLE{Mertens25,
       author = {{Mertens}, F.~G. and {Mevius}, M. and {Koopmans}, L.~V.~E. and {Offringa}, A.~R. and {Zaroubi}, S. and {Acharya}, A. and {Brackenhoff}, S.~A. and {Ceccotti}, E. and {Chapman}, E. and {Chege}, K. and {Ciardi}, B. and {Ghara}, R. and {Ghosh}, S. and {Giri}, S.~K. and {Hothi}, I. and {H{\"o}fer}, C. and {Iliev}, I.~T. and {Jeli{\'c}}, V. and {Ma}, Q. and {Mellema}, G. and {Munshi}, S. and {Pandey}, V.~N. and {Yatawatta}, S.},
        title = "{Deeper multi-redshift upper limits on the Epoch of Reionization 21-cm signal power spectrum from LOFAR between z=8.3 and z=10.1}",
      journal = {\aap},
         year = 2025,
       volume = {698},
          eid = {A186},
        pages = {A186},
          doi = {10.1051/0004-6361/202554158},
archivePrefix = {arXiv},
       eprint = {2503.05576},
 primaryClass = {astro-ph.CO},
       adsurl = {https://ui.adsabs.harvard.edu/abs/2025arXiv250305576M}
}

@ARTICLE{Nunhokee25,
       author = {{Nunhokee}, C.~D. and {Null}, D. and {Trott}, C. M and {Barry}, N. and {Qin}, Y. and {Wayth}, R.~B. and {Line}, J.~L.~B. and {Jordan}, C.~H. and {Pindor}, B. and {Cook}, J.~H. and {Bowman}, J. and {Chokshi}, A. and {Ducharme}, J. and {Elder}, K. and {Guo}, Q. and {Hazelton}, B. and {Hidayat}, W. and {Ito}, T. and {Jacobs}, D. and {Jong}, E. and {Kolopanis}, M. and {Kunicki}, T. and {Lilleskov}, E. and {Morales}, M.~F. and {Pober}, J.~C. and {Selvaraj}, A. and {Shi}, R. and {Takahashi}, K. and {Tingay}, S.~J. and {Webster}, R.~L. and {Yoshiura}, S. and {Zheng}, Q.},
        title = "{Limits on the 21 cm Power Spectrum at $z=6.5$--$7.0$ from Murchison Widefield Array Observations}",
      journal = {\apj},
         year = 2025,
       volume = {989},
       number = {1},
        pages = {57},
          doi = {10.3847/1538-4357/adda45},
archivePrefix = {arXiv},
       eprint = {2505.09097},
 primaryClass = {astro-ph.CO},
}

@ARTICLE{Ong25,
       author = {{Ong}, Dily Duan Yi and {Handley}, Will},
        title = "{unimpeded: A Public Grid of Nested Sampling Chains for Cosmological Model Comparison and Tension Analysis}",
      journal = {\jcap},
         year = 2026,
       volume = {2026},
       number = {9},
        pages = {026},
          doi = {10.1088/1475-7516/2026/09/026},
archivePrefix = {arXiv},
       eprint = {2511.04661},
 primaryClass = {astro-ph.CO}
}

@ARTICLE{Qin25,
       author = {{Qin}, Yuxiang and {Mesinger}, Andrei and {Prelogovi{\'c}}, David and {Becker}, George and {Bischetti}, Manuela and {Bosman}, Sarah and {Davies}, Frederick and {D'Odorico}, Valentina and {Gaikwad}, Prakash and {Haehnelt}, Martin and {Keating}, Laura and {Lai}, Samuel and {Ryan-Weber}, Emma and {Satyavolu}, Sindhu and {Walter}, Fabian and {Zhu}, Yongda},
        title = "{Percent-level timing of reionisation: Self-consistent, implicit-likelihood inference from XQR-30+ Ly{\ensuremath{\alpha}} forest data}",
      journal = {\pasa},
         year = 2025,
        month = apr,
       volume = {42},
          eid = {e049},
        pages = {e049},
          doi = {10.1017/pasa.2025.35},
archivePrefix = {arXiv},
       eprint = {2412.00799},
 primaryClass = {astro-ph.CO},
       adsurl = {https://ui.adsabs.harvard.edu/abs/2025PASA...42...49Q}
}

@ARTICLE{Somerville25,
       author = {{Somerville}, Rachel S. and {Yung}, L.~Y. Aaron and {Lancaster}, Lachlan and {Menon}, Shyam and {Sommovigo}, Laura and {Finkelstein}, Steven L.},
        title = "{Density-modulated star formation efficiency: implications for the observed abundance of ultraviolet luminous galaxies at z > 10}",
      journal = {\mnras},
         year = 2025,
        month = dec,
       volume = {544},
       number = {4},
        pages = {3774-3798},
          doi = {10.1093/mnras/staf1824},
archivePrefix = {arXiv},
       eprint = {2505.05442},
 primaryClass = {astro-ph.GA},
       adsurl = {https://ui.adsabs.harvard.edu/abs/2025MNRAS.544.3774S}
}

@ARTICLE{Ventura25,
       author = {{Ventura}, Emanuele M. and {Qin}, Yuxiang and {Balu}, Sreedhar and {Wyithe}, J. Stuart B.},
        title = "{Semi-analytic modelling of Pop. III star formation and metallicity evolution - II. Impact on 21cm power spectrum}",
      journal = {\mnras},
         year = 2025,
        month = jun,
       volume = {540},
       number = {1},
        pages = {483-497},
          doi = {10.1093/mnras/staf699},
archivePrefix = {arXiv},
       eprint = {2502.08971},
 primaryClass = {astro-ph.CO},
       adsurl = {https://ui.adsabs.harvard.edu/abs/2025MNRAS.540..483V}
}

@ARTICLE{Weibel25,
       author = {{Weibel}, Andrea and {Jespersen}, Christian Kragh and {Oesch}, Pascal A. and {Williams}, Christina C. and {Bezanson}, Rachel and {Brammer}, Gabriel and others},
        title = "{Exploring Cosmic Dawn with PANORAMIC. II. Cosmic Variance and Galaxy Clustering at $z\sim10$}",
      journal = {\apj},
         year = 2026,
       volume = {1005},
       number = {1},
        pages = {118},
          doi = {10.3847/1538-4357/ae7716},
archivePrefix = {arXiv},
       eprint = {2512.14212},
 primaryClass = {astro-ph.GA}
}

@ARTICLE{Whitler25,
       author = {{Whitler}, Lily and {Stark}, Daniel P. and {Topping}, Michael W. and {Robertson}, Brant and {Rieke}, Marcia and {Hainline}, Kevin N. and {Endsley}, Ryan and {Chen}, Zuyi and {Baker}, William M. and {Bhatawdekar}, Rachana and {Bunker}, Andrew J. and {Carniani}, Stefano and {Charlot}, St{\'e}phane and {Chevallard}, Jacopo and {Curtis-Lake}, Emma and {Egami}, Eiichi and {Eisenstein}, Daniel J. and {Helton}, Jakob M. and {Ji}, Zhiyuan and {Johnson}, Benjamin D. and {P{\'e}rez-Gonz{\'a}lez}, Pablo G. and {Rinaldi}, Pierluigi and {Tacchella}, Sandro and {Williams}, Christina C. and {Willmer}, Christopher N.~A. and {Willott}, Chris and {Witstok}, Joris},
        title = "{The z {\ensuremath{\gtrsim}} 9 Galaxy UV Luminosity Function from the JWST Advanced Deep Extragalactic Survey: Insights into Early Galaxy Evolution and Reionization}",
      journal = {\apj},
         year = 2025,
        month = oct,
       volume = {992},
       number = {1},
          eid = {63},
        pages = {63},
          doi = {10.3847/1538-4357/adfddc},
archivePrefix = {arXiv},
       eprint = {2501.00984},
 primaryClass = {astro-ph.GA},
       adsurl = {https://ui.adsabs.harvard.edu/abs/2025ApJ...992...63W}
}

@ARTICLE{Breitman26,
       author = {{Breitman}, Daniela and {Mesinger}, Andrei and {Murray}, Steven G. and {Nikoli{\'c}}, Ivan and {Trotta}, Roberto},
        title = "{21cmEMUv3: a hybrid diffusion-LSTM emulator of 21cmFAST summary observables}",
      journal = {arXiv e-prints},
         year = 2026,
        month = jun,
          eid = {arXiv:2606.00219},
        pages = {arXiv:2606.00219},
archivePrefix = {arXiv},
       eprint = {2606.00219},
 primaryClass = {astro-ph.CO},
       adsurl = {https://ui.adsabs.harvard.edu/abs/2026arXiv260600219B}
}

@ARTICLE{Chemerynska26,
       author = {{Chemerynska}, Iryna and {Atek}, Hakim and {Furtak}, Lukas J. and {Chisholm}, John and {Endsley}, Ryan and {Kokorev}, Vasily and {Rosdahl}, Joki and {Blaizot}, Jeremy and {Adamo}, Angela and {Bouwens}, Rychard and {Fujimoto}, Seiji and {Korber}, Damien and {Mason}, Charlotte and {McQuinn}, Kristen B.~W. and {Mu{\~n}oz}, Julian B. and {Natarajan}, Priyamvada and {Nelson}, Erica and {Oesch}, Pascal A. and {Pan}, Richard and {Richard}, Johan and {Saldana-Lopez}, Alberto and {Schaerer}, Daniel and {Volonteri}, Marta and {Zitrin}, Adi and {Berg}, Danielle A. and {Claeyssens}, Ad{\'e}la{\"\i}de and {Dessauges-Zavadsky}, Miroslava and {Jecmen}, Michelle and {Labb{\'e}}, Ivo and {Naidu}, Rohan and {Trebitsch}, Maxime},
        title = "{The first GLIMPSE of the faint galaxy population at Cosmic Dawn with JWST: The evolution of the ultraviolet luminosity function across z {\ensuremath{\sim}} 9{\ensuremath{-}}15}",
      journal = {\mnras},
         year = 2026,
        month = feb,
       volume = {546},
       number = {2},
          eid = {staf2267},
        pages = {staf2267},
          doi = {10.1093/mnras/staf2267},
archivePrefix = {arXiv},
       eprint = {2509.24881},
 primaryClass = {astro-ph.GA},
       adsurl = {https://ui.adsabs.harvard.edu/abs/2026MNRAS.546f2267C}
}

@ARTICLE{CruzMartinez26,
       author = {{Cruz-Martinez}, Juan and {Cuesta-Lazaro}, Carolina and {Held}, Alexander and {Kagan}, Michael},
        title = "{Unknown Unknowns: Model Misspecification in Machine Learning for Physics}",
      journal = {arXiv e-prints},
         year = 2026,
        month = aug,
          eid = {arXiv:2608.13633},
        pages = {arXiv:2608.13633},
          doi = {10.48550/arXiv.2608.13633},
archivePrefix = {arXiv},
       eprint = {2608.13633},
 primaryClass = {physics.data-an},
       adsurl = {https://ui.adsabs.harvard.edu/abs/2026arXiv260813633C}
}

@ARTICLE{Kar26,
       author = {{Kar}, Abhijnan and {Alam}, Shadab and {Silk}, Joseph},
        title = "{Beyond Extreme Burstiness: Evolving Star Formation Efficiency as the Key to Early Galaxy Abundance}",
      journal = {\apj},
         year = 2026,
        month = jan,
       volume = {996},
       number = {1},
          eid = {103},
        pages = {103},
          doi = {10.3847/1538-4357/ae1f8b},
archivePrefix = {arXiv},
       eprint = {2507.20606},
 primaryClass = {astro-ph.GA},
       adsurl = {https://ui.adsabs.harvard.edu/abs/2026ApJ...996..103K}
}

@ARTICLE{Krishna26,
       author = {{Krishna}, Anoop and {Moorkanat}, Deepthi and {Hiten} and {Mondal}, Rajesh},
        title = "{Constraining the neutral hydrogen fraction during reionization: Cross-simulation inference using power spectrum and bispectrum}",
      journal = {\mnras},
         year = 2026,
       volume = {550},
       number = {2},
        pages = {stag1152},
          doi = {10.1093/mnras/stag1152},
archivePrefix = {arXiv},
       eprint = {2603.00531},
 primaryClass = {astro-ph.CO}
}

@ARTICLE{McLeod26,
       author = {{McLeod}, D.~J. and {Dunlop}, J.~S. and {McLure}, R.~J. and {Donnan}, C.~T. and {Begley}, R. and {Antonogiannaki}, S. and {Magee}, D. and {Illingworth}, G.~D. and {Arrabal Haro}, P. and {Bondestam}, C. and {Carnall}, A.~C. and {Cullen}, F. and {Dickinson}, M. and {Ellis}, R.~S. and {Frye}, B.~L. and {Golawska}, H. and {Grogin}, N.~A. and {Holst}, I.~J.~B. and {Kamieneski}, P.~S. and {Leung}, H.-H. and {Liu}, F.-Y. and {Stanton}, T.~M. and {Tittley}, E.~R.},
        title = "{A search for the first galaxies across $>0.6$ deg$^2$ of JWST imaging: new evidence for a rapid decline in star-formation activity at $z>12$}",
      journal = {arXiv e-prints},
         year = 2026,
        month = apr,
          eid = {arXiv:2604.16666},
        pages = {arXiv:2604.16666},
          doi = {10.48550/arXiv.2604.16666},
archivePrefix = {arXiv},
       eprint = {2604.16666},
 primaryClass = {astro-ph.GA},
       adsurl = {https://ui.adsabs.harvard.edu/abs/2026arXiv260416666M}
}

@ARTICLE{Naidu26,
       author = {{Naidu}, Rohan P. and {Oesch}, Pascal A. and {Brammer}, Gabriel and {Weibel}, Andrea and {Li}, Yijia and {Matthee}, Jorryt and {Chisolm}, John and {Pollock}, Clara L. and {Heintz}, Kasper E. and {Johnson}, Benjamin D. and {Shen}, Xuejian and {Hviding}, Raphael E. and {Leja}, Joel and {Tacchella}, Sandro and {Ganguly}, Arpita and {Witten}, Callum and {Atek}, Hakim and {Belli}, Siro and {Bose}, Sownak and {Bouwens}, Rychard and {Dayal}, Pratika and {Decarli}, Roberto and {de Graaff}, Anna and {Fudamoto}, Yoshinobu and {Giovinazzo}, Emma and {Greene}, Jenny E. and {Illingworth}, Garth and {Inoue}, Akio K. and {Kane}, Sarah G. and {Labbe}, Ivo and {Leonova}, Ecaterina and {Marques-Chaves}, Rui and {Meyer}, Roman A. and {Nelson}, Erica J. and {Roberts-Borsani}, Guido and {Schaerer}, Daniel and {Simcoe}, Robert A. and {Stefanon}, Mauro and {Sugahara}, Yuma and {Toft}, Sune and {van der Wel}, Arjen and {van Dokkum}, Pieter and {Walter}, Fabian and {Watson}, Darrach and {Weaver}, John R. and {Whitaker}, Katherine E.},
        title = "{A Cosmic Miracle: A Remarkably Luminous Galaxy at zspec = 14.44 Confirmed with JWST}",
      journal = {The Open Journal of Astrophysics},
         year = 2026,
        month = jan,
       volume = {9},
        pages = {56033},
          doi = {10.33232/001c.156033},
archivePrefix = {arXiv},
       eprint = {2505.11263},
 primaryClass = {astro-ph.GA},
       adsurl = {https://ui.adsabs.harvard.edu/abs/2026OJAp....956033N}
}

@ARTICLE{Narayanan26,
       author = {{Narayanan}, Desika and {Torrey}, Paul and {Stark}, Daniel and {Chisholm}, John and {Finkelstein}, Steven and {Garcia}, Alex and {Kelley-Derzon}, Jessica and {Marinacci}, Federico and {Sales}, Laura and {Savitch}, Ethan and {Vogelsberger}, Mark and {Zimmerman}, Dhruv},
        title = "{The Growth of Dust in Galaxies in the First Billion Years with Applications to Blue Monsters}",
      journal = {The Open Journal of Astrophysics},
         year = 2026,
        month = apr,
       volume = {9},
        pages = {59986},
          doi = {10.33232/001c.159986},
archivePrefix = {arXiv},
       eprint = {2509.18266},
 primaryClass = {astro-ph.GA},
       adsurl = {https://ui.adsabs.harvard.edu/abs/2026OJAp....959986N}
}

@ARTICLE{NikolicMason26,
       author = {{Nikoli{\'c}}, Ivan and {Mason}, Charlotte A.},
        title = "{Constraining the galaxy-halo connection at $z>10$ using the neighbors of UV-bright galaxies}",
      journal = {arXiv e-prints},
         year = 2026,
        month = sep,
          eid = {arXiv:2609.33817},
        pages = {arXiv:2609.33817},
archivePrefix = {arXiv},
       eprint = {2609.33817},
 primaryClass = {astro-ph.GA},
       adsurl = {https://ui.adsabs.harvard.edu/abs/2026arXiv260933817N}
}

@ARTICLE{Pietschke26,
       author = {{Pietschke}, Yannic and {Heneka}, Caroline and {Ore}, Ayodele and {Meriot}, Romain},
        title = "{Cross-simulator transfer with foundation model summaries: Towards robust SKA-era reionization inference}",
      journal = {arXiv e-prints},
         year = 2026,
        month = aug,
          eid = {arXiv:2608.26354},
        pages = {arXiv:2608.26354},
          doi = {10.48550/arXiv.2608.26354},
archivePrefix = {arXiv},
       eprint = {2608.26354},
 primaryClass = {astro-ph.CO},
       adsurl = {https://ui.adsabs.harvard.edu/abs/2026arXiv260826354P}
}

@ARTICLE{Lemos21,
       author = {{Lemos}, P. and {Raveri}, M. and {Campos}, A. and {Park}, Y. and {Chang}, C. and {Weaverdyck}, N. and {Huterer}, D. and {Liddle}, A.~R. and {Blazek}, J. and {Cawthon}, R. and others},
        title = "{Assessing tension metrics with dark energy survey and Planck data}",
      journal = {\mnras},
         year = 2021,
       volume = {505},
       number = {4},
        pages = {6179-6194},
          doi = {10.1093/mnras/stab1670},
archivePrefix = {arXiv},
       eprint = {2012.09554},
 primaryClass = {astro-ph.CO}
}

@ARTICLE{YeLin25,
       author = {{Ye}, Gen and {Lin}, Shi-Jie},
        title = "{On the tension between DESI DR2 BAO and CMB}",
      journal = {arXiv e-prints},
         year = 2025,
        month = may,
          eid = {arXiv:2505.02207},
archivePrefix = {arXiv},
       eprint = {2505.02207},
 primaryClass = {astro-ph.CO}
}

@ARTICLE{Ong25b,
       author = {{Ong}, Dily Duan Yi and {Yallup}, David and {Handley}, Will},
        title = "{A Bayesian Perspective on Evidence for Evolving Dark Energy}",
      journal = {arXiv e-prints},
         year = 2025,
        month = nov,
          eid = {arXiv:2511.10631},
archivePrefix = {arXiv},
       eprint = {2511.10631},
 primaryClass = {astro-ph.CO}
}

@ARTICLE{Ong26,
       author = {{Ong}, Dily Duan Yi and {Yallup}, David and {Handley}, Will},
        title = "{The Bayesian view of DESI DR2 with unimpeded: Evidence and tension in a combined analysis with CMB and supernovae across cosmological models}",
      journal = {arXiv e-prints},
         year = 2026,
        month = mar,
          eid = {arXiv:2603.05472},
archivePrefix = {arXiv},
       eprint = {2603.05472},
 primaryClass = {astro-ph.CO}
}

@ARTICLE{Trotta08,
       author = {{Trotta}, Roberto},
        title = "{Bayes in the sky: Bayesian inference and model selection in cosmology}",
      journal = {Contemporary Physics},
         year = 2008,
       volume = {49},
       number = {2},
        pages = {71-104},
          doi = {10.1080/00107510802066753},
archivePrefix = {arXiv},
       eprint = {0803.4089},
 primaryClass = {astro-ph}
}

@ARTICLE{Starkman08,
       author = {{Starkman}, Glenn D. and {Trotta}, Roberto and {Vaudrevange}, Pascal M.},
        title = "{Introducing doubt in Bayesian model comparison}",
      journal = {arXiv e-prints},
         year = 2008,
        month = nov,
          eid = {arXiv:0811.2415},
          doi = {10.48550/arXiv.0811.2415},
archivePrefix = {arXiv},
       eprint = {0811.2415},
 primaryClass = {physics.data-an}
}

\end{document}